\documentclass[longauth]{aa} 

\usepackage[varg]{txfonts}
\usepackage{xcolor}

\usepackage{graphicx}
\usepackage{txfonts}
\begin{document}

   \title{The rise and fall of the iron-strong nuclear transient PS16dtm}


   \author{T. Petrushevska
          \inst{\ref{CAC}}
          \and
          G. Leloudas\inst{\ref{DTU}}
                   \and
         D. Ili\'{c}\inst{\ref{Belgrade},\ref{HAmburg}}
                \and 
          M. Bronikowski\inst{\ref{CAC}} 
                 \and
          P. Charalampopoulos\inst{\ref{DTU},\ref{Turku}}
                   \and
         G. K. Jaisawal\inst{\ref{DTU}}
                   \and 
          E. Paraskeva \inst{\ref{Greece1},\ref{Greece2}}
         \and
          M. Pursiainen\inst{\ref{DTU}}
                    \and
         N. Raki\'{c}\inst{\ref{Belgrade},\ref{BanjaLuka}}  
                   \and
          S. Schulze\inst{\ref{OKCphysics}}
                    \and
          K. Taggart\inst{\ref{SantaCruz}}
	\and
          C. K. Wedderkopp\inst{\ref{DTU}}
          \and
          J. P. Anderson\inst{\ref{ESOchile}}  \and
          T. de Boer\inst{\ref{Hawai}}
          \and K. Chambers\inst{\ref{Hawai}} \and
T. W. Chen \inst{\ref{OKCastro}}  \and 
G. Damljanovi\'{c} \inst{\ref{BelgradeObs}} \and
M. Fraser \inst{\ref{Dublin4}}  \and
H. Gao \inst{\ref{Hawai}}          \and
          A. Gomboc \inst{\ref{CAC}}\and
M. Gromadzki \inst{\ref{Warwaw}}  \and
N. Ihanec \inst{\ref{Warwaw},\ref{ESOchile}} \and
K. Maguire \inst{\ref{Dublin}}           \and 
B. Mar\v{c}un\inst{\ref{CAC}}    \and
T. E. M\"uller-Bravo \inst{\ref{Barcelona1},\ref{Barcelona2}}  \and
M. Nicholl \inst{\ref{Birmingham}}  \and
F. Onori \inst{\ref{Teramo}}  \and
T. M. Reynolds \inst{\ref{dawn}}  \and
S. J. Smartt \inst{\ref{Belfast}}  \and
J. Sollerman \inst{\ref{OKCastro}}  \and
K. W. Smith \inst{\ref{Belfast}}  \and
T. Wevers \inst{\ref{ESOchile}}  \and
\L{}. Wyrzykowski \inst{\ref{Warwaw}}
          }
   \institute{Center for Astrophysics and Cosmology, University of Nova Gorica, Vipavska 11c, 5270 Ajdov\v{s}\v{c}ina, Slovenia.\\
              \email{tanja.petrushevska@ung.si}\label{CAC}
         \and
            DTU Space, National Space Institute, Technical University of Denmark, Elektrovej 327, 2800 Kgs. Lyngby, Denmark.\\
             \email{giorgos@space.dtu.dk}\label{DTU}
          \and{Department of Astronomy, University of Belgrade - Faculty of Mathematics, Studentski trg 16, 11000 Belgrade, Serbia.} \\ \email{dilic@matf.bg.ac.rs}\label{Belgrade}
                    \and{Hamburger Sternwarte, Universitat Hamburg, Gojenbergsweg 112, 21029 Hamburg, Germany.}\label{HAmburg} 
                    \and{Tuorla Observatory, Department of Physics and Astronomy, University of Turku, FI-20014 Turku, Finland
}\label{Turku}
                    \and{IAASARS, National Observatory of Athens, 15236 Penteli, Greece}\label{Greece1} \and{Department of Astrophysics, Astronomy \& Mechanics, Faculty of Physics,
National and Kapodistrian University of Athens, 15784 Athens, Greece.}\label{Greece2}
\and{Faculty of Natural Sciences and Mathematics, University of Banjaluka, Mladena Stojanovi\'{c}a 2, 78000 Banjaluka, Bosnia and Herzegovina.}\label{BanjaLuka}
\and{Oskar Klein Centre, Department of Physics, Stockholm University, SE-10691 Stockholm, Sweden.}\label{OKCphysics}
          \and{Department of Astronomy and Astrophysics, University of California, Santa Cruz, CA 95064, USA.}\label{SantaCruz}
                    \and{European Southern Observatory, Alonso de C\'ordova 3107, Casilla 19, Santiago, Chile.}\label{ESOchile}
              \and{Institute for Astronomy, University of Hawaii, 2680 Woodlawn Drive, Honolulu HI 96822, USA.}\label{Hawai}                
          \and{Oskar Klein Centre, Department of Astronomy, Stockholm University, AlbaNova, SE-10691 Stockholm, Sweden.}\label{OKCastro}                    
          \and{Astronomical Observatory, Volgina 7, 11060 Belgrade, Serbia.}\label{BelgradeObs}
     \and{School of Physics, Trinity College Dublin, The University of Dublin, Dublin 2, Ireland.}\label{Dublin}     
         \and{Astronomical Observatory, University of Warsaw, Al. Ujazdowskie 4, 00-478 Warszawa, Poland.}\label{Warwaw} 
                   \and{Institute of Space Sciences (ICE, CSIC), Campus UAB, Carrer de Can Magrans, s/n, E-08193 Barcelona, Spain.}\label{Barcelona1}
          \and{Institut d’Estudis Espacials de Catalunya (IEEC), E-08034 Barcelona, Spain.}\label{Barcelona2}
                    \and{Birmingham Institute for Gravitational Wave Astronomy and School of Physics and Astronomy, University of Birmingham, Birmingham B15 2TT, UK.}\label{Birmingham}
                    \and{INAF-Osservatorio Astronomico d'Abruzzo, via M. Maggini snc, I-64100 Teramo, Italy.}\label{Teramo}
                    \and{Cosmic DAWN centre, Niels Bohr Institute, University of Copenhagen, R{\aa}dmandsgade 62-64, 2200, Copenhagen, Denmark.}\label{dawn}
          \and{Astrophysics Research Centre,  School of Mathematics and Physics,  Queen's University Belfast,  Belfast, BT7 1NN,  Northern Ireland, UK.}\label{Belfast}
          \and{School of Physics, O'Brien Centre for Science North, University College Dublin, Belfield, Dublin 4, Ireland}\label{Dublin4}
             }
   \date{Received XX; accepted XX}

 
  \abstract
   {Thanks to the advent of large-scale optical surveys, a diverse set of flares from the nuclear regions of galaxies has recently been discovered. These include
   the disruption of stars by supermassive black holes at the centers of galaxies - nuclear transients known as tidal disruption events (TDEs). Active galactic nuclei (AGN) can show extreme changes in the brightness and emission line intensities, often referred to as changing-look AGN (CLAGN).  Given the physical and observational similarities, the interpretation and distinction of nuclear transients as CLAGN or TDEs remains difficult. One of the obstacles of making progress in the field is the lack of well-sampled data of long-lived
nuclear outbursts in AGN.}
   {Here, we study PS16dtm, a nuclear transient in a Narrow Line Seyfert 1 (NLSy1) galaxy, which has been proposed to be a TDE candidate. Our aim is to study the spectroscopic and photometric properties of PS16dtm, in order to better understand the outbursts originating in NLSy1 galaxies.
  }
  {Our extensive multiwavelength follow-up that spans around $2000$ days includes photometry and spectroscopy in the UV/optical, as well as mid-infrared (MIR) and X-ray observations. Furthermore, we improved an existing semiempirical model in order to reproduce the spectra and study the evolution of the spectral lines. 
  }
   {The UV/optical light curve shows a double peak at $\sim50$ and $\sim100$ days after the first detection, and it declines and flattens afterward, reaching preoutburst levels after 2000 days of monitoring. The MIR light curve rises almost simultaneously with the optical, but unlike the UV/optical which is approaching the preoutburst levels in the last epochs of our observations, the MIR emission is still rising at the time of writing.
   The optical spectra show broad Balmer features and the strongest broad Fe II emission  ever detected in a nuclear transient. This broad Fe II emission was not present in the archival preoutburst spectrum and almost completely disappeared +1868 days after the outburst.
   We found that the majority of the flux of the broad Balmer and Fe II lines is produced by photoionization.
    We detect only weak X-ray emission in the 0.5-8 keV band at the location of PS16dtm, at +848, +1130, and +1429 days past the outburst. This means that the X-ray emission continues to be lower by at least an order of magnitude, compared to archival, preoutburst measurements. }
   {We confirm that the observed properties of PS16dtm are difficult to reconcile with normal AGN variability. The TDE scenario continues to be a plausible explanation for the observed properties, even though PS16dtm shows differences compared to TDE in quiescent galaxies. We suggest that this event is part of a growing sample of TDEs that show broad Balmer line profiles and Fe II complexes. We argue that the extreme variability seen in the AGN host due to PS16dtm may have easily been misclassified as a CLAGN, especially if the rising part of the light curve had been missed. 
   This implies that some changing look episodes in AGN may be triggered by TDEs. Imaging and spectroscopic data of AGN with good sampling are needed to enable testing of possible physical mechanisms behind the extreme variability in AGN. }


   \maketitle
%

\section{Introduction}\label{sec:intro}
In recent years, the detection of a variety of nuclear transients has become attainable due to the advent of large-scale optical robotic surveys such as the Asteroid Terrestrial-impact Last Alert System   \citep[ATLAS;][]{2011PASP..123...58T}, the All-Sky Automated Survey for Supernovae \citep[ASAS-SN;][]{2014ApJ...788...48S},  the Pan-STARRS Survey for Transients \citep[PSST;][]{2016arXiv161205560C}, the Gaia Photometric Science Alerts program \citep{2021A&A...652A..76H}, the
intermediate Palomar Transient Factory \citep[iPTF;][] {2013ATel.4807....1K}, and, its successor, the Zwicky Transient Facility \citep[ZTF;][]{2019PASP..131f8003B}. It has long been known that galaxies with an active galactic nucleus (AGN), where matter is accreted onto the central supermassive black hole (SMBH), show variability in their brightness \citep[e.g.,][]{1967ApJ...150L..67F}. Apart from regular, low-level stochastic variability, some AGN occasionally show exceptionally large changes in the luminosity, spectral shape, and/or X-ray absorption \citep[e.g.,][]{Frederick_2019,MacLeod_2019,Sanchez_2018,Zhang_2021,2022MNRAS.511...54H, 2022ApJ...933..180G, 2022ApJ...925...50R}. The most notable of these changes is when a Seyfert 1 type AGN transforms into a Seyfert 2 AGN and vice versa, the so-called changing-look AGN (CLAGN). The physical explanation of this phenomenon is still debated. It has been argued that the variability arises in the intrinsic changes in the accretion state of the SMBH, which include several scenarios \citep{2018MNRAS.480.3898N,2020A&A...641A.167S,Stern_2018,2022ApJ...925...84M}. A number of works have argued that there might be more than one mechanism that explains CLAGN, given the diversity in the timescale of changes, and their amplitude \citep[e.g.,][]{Wang_2012,2015A&A...581A..17C,2017IAUS..324..168K,2019ApJ...883...94T,2022AN....34310065S}.

One important class of optical transients occurring in the nuclear regions of the host galaxies are the flares from tidally disrupted stars by SMBHs \citep{1988Natur.333..523R,Phinney_1989}. They are particularly interesting since there are several suggestions of using tidal disruption events (TDEs) to study SMBH properties such as its mass \citep{2019ApJ...872..151M} and perhaps even more elusive, its spin \citep{Leloudas_2016,2019MNRAS.487.4790G}. Optically discovered TDEs differ between each other in the spectroscopic features, shape, and timescale of the light curve, and have been detected in both quiescent \citep[e.g.,][]{Gezari_2021} and active galaxies \citep[e.g.,][]{2017MNRAS.465L.114W,2021ApJ...920...56F}. The discoveries of TDEs in AGN were initiated later compared to TDEs in quiescent galaxies, partly because, initially, transients in galaxies with known AGN have been excluded in optical surveys from follow-up investigation, so as to avoid being overwhelmed by spurious candidates. Furthermore, TDEs arising in a galaxy with an existing AGN are more difficult to identify simply because the nucleus itself can be a variable source. 

In recent years, transient phenomena in galaxies with AGN have been observed, with extraordinary changes in photometric and spectroscopic properties on short timescales \citep[e.g.,][]{2021ApJ...920...56F}. However, the physical mechanism to explain these phenomena is unclear. One example is the transient AT~2018dyk, initially classified as a TDE \citep{2018ATel11953....1A}; however, a subsequent study argued that it is a CLAGN \citep{Frederick_2019}. Another example is the transient CSS100217 which \citet{Drake_2011} argued was a supernova, while \citet{2021ApJ...920...56F} favored the AGN scenario. Furthermore, \citet{2022A&A...660A.119Z} and \citet{Cannizzaro_2022} suggested that it could be explained as a TDE, after gathering data for more than 10 years. The reason for this ambiguity is that, at present, there is no single observational signature that disentangles, unambiguously, the physical mechanism that causes the luminous {phenomena}  \citep[see][for a recent review]{2021SSRv..217...54Z}. Since a TDE happening in AGN can also cause intrinsic change in the accretion disk, it can also provide one of the channels to explain CLAGN \citep[see e.g.,][]{Kool_2020,Cannizzaro_2020,Zhang_2021,2022ApJ...930...12H}. An ongoing issue in understanding long-lived nuclear outbursts in AGN is, on the one hand, the low numbers of such events, and on the other hand, often the data are sparsely sampled from the long-lived nuclear outbursts, so it does not allow for one to study them in detail.

Here, we focus on the nuclear transient, PS16dtm, for which we gathered an extensive dataset over a time scale of almost six years. PS16dtm was first classified as supernova Type IIn \citep{2016ATel.9417....1T,2016ATel.9843....1D}, but further in-depth analysis based on follow-up observations concluded that it is more consistent with a TDE interpretation \citep[][hereafter B17]{Blanchard_2017}. B17 presented observations up to 200 days after it was detected. In this paper, we present photometric and spectroscopic data for a period of $\sim2000$ days after the first detection of PS16dtm. 
As we subsequently show in this paper, PS16dtm is perhaps the nuclear transient with the most dramatic increase of Fe II emission after the outburst, although there are other examples, albeit not with such strong Fe II emission, such as J123359.12+084211.5 \citep{MacLeod_2019}, AT~2019dsg \citep{Cannizzaro2021}, AT~2018fyk \citep{Wevers_2019}, CSS100217 \citep{Drake_2011}, and PS1-10adi \citep{2017NatAs...1..865K}. We also show that PS16dtm exhibits strong MIR emission similar to other nuclear flares in hosts with AGN. Intriguingly, the X-ray emission continues to be dimmed compared the archival preoutburst state, despite that UV/optical photometry and spectroscopy are approaching the preoutburst levels.

The paper is organized as follows. In Sect.~\ref{sec:previous}, we summarize the findings on PS16dtm that have been previously published. In Sect.~\ref{sec:obs} we present our observations,  while in Sect.~\ref{sec:analys} we perform the analysis. The discussion is found in Sect.~\ref{sec:disc} and finally, in Sect.~\ref{sec:conc} we present our conclusions. We use the AB magnitude system throughout this work. Furthermore, we assume a cosmology with $H_0=67$ km s$^{-1}$ Mpc$^{-1}$, $\Omega_m =0.32$ and $\Omega_{\Lambda} =0.68$ as in B17, which implies a luminosity distance of 381 Mpc to PS16dtm. The Galactic extinction along the line of sight to PS16dtm is $\rm A_V=0.069\pm 0.001$ mag \citep{2011ApJ...737..103S}.

\section{Summary of previous work on PS16dtm} \label{sec:previous}
In this section, we summarize previously published work on PS16dtm.\footnote{PS16dtm is registered as SN~2016ezh on the Transient Name Server\url{https://www.wis-tns.org/object/2016ezh}. We use the discovery name throughout this paper as the supernova interpretation has been disfavored.}
It was discovered by PSST on 12 August 2016 (MJD 57612) \citep{2016TNSTR.562....1C}, and it was also observed independently by ATLAS and ASSA-SN. The host galaxy of PS16dtm, SDSSJ015804.75-005221.8, is a narrow-line Seyfert 1 (NLSy1) galaxy at $z=0.0804$ with a SMBH with $\sim 10^6M_{\odot}$, measured by the velocity dispersion method \citep{2011ApJ...739...28X} and from the AGN luminosity and the radius of the broad-line region (BLR) (B17). B17 reported the optical and UV light curves of  PS16dtm which brightened approximately two magnitudes above the archival host brightness in $\sim 50$ days, with two recognizable  peaks distanced $\sim 50$ days between them. Furthermore, the light curves showed little color evolution, and stayed approximately at the Eddington luminosity of the SMBH. 

B17 considered scenarios of a variable AGN, a Type IIn supernova, and a TDE as possible explanations for PS16dtm. They concluded that the small evolution in brightness and color seen in PS16dtm, is not typical for supernovae where cooling is expected due to expansion and radiative losses, but it is similar to that of TDEs. Another argument in favor of the TDE scenario put forward by B17 is the short timescale of the rising part of the PS16dtm light curve ($\sim 50$ days) and the amplitude (2 magnitudes rise over the host level). The typical variability amplitudes of AGN are low in large AGN samples and they vary on timescales of years \citep[$\sim 0.1-0.2$ magnitudes per year, see e.g.][]{Sanchez_2018}. 
Another interesting aspect of PS16dtm is that after the outburst, X-ray emission from the nucleus decreased by at least an order of magnitude, compared to archival measurements \citep{Pons_2014}. B17 argued that neither the supernova scenario nor AGN variability can explain the X-ray dimming, but that it could be explained by the accreted stellar debris that obscures the X-ray emitting region of the AGN accretion disk. They proposed a simple model which assumes a face-on orientation for the preexisting, X-ray emitting disk, and a nearly edge-on orientation for the disk newly formed in the disruption, which also blocks the X-ray emission. They argued that such a geometrical configuration can also explain the spectral properties of both the debris disk and the host galaxy. 

The spectra shown in B17 are dominated by hydrogen Balmer and Fe II emission lines. B17 compared the PS16dtm spectra to those of Type IIn supernovae, and found no reasonable matches. B17 interpreted the spectra as being more similar to those of some NLSy1 galaxies, as optical Fe II (4000-5400 \AA) emission are common features in NLSy1 spectra.

PS16dtm also flared in the mid-infrared (MIR) and it was detected by WISE as part of the NEOWISE survey \citep[][J17; hereafter]{2017ApJ...850...63J}. J17 reported three NEOWISE detections starting 11 days before the optical ASSA-SN survey and extending to 327 days after.
The authors conclude that the MIR flare is consistent with a dust echo, despite the fact that the MIR data shows no delay with respect to the optical. They estimated  that the inner radius of the preexisting dust torus increased from $\sim 10$ light days to $\sim 70$ light days, where the emptied region was replaced with a gas torus. They also argued that the detected Fe II emission lines are produced in the gas coming from the evaporated dust due to the strong radiation field.

The peculiarity of PS16dtm has prompted other authors to examine what powers its luminosity. \citet{2021ApJ...920...56F}, who looked at NLSy1 transients from the literature, put PS16dtm in the class of TDE with strong Balmer and Fe II complexes, despite that it does not satisfy all their defined requirements for TDE classification. \citet{Moriya_2017} argued that PS16dtm can be explained by changes related to the AGN, and not as a result of tidal disruption of a star by the SMBH. They explain it as an interaction between  accretion disk winds and clouds in the BLRs surrounding them. 
They also argue that the observed broad ($\sim10,000$ $\rm km\  s^{-1}$) Mg II absorption in the UV (B17) could be related to the fast SMBH disk wind. Nevertheless, their model predicts that the emission timescale is $\sim110$ days, while PS16dtm has stayed bright above the host-galaxy baseline for much longer, as will be shown in the next sections.

\section{Observations} \label{sec:obs}
\begin{figure*}
\includegraphics[width=16cm]{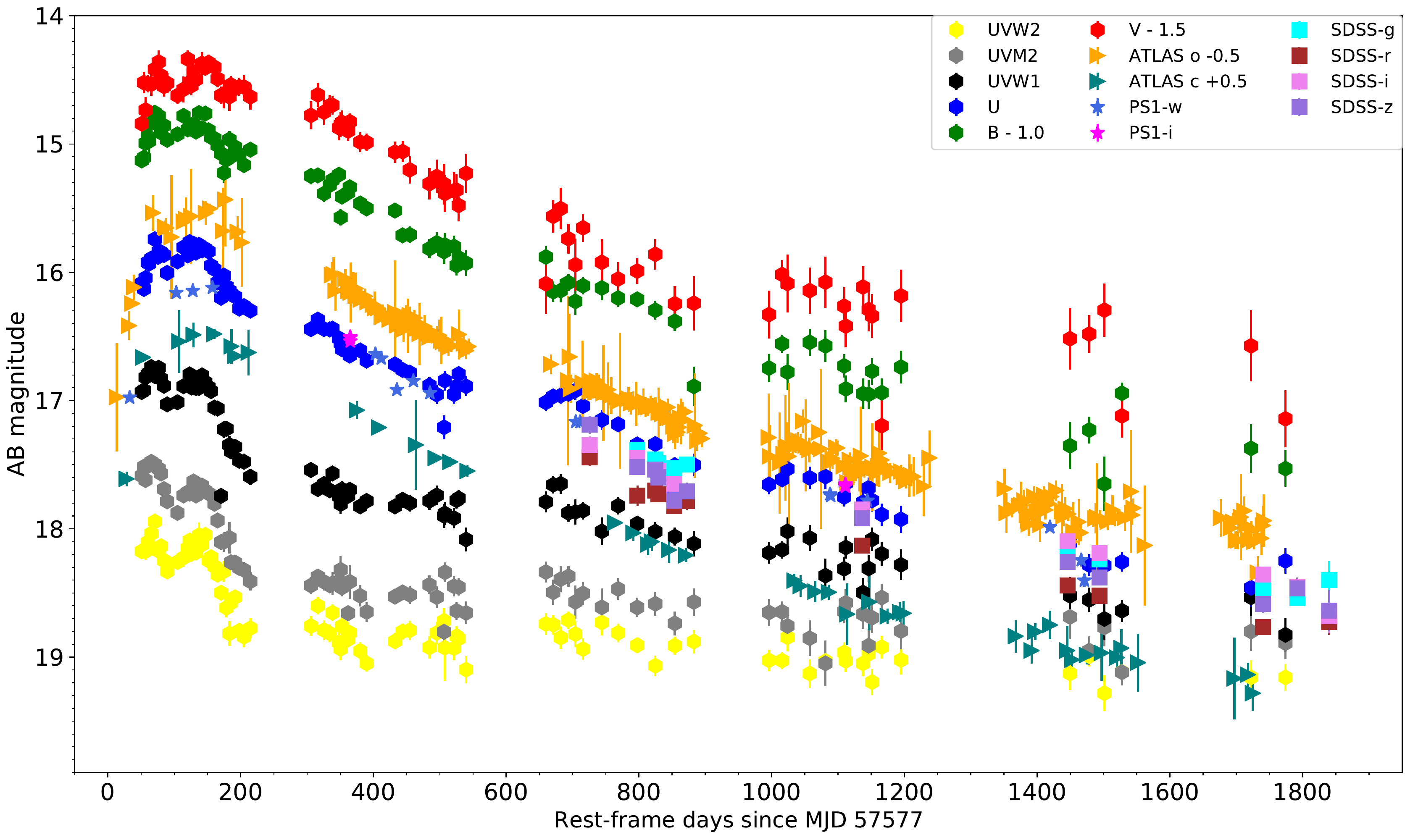}
\caption{Photometric evolution of PS16dtm. The photometry has been corrected for the Galactic extinction and the host contribution has been subtracted. The reference MJD for the outburst is 57577 obtained from a second order polynomial fit to the ATLAS $o$ photometry in the rising part of the light curve. The photometry has been arbitrarily shifted in the y axis for easier viewing, as indicated in the legend.}
\label{photometry}
\end{figure*}

In this section, we present new observations of PS16dtm out to $\sim2000$ days after discovery. We also present ATLAS and PSST data of PS16dtm for the first time. Notably, ATLAS detected PS16dtm before the PSST and ASSA-SN surveys, and this enabled us to roughly estimate the time of outburst, as it will be shown in Sec. \ref{sec:atlas}. The evolution of the host-subtracted photometry corrected for Milky Way extinction is shown in Figs.~\ref{photometry} and \ref{wise_plus_optical}. The gaps in the data (from February to August) are when PS16dtm was behind the Sun. We note that at $\sim2000$ days after the discovery, PS16dtm is fading and the emission is approaching preoutburst, host levels \citep[see][for measured and synthetically computed PS16dtm host magnitudes]{2021ApJ...910...83H}. All our photometry will be available through the WISeREP archive\footnote{\url{https://www.wiserep.org}} \citep{Yaron_2012}.

\subsection{\textit{Swift} observations}
PS16dtm was monitored by the Neil Gehrels \textit{Swift} Observatory  \citep{2004ApJ...611.1005G} with the UV/Optical Telescope \citep[UVOT][]{2005SSRv..120...95R} in six filters; three optical ($V$ at 5468 $\AA$, $B$ at 4392 $\AA$, $U$ at 3465 $\AA$), and three near-UV ($UVW1$ at 2600 $\AA$, $UVM2$ at 2246 $\AA$, and $UVW2$ at 1928 $\AA$). There are 83 epochs of UVOT/\textit{Swift} observations spanning from MJD 57632 to MJD 59493.
The \textit{Swift}/UVOT observations were reduced following the standard pipeline from HEAsoft\footnote{\url{https://heasarc.gsfc.nasa.gov/docs/software/lheasoft/}}. The photometry was extracted using the tool \emph{UVOTSOURCE} and a source extraction region with a radius of 5". We corrected for Galactic extinction and subtracted the contribution from the host galaxy, for which we used the PS16dtm host galaxy magnitudes from \citet{2021ApJ...910...83H}. \citet{2021ApJ...910...83H} provided corrections to the NUV photometry of TDEs published after 2015 in the literature, which were using Swift/UVOT. Their work was motivated by an update by the Swift team to the UVOT calibration to correct for the loss of sensitivity over time. \citet{2021ApJ...910...83H} fitted archival multiwavelength photometry from GALEX, 2MASS, SDSS and WISE of the host galaxy and modeled the spectral energy distribution (SED) of the host galaxy, from which they updated the magnitudes in the NUV UVOT filters. Therefore, the NUV Swift/UVOT magnitudes of the PS16dtm host galaxy are different to those used in B17, leading to an average difference in the host-subtracted photometry of PS16dtm compared to the ones in B17 by $0.58$ magnitudes for $UVW2$ and by $0.02-0.07$ magnitudes for $U$, $B$, $V$, $UVW1$ and $UVM2$.

Simultaneous with the \textit{Swift}/UVOT observations,  PS16dtm was observed with the \textit{Swift} X-ray Telescope (XRT). After building the XRT light-curve using the online tool \footnote{\url{https://www.swift.ac.uk/user_objects/}}
we conclude that there was no confident detection by \textit{Swift}/XRT and we can only place upper-limits to the X-ray emission, similar to those presented in B17 ($F_X<1\times$10$^{-14}$~erg~s$^{-1}$~cm$^{-2}$ and $L_X <1.7\times$10$^{41}$~erg~s$^{-1}$).

\subsection{Optical ground-based photometry}

\subsubsection{ATLAS}\label{sec:atlas}
ATLAS is a robotic survey that uses at least two identical 50-centimeter telescopes, located at 
Haleakala and Mauna Loa observatories in Hawaii. PS16dtm was observed by ATLAS in two bands, $c$ (\textit{cyan}) and $o$ (\textit{orange}), that cover the wavelength range of $4200-6500$ and $5600- 8200$ $\AA$, respectively. The ATLAS image processing is done with a fully automated pipeline that performs flat fielding, astrometric calibration, and photometric calibration. The photometry of PS16dtm is publicly available online, of which we used the reduced photometry from the latest ATLAS data release \citep{2018PASP..130f4505T, 2020PASP..132h5002S}. Since the reference images contain the flux from PS16dtm, we calculated the mean flux between MJD 57250 and 57500 (prediscovery epochs) and 
subtract the flux from all the data subsequently.  As visible from Fig.~\ref{wise_plus_optical}, the first ATLAS detection of PS16dtm was made at MJD 57591.6, which is $\sim6$ days before the first detection reported by ASSA-SN and $\sim20$ days before Pan-STARRS. We used a second order polynomial to fit the points in the rising part of the light curve and found the intersection with the host level. We found that the outburst likely happened around MJD 57577, with an uncertainty of $\sim 10$ days. Throughout the paper we use this date as a reference epoch. We note that this is slightly different than the reference epoch used by B17, which was arbitrarily set to MJD 57600.

\subsubsection{Pan-STARRS}

Pan-STARRS1 uses a 1.8 m telescope and it is located at Haleakala, Hawaii. After  Pan-STARRS1 discovered PS16dtm on 12 August 2016, it also provided photometry from the follow-up observations in two broad filters, the $w_{\rm PS1}$  and $i_{\rm PS1}$ bands, which have wavelength ranges 4000-8300 $\AA$ and 6800-8300 $\AA$, respectively \citep{Tonry+2012}. Images obtained by the Pan-STARRS1 system are processed automatically with the Image Processing Pipeline and transient sources are identified  through analysis of difference images, created by subtracting a template from the observed image taken as part of the search for the counterpart \citep[see details in][]{Magnier2013,2015ATel.7153....1H}.  
	
\subsubsection{Liverpool telescope}

We collected ten epochs of multiband ($griz$) imaging of PS16dtm with the IO:O instrument at the robotic 2-m Liverpool telescope at the Roque de los Muchachos Observatory, Spain \citep{2004SPIE.5489..679S}. The images were reduced with the IO:O\footnote{https://telescope.livjm.ac.uk/TelInst/Inst/IOO/} pipeline and were subtracted against PSST \citep[][]{Tonry+2012} reference imaging, leaving only the transient light. Then PSF photometry was done on the source if it is detected after the subtraction and the photometry was calibrated relative to PSST photometric standards. 

\subsection{NEOWISE MIR photometry}

\begin{figure}
\includegraphics[width=8.5cm]{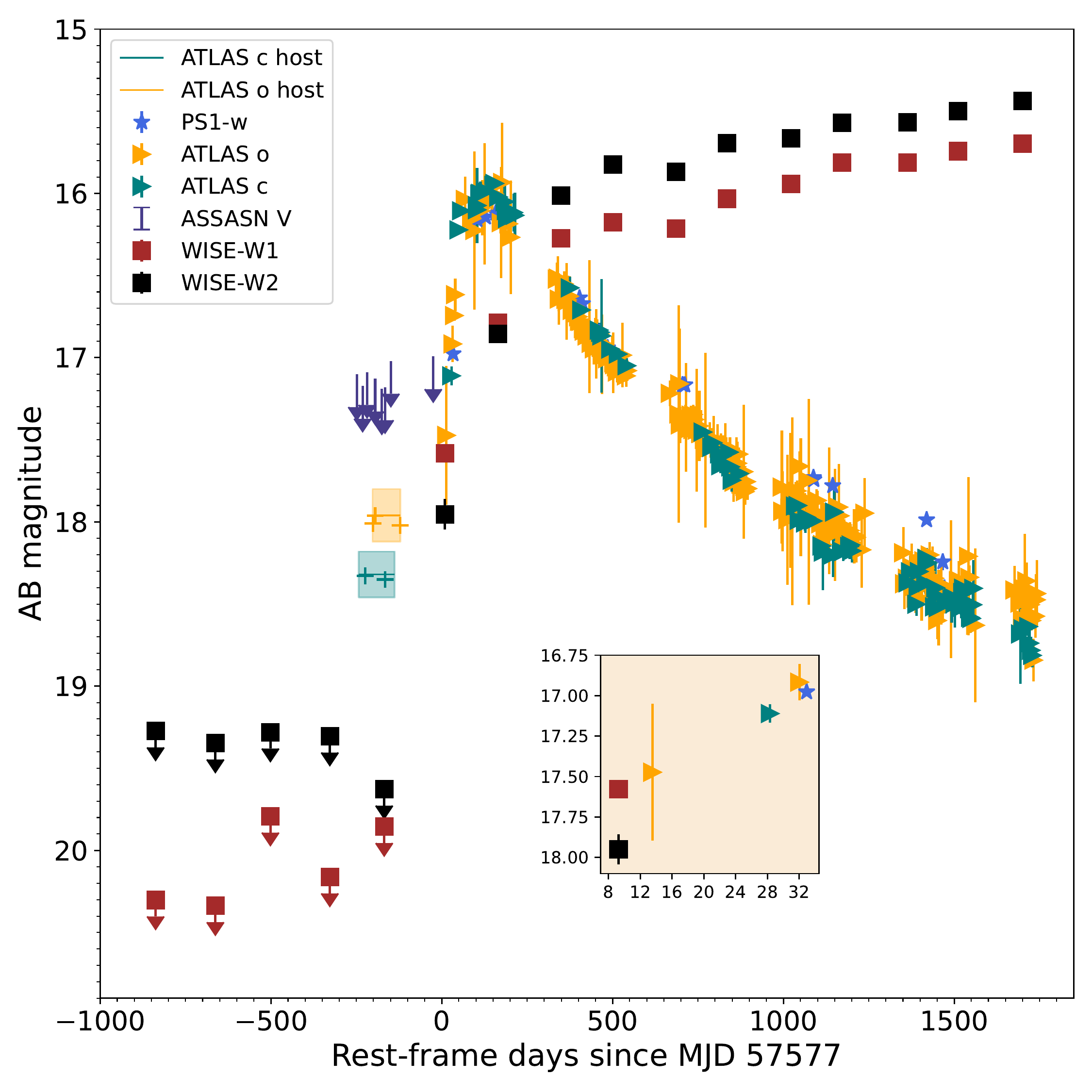}
\caption{Optical (ATLAS and PSST) and mid-infrared NEOWISE light curves of PS16dtm. The photometry has been extinction corrected for Milky Way dust and host subtracted. The limits from ASSASN in V band are also shown. The WISE error-bars are smaller than the symbols. The inset shows a zoom-in at the epochs around first detection.}
\label{wise_plus_optical}
\end{figure}

NEOWISE is a project by the Wide-field Infrared Survey Explorer, WISE \citep{2010AJ....140.1868W}, which surveys the sky in 3.4 ($W1$) and 4.6 $\mu m$ ($W2$)  \citep{2014ApJ...792...30M}. We retrieved the photometry from the IRSA public data archive\footnote{\url{https://irsa.ipac.caltech.edu/cgi-bin/Gator/nph-scan?mission=irsa&submit=Select&projshort=WISE}}. We first computed the variance-weighted average for 1-day bins and filter for all epochs after PS16dtm was detected. To remove the host contribution from the transient light curve, we computed the variance-weighted average of all preoutburst data and subtracted the host contribution from the transient light curve.  In Fig.~\ref{wise_plus_optical} we show the host-subtracted MIR light curve together with the optical data from ATLAS and PSST. The first detection by NEOWISE is on MJD 57587 (2016 July 18), it is $\sim4$ days before the first ATLAS detection of PS16dtm which was made at MJD 57591.6 and it is the earliest optical detection, to our knowledge. With the ATLAS data reported here, the gap between the first PS16dtm optical and MIR outburst are closer that previously shown in J17, which presented the ASSA-SN data only.

\subsection{NTT/EFOSC2 spectroscopy}

\begin{figure*}
\includegraphics[width=15.5cm]{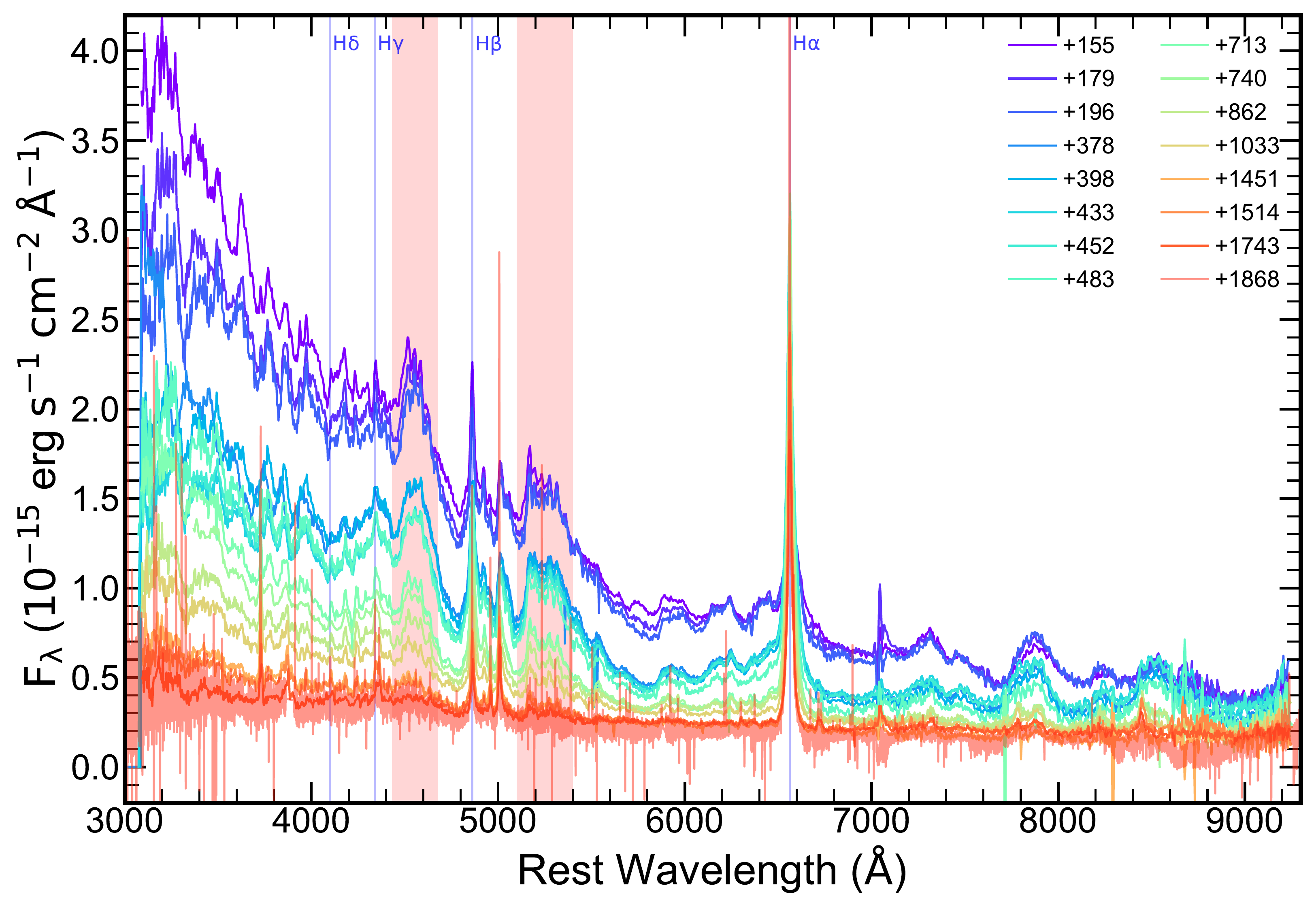}
\caption{Temporal evolution of the spectra taken with NTT/EFOSC2 (and the final X-shooter spectrum at +1868 days). The spectra are dominated by the Balmer series and a plethora of Fe II lines. The spectra are colored as indicated in the legend and the phases refer to rest-frame days after MJD 57577. Vertical solid lines indicate the position of hydrogen Balmer lines whereas the shaded pink areas mark the location of the strongest Fe II templates.}
\label{fig_spectra}
\end{figure*}

We carried out observations of PS16dtm within the Public ESO Spectroscopic Survey for Transient Objects \citep[PESSTO;][]{2015A&A...579A..40S}, and its continuations, ePESSTO and ePESSTO+.  
PESSTO uses the low-resolution ESO Faint Object Spectrograph and Camera v.2 (EFOSC2) on the New Technology Telescope (NTT) in La Silla Observatory, Chile. We have collected NTT/EFOSC2 spectra of PS16dtm at 16 different epochs, using various settings including the standard grims 11, 13 and 16 \citep{2015A&A...579A..40S}. 
The spectra were reduced in a standard manner with the aid of the PESSTO pipeline \citep{2015A&A...579A..40S}, to apply bias and flat-field corrections, determine a wavelength solution, and calibrate the relative flux with a standard star observed in the same setup. Following common practice, the intraday spectra of grism 11 and grism 16 were combined, since they are characterized by almost the same spectral resolution. The spectra were corrected for the Galactic extinction and the absolute flux calibration was improved by scaling the spectra with the aid of photometry. A spectroscopic log can be found in Table \ref{ePESSTO} and the spectral series is shown in Fig.~\ref{fig_spectra}. All reduced spectra will be available through the WISeREP archive\footnote{\url{https://www.wiserep.org}} \citep{Yaron_2012}.

\begin{table*}
\begin{center}
\caption{Spectroscopic observations of PS16dtm} \label{ePESSTO}
\begin{tabular}{lccccc}
\hline \hline
Date & MJD  & Phase$^*$ & Wavelength range & Exposure time & Grism	 \\
				&  &  &   \\
\hline

2016-12-22 & 57744 & +155 & 3380--10320 & 1500 & gr11 \& gr16 \\
2017-01-17 & 57770 & +179 & 3380--10320 & 900 & g11 \& gr16 \\
2017-02-05 & 57789 & +196 & 3380--10320 & 900 &g11 \& gr16 \\
2017-08-20 & 57985 & +378 & 3380--10320 & 1500 & g11 \& gr16 \\
2017-09-11 & 58007 & +398 & 3380--10320 & 1500 & g11 \& gr16\\
2017-10-19 & 58045 & +433 & 3380--10320 & 1500 & g11 \& gr16\\
2017-11-09 & 58065 & +452 & 3380--10320 & 1500 & g11 \& gr16 \\
2017-12-13 & 58099 & +483 & 3380--10320 & 1500 & g11 \& gr16\\
2018-08-17 & 58347 & +713 & 3380--10320 & 1500 & g11 \& gr16\\
2018-09-16 & 58377 & +740 & 3685--9315  & 1500 & g13 \\
2019-01-25 & 58508 & +862 & 3380--10320 & 1500 & g11 \& gr16 \\
2019-07-29 & 58693 & +1033 & 3380--7520 & 1800 & g11 \\
2020-10-23 & 59145 & +1451 & 3380--10320 & 1800 & g11 \& gr16 \\
2020-12-30 & 59213 & +1514 & 3380--10320 & 1800 & g11 \& gr16 \\
2021-09-04 & 59460 & +1743 & 3380-10320 & 2700/1800 & g11 \& 2xgr16 \\
2022-01-16 & 59595 & +1868 & 3000-24800 &  & UVB, VIS, NIR \\
\hline
\end{tabular}
\tablefoot{$^*$ In the rest frame with respect to the estimated time of outburst MJD 57577. }
\end{center}
\end{table*}

\subsection{VLT/X-shooter spectrum}

We also observed PS16dtm  with X-shooter \citep{2011A&A...536A.105V} on the Very Large Telescope (VLT), on MJD 59595, that is at +1868 rest-frame days. X-shooter is a  medium resolution spectrograph covering the wavelength range from 3000 to 24800 $\AA$ in three spectroscopic arms.  
We used slit widths of 1.0", 0.9" and 0.9" for the UVB, VIS and NIR arms respectively, resulting in  nominal resolutions of R=5400, 8900 and 5600.
The data were reduced by employing the X-shooter pipeline in the EsoReflex GUI environment \citep{2013A&A...559A..96F}, as implemented in \citet{2019A&A...623A..92S}.

\subsection{X-ray observations}\label{sec:xray}

The {\it Chandra X-ray Observatory (CXO)} observed PS16dtm on three epochs on 2019 January 10 (MJD 58493), 2019 November 11 (MJD 58798), and 2020 September 29 (MJD 59121)\footnote{P.I. Blanchard P., ID 21460, 21461 and 22618} for a net exposure of 10, 10, and 20 ks, respectively. We reduced these IDs with the {\tt CIAO} v4.14 software \citep{2006SPIE.6270E..1VF} in presence of the latest calibration files using the {\tt chandra\_repro} pipeline. Standard filtering and analysis methods are further applied to Advanced CCD Imaging Spectrometer mode data in our study.
We found weak X-ray emission in the 0.5-8 keV band at the location of PS16dtm. The source is moderately present in a 1 arcsec region at a significance level of 2.6, 1.6, and 2.5$\sigma$ level in the first, second, and third {\it Chandra} epochs with a count rate of about 4$\times$10$^{-4}$, 2.3$\times$10$^{-4}$, and 2.7$\times$10$^{-4}$~counts~s$^{-1}$, respectively. No significant emission, however, is detected in the 0.5-1 keV range. 

We subsequently performed the spectral analysis to examine the long-term X-ray flux evolution of the host galaxy and its connection to the transient. As the number of detected source photons is limited to $<$5 in each observation, all three {\it Chandra} spectra were stacked together to carry out the analysis using the \emph{Cash} statistics in XSPEC \citep[]{Arnaud1996ASPC..101...17A}. Using a redshifted power-law model with a column density fixed at 2.5$\times$10$^{20}$~cm$^{-2}$ \citep[][B17]{Pons_2014}, the photon index is found to be  0.8$_{-0.1}^{+8.5}$ along with a fit null hypothesis probability of 0.9 at four degrees of freedom. The errors are calculated for a 90\% confidence interval using the Markov chain Monte Carlo simulation method in XSPEC. The 0.3-10 keV unabsorbed source flux is (1.1$\pm$0.8)$\times$10$^{-14}$ erg~s$^{-1}$~cm$^{-2}$ that is obtained by the \emph{cflux} convolution model. Similarly, in place of the power-law component, a red-shifted blackbody model at the above-fixed column density is also tested on the {\it Chandra} data. We found a blackbody temperature of 1.5$_{-0.5}^{+2.0}$ keV along with a null hypothesis probability of  0.8 at four degrees of freedom. The 0.3-10 keV unabsorbed model flux is estimated to be (8$\pm$3)$\times$10$^{-15}$ erg~s$^{-1}$~cm$^{-2}$ in the latter case. 
Based on our spectroscopy, we conclude that the source flux remains close to the last measurement (upper limit of 1$\times$10$^{-14}$ erg~s$^{-1}$~cm$^{-2}$) provided by XRT in 2016 and 2017 (B17). In 2019 and 2020 the system was still present at an order of magnitude lower than the 2005 {\it XMM-Newton} detection \citep{Pons_2014}  before the transient.

\section{Analysis} \label{sec:analys}
\subsection{Light curve and SED analysis}\label{sec:blackbody}

The data presented here, make PS16dtm one of the few nuclear outbursts that have been observed with a photometric and spectroscopic campaign that spans over six years of the transient activity.  After the first peak, at which PS16dtm reached the absolute magnitude of $M_V= -22.0$ mag in the UVOT $V$ band, the luminosity dropped for $\sim 50$ days, after which it rose again to a second peak and stayed at almost plateau level for $\sim100$ days. Then the flux decreased steadily with time at all wavelengths, as seen in Fig.~\ref{photometry}. However, the decline is most pronounced up to $+270$ days in the NUV/UVOT filters, afterward the decline is slower. Interestingly, there have been few other TDEs which have shown double peaks in the light curve, such as ASSASN-15lh \citep{Leloudas_2016} and AT~2018fyk \citep{Wevers_2019}, but the physical process behind this feature is still not clear.

In order to explore the physical parameters of the transient, we attempted to fit a blackbody to SEDs constructed from Milky Way extinction-corrected and host-subtracted photometry. We used the Markov chain Monte Carlo (MCMC) method with a publicly available code\footnote{\url{https://github.com/nblago/utils}} that uses  the Python package {\tt emcee} (v\,3.0.2) \citep{ForemanMackey2013}. It was already pointed out in B17 that a simple blackbody is not a good fit to the SED, which they constructed with the first 200 days of PS16dtm data. Therefore, B17 fitted a blackbody by excluding the photometry from the three NUV/UVOT filters by arguing that there must be a significant obscuring material in the line of sight. This implies that true luminosity could be higher, but they argued that it should not be more than a factor 2 (that is not higher than $5\times10^{44}$~erg s$^{-1}$).  We also note the updated \textit{Swift}/UVOT calibration correction from \citet{2021ApJ...910...83H} which only affected the NUV filters, making them fainter compared to B17 measurements. This means that this problem persists and it is even more pronounced than from what B17 estimated.

\begin{figure}
\includegraphics[width=8.5 cm]{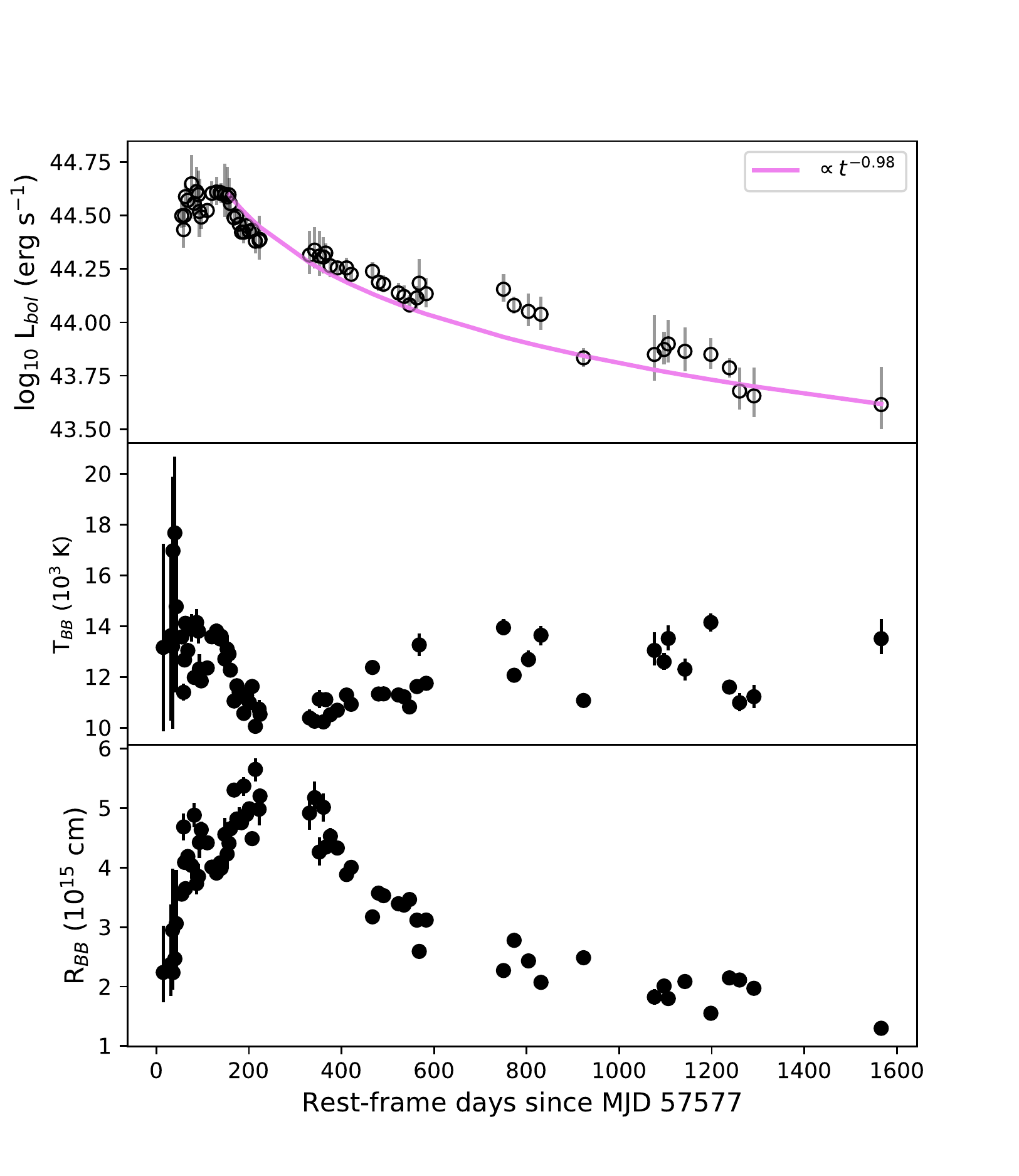}
\caption{Evolution of the bolometric luminosity, temperature and radius for PS16dtm, by fitting a blackbody to the Galactic extinction and host-corrected photometry, excluding the photometry in the three NUV UVOT filters and requiring at least 4 photometric measurements in each epoch. The purple line in the upper insert represents the best fit of the fading light curve with a power-law profile, where the power-law index is fit freely, $L= L_o(t-t_o)^{-\alpha}$.   
}
\label{BB_noUV}
\end{figure}

We repeated the same exercise of fitting a blackbody to our photometry without the NUV/UVOT bands, requiring at least four photometric measurements in each epoch. The results are shown in Fig.~\ref{BB_noUV}. Keeping in mind this problem with likely underestimated temperature due to the omission of NUV/UVOT bands, the blackbody fits would indicate an initial rise in the temperature to $\sim15000$ K contemporaneous with the rise in the luminosity. After that, the temperature drops to $\sim10000$ K, and is held relatively constant. After $\sim500$ rest-frame days, the blackbody fit to the SED appears increasingly erratic, as at later epochs when the host-subtracted photometry of PS16dtm becomes fainter, thus noisier.

We explored the possibility that the inability to fit a single blackbody to the NUV/optical data is caused by the extinction in the line of sight by the circumnuclear dust near the SMBH, despite that there is no clear consensus regarding the wavelength dependence of the extinction in the line of sight to AGN. However, some studies based on individual reddened AGN have suggested they have an extinction similar to that measured in the Small Magellanic Cloud (SMC).\footnote{We note that this implies the presence of small dust grains, in contrast to what other authors have argued, that the largest grains are those able to survive in the vicinity of the SMBH.}  SMC extinction strongly affects the UV part of the SED \citep{2005ApJ...627L.101W} and compared to the Milky Way or the one used for starburst galaxies \citep{Calzetti2001}, SMC average extinction curve has essentially no 2175 $\AA$ bump and it has a strong far-UV rise \citep[see e.g.,][for a recent review]{2020ARA&A..58..529S}. We attempted to fit blackbody SEDs reddened with the SMC-like extinction curve \citep{2003ApJ...594..279G} where we add $E(B-V)$ as a free parameter. It was possible to fit the SED to a blackbody with $T\sim 2 - 4\cdot10^5$ K and a bolometric luminosity of $L \sim 10^{48} - 10^{49}$ erg $s^{-1}$, at a corresponding blackbody radius of $R \sim 1 \cdot 10^{15}$ cm, and SMC-like extinction curve with $E(B-V)$ $\sim0.5$ mag. This best fit model is clearly nonphysical, with such high temperatures, which would also imply an extremely bright X-ray source. In fact, this solution for sufficiently large (that is above 20000 K) temperatures corresponds to a degeneracy between temperature and radius, where all the measurement points lie on the high-wavelength slope of the distribution, and for any arbitrarily large temperature there exists a radius that produces the observed light curve.  In conclusion, a single blackbody does not provide a satisfactory fit to the SED, even when we included extinction. However, the fact that the SED cannot be described with a single blackbody does not exclude the interpretation of PS16dtm as a TDE. According to the radiative transfer calculations by \citet{Roth_2016} and \citet{2022ApJ...937L..28T}, the optical/UV continuum of TDEs is not necessarily described by a single blackbody. In addition, it is possible that the SEDs of TDEs in AGN are different compared to those in inactive galaxies due to the preexisting AGN disk, as suggested by the simulations in \citet{Chan2021}.

We also fitted the fading light curve with a power-law profile, where the power-law index is fit freely, $L= L_o(t-t_o)^{-\alpha}$. Our best fit parameter, $\alpha\sim{5/7}$ for the NUV/UVOT light curves for the initial ($158<t<270$) days, while $\sim270$ days after the outburst, it follows a $\alpha\sim{1/6}$ decay. For the bolometric luminosity, the best-fit power-law has the index $\alpha \sim{0.98}$  (see Fig.~\ref{BB_noUV}). Interestingly, a similar decline trend is also seen in the bolometric light curve of the TDE AT~2017gge ($L\propto t^{-1}$, \citealt{Onori_2022}), which also shows a MIR echo and coronal emission lines in the late-time spectra, although more numerous and more intense than what it has been detected in PS16dtm, as shown in the next sections. Nevertheless, the luminosity evolution of PS16dtm cannot be simply explained by the $t^{-5/3}$ decline as expected from simulations of the fallback rate for the stellar debris onto the SMBH \citep{1988Natur.333..523R,2009ApJ...698.1367G}. At the moment, the origin of the optical emission from TDEs remains a puzzle, but it might be powered by an inner accretion disk or by shocks of the intersecting debris streams \citep[see][for a recent review]{van_Velzen_2020}. In the first months, optical and UV light curves of TDEs in quiescent galaxies often decline as  $t^{-5/3}$, but after a few months, models that include reprocessing of the disk emission by outer debris predict weaker temperature evolution, are expected to show weaker temperature evolution and to follow a flatter decline as $t^{-5/12}$ \citep{2011MNRAS.410..359L}.
Furthermore, if the  disruption of the star is only partial, the expected fallback rate could also be shallower  \citep{2013ApJ...767...25G}. Interestingly, \citet{2019ApJ...878...82V}, by observing TDEs at late times in the far-UV (FUV), found that for the light curves of TDEs from low-mass SMBHs ($\rm M_{BH}<10^{6.5}M_{\odot}$), the early-time decay follow a $t^{-5/3}$ power-law decline, but the later-time evolution is much shallower. They conclude that this could be the sign of different disk emission mechanisms operating at early and late times. They argue that unobscured accretion disk models, rather than reprocessing and circularization paradigms, can explain the late-time FUV emission.

\subsection{Blackbody fits with the NEOWISE MIR photometry.}
As mentioned in the previous sections, PS16dtm exhibits MIR emission almost simultaneously with the optical one. In Fig.~\ref{wise_plus_optical} we show that the first NEOWISE detection is $\sim4$ days before the first ATLAS detection.  In the first 500 days, PS16dtm shows a steep rise in the MIR light curve. After that, it keeps rising, albeit at a much slower rate. Another noticeable trait from Fig.~\ref{wise_plus_optical} is that not only the rise in brightness, but also the change in color that happens in the first epochs, that is at +9 days $W1-W2$ has a negative value of $-0.37$, then at +165 days it is $-0.07$, and in the epochs afterward to the last epoch, from +349 to +1700 days, $W1-W2$ remains steady with a positive value of $0.24 - 0.35$ mag.

\begin{figure}
\includegraphics[width=8.7cm]{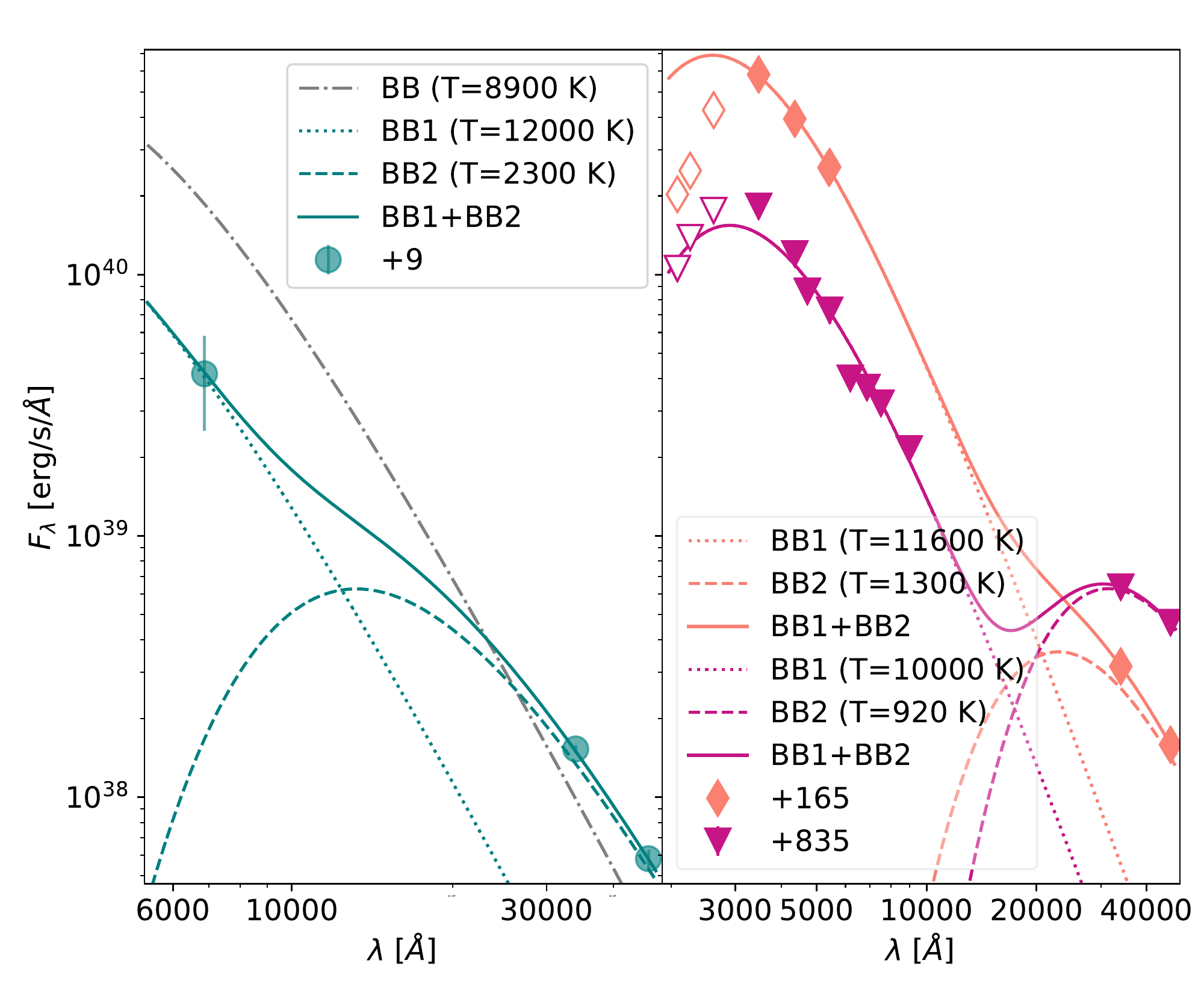}
\caption{Examples of three epochs of blackbody fits to the photometry that includes the NEOWISE data. In the left panel, the best fit by using one blackbody is shown. For the first epoch of the MIR detection, together with the quasi-simultaneous ATLAS photometry, the best fit temperature is $\sim8900$ K. A better fit is obtained by using a double blackbody model, and the resulting blackbody temperatures are shown. In the right panel, the fits of the two-blackbody model are shown to the photometry of two epochs, +165 and +835 days where the NUV bands are plotted, but have not been included in the fitting procedure.}
\label{two_bb}
\end{figure}

First, we attempted to fit a single blackbody to the data by including the optical and the MIR photometry at quasi simultaneous epochs. As visible from Fig.~\ref{wise_plus_optical}, the first MIR point has a very different color than the rest of the MIR light curve, so we examined the possibility that it is compatible with the Rayleigh-Jeans tail of a single blackbody. As shown in the example in Fig.~\ref{two_bb}, the fit would indicate a temperature of 8900 K at the earliest epochs after the outburst, however the single blackbody is a poor fit to the data. Second, we attempted to fit a double-blackbody model to the optical and the MIR NEOWISE photometry, as it was done in J17, for the first three NEOWISE epochs.
The first NEOWISE epoch at +9 days,  yields $\sim 2300$ K for the dust blackbody temperature, while the second epoch at +165 days, yields $\sim 1300$ K, and the epochs afterward (from +349 to +1700 days) settle at $\sim 900-1000$ K. J17 interpreted the MIR emission as light coming from the dust thermal emission heated by the TDE; in the first epoch, the temperature is above the sublimation temperature; after the second epoch, the dust temperature dropped below the sublimation temperature, the MIR emission is from larger distances from the SMBH, so the dust in the inner region must be optically thin.

\subsection{Spectroscopic analysis}

\subsubsection{Summary of the spectral features presented in B17}

Before we proceed with our spectroscopic analysis, we summarize here the findings presented in B17 in which they used ten spectra that spanned from +54 to +198 days:
\begin{itemize}
\item The main spectral features of PS16dtm are the multicomponent hydrogen Balmer lines and Fe II lines, visible at all epochs.
    \item The $\rm H\alpha$ exhibited a complex asymmetric profile consisting of a broad component and a superimposed narrower component that has a slight shoulder on the blue side of the peak.
    \item Several broad features appear near the [O II] $\lambda\lambda7320$, 7330, [Ca II] $\lambda\lambda7291$, 7324, which they interpret as a blend of several Fe II lines. The Fe II lines became stronger with time.
\item In the NIR part of the spectra, they noticed several broad emission features with asymmetric profiles, mainly Fe II lines, out to 1.2 $\mu m$, beyond which the spectrum is relatively featureless, 
\item Paschen $\gamma$ and other lines of the Paschen series were detected. Paschen $\alpha$ and $\beta$ were not detected because they fall in the NIR telluric bands.
\item The Ca II triplet shows an unusual shape where the central line in the triplet is much stronger than the others.
\item Blueward of $\sim 4500$ $\AA$ the spectrum shows a very complex combination of narrow features superimposed on broad features. They identify [O II] $\lambda\lambda3727$, Balmer lines, and additional Fe II lines.
\item In their UV spectrum obtained with HST, they found evidence for absorption, in particular, broad Mg II $\lambda\lambda2800$ absorption lines were detected, with $\sim10,000$ $\rm km\,s^{-1}$, that can perhaps indicate the presence of an outflow.
\end{itemize}

B17 also studied the temporal evolution of the hydrogen Balmer lines $\rm H\alpha$ and $\rm H\beta$. They fitted the spectral lines with the sum of three Gaussians. They reported that the widths of the intermediate component  of $\rm H\alpha$ is around 750 $\rm km\,s^{-1}$ that narrows as a function of time, going from 900 $\rm km\,s^{-1}$ to 600 $\rm km\,s^{-1}$. For the broad component, they found 3500 $\rm km\,s^{-1}$ which increases to 4000 $\rm km\,s^{-1}$ in the decline phase of the light curve. They found that the narrow component is 100 $\rm km\,s^{-1}$ from their spectrum with the highest resolution. For their low-resolution spectra the narrow component is unresolved, so they fix it at the instrumental resolution.  Furthermore, they found that the width in the earliest epoch of their intermediate component was similar to the width of the preexisting broad component in the archival host spectra. For this reason, B17 argues that the intermediate component that they measure in PS16dtm might be associated with the BLR of the NLSy1 host galaxy. The flux of the narrow component of $\rm H\alpha$ has remained unchanged relative to the host spectrum. 

\subsubsection{Spectral evolution to 1868 rest-frame days after the outburst}

\begin{figure}
\includegraphics[width=8.5cm]{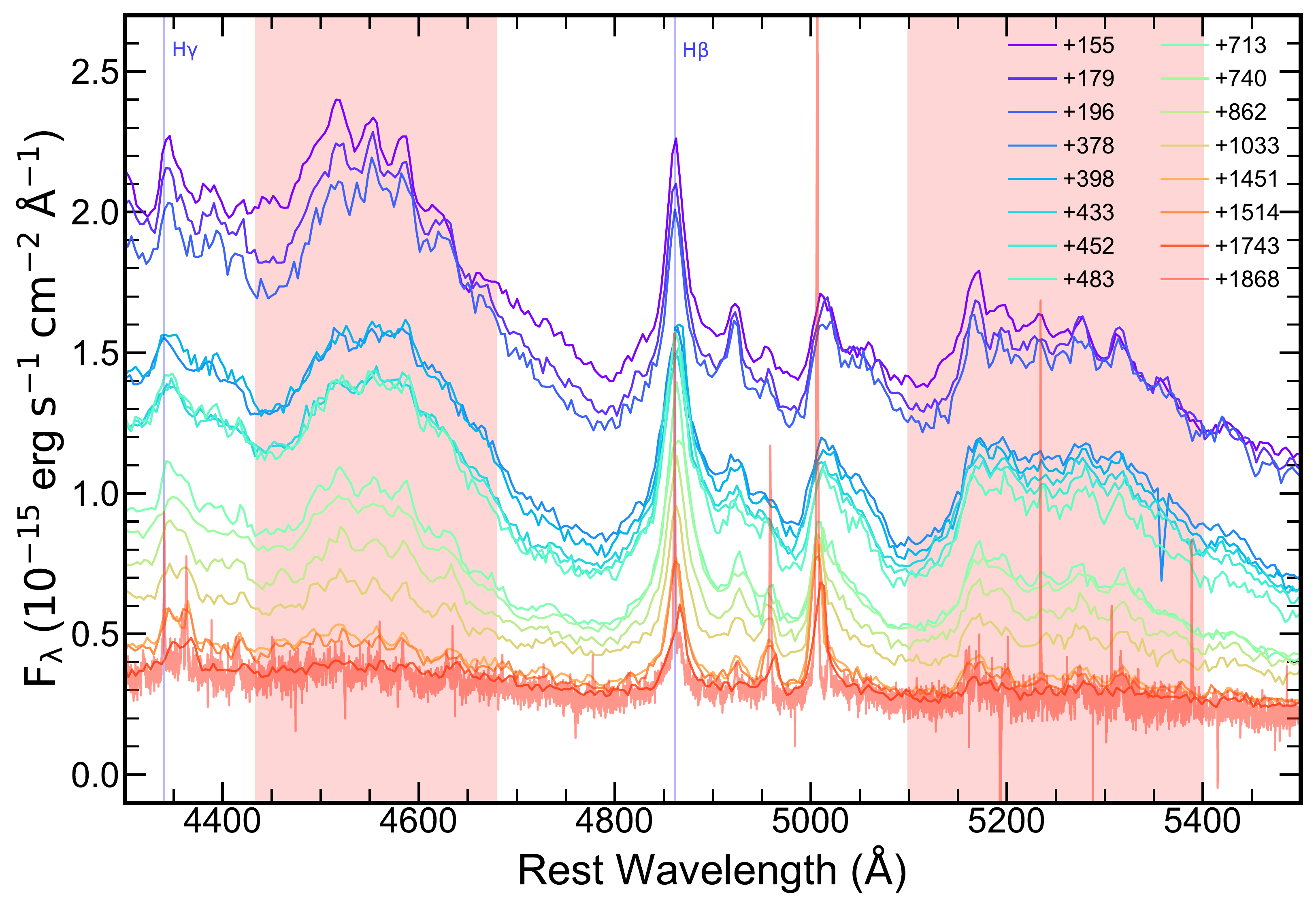}
\caption{Zoomed-in view of the spectra containing the $\rm H_\beta$ and the Fe II multiplets. The phase indicated in the legend refer to the rest-frame days after MJD 57577. Vertical solid lines indicate the position of hydrogen Balmer lines whereas the shaded pink areas mark the location of the strongest Fe II features.}
\label{Hbeta}
\end{figure}

\begin{figure}
\includegraphics[width=9cm]{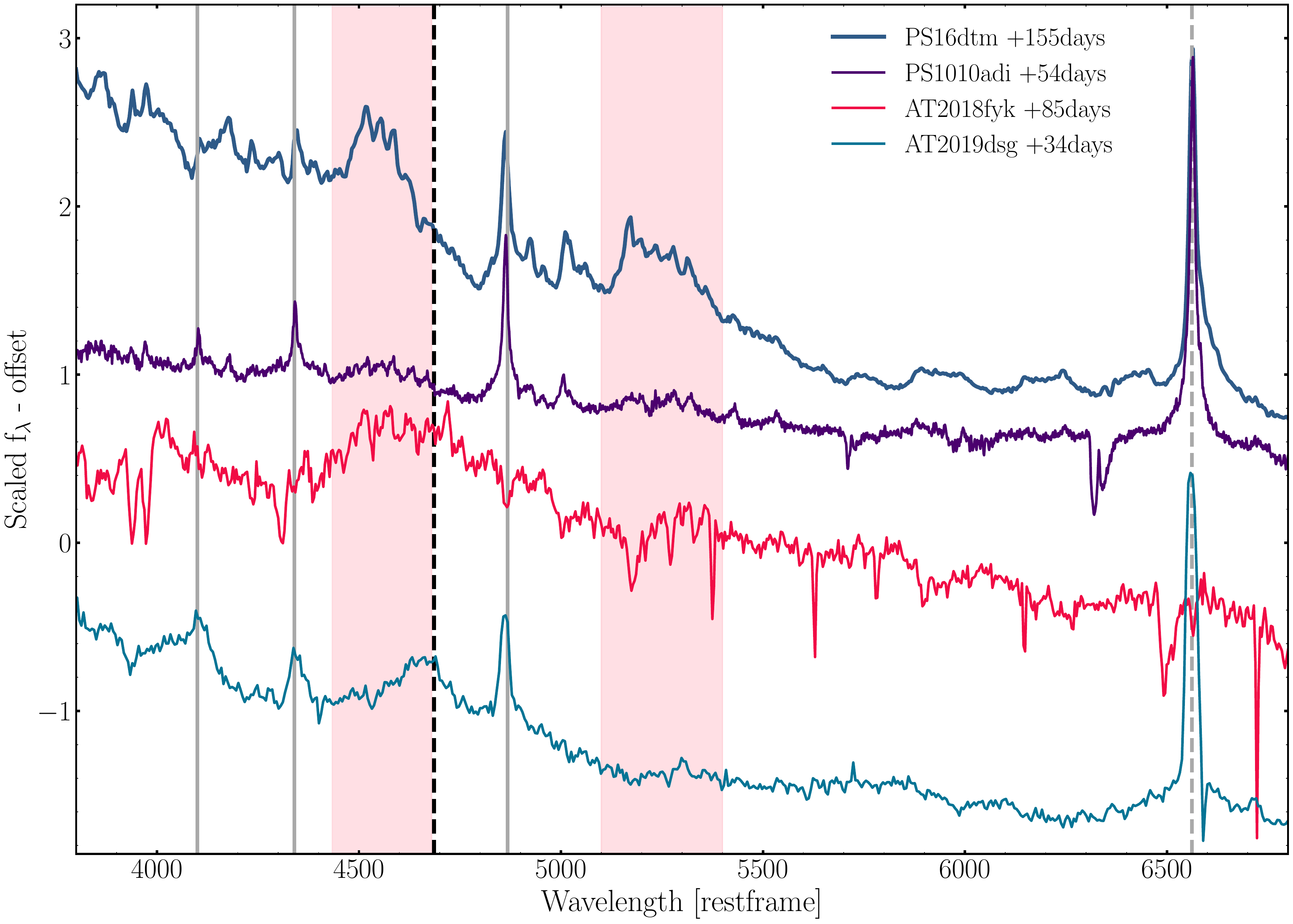}
\caption{Spectroscopic comparison of PS16dtm with TDEs from the literature where Fe II emission has been identified: PS1-10adi \citep{2017NatAs...1..865K}, AT~2018fyk \citep{Wevers_2019}, and AT~2019dsg \citep{Cannizzaro2021}. Vertical solid lines indicate the position of hydrogen Balmer lines, the dashed line denotes He II $\lambda$4686, whereas the shaded pink areas mark the location of the strongest Fe II complexes. The spectra of the three comparison TDEs have been scaled up and offset for display.
}
\label{fig:iron_strong}
\end{figure}

Now we turn to the analysis of our spectra, 16 of which were taken with the low-resolution NTT/EFOSC spectrograph at +155 to +1743 rest-frame days, and one medium-resolution spectrum with the VLT/X-shooter at +1868 days. The spectra are shown in Fig.~\ref{fig_spectra}. Similar to B17 (see their Fig.~7), we also found that the main spectral characteristics are the hydrogen Balmer lines and the Fe II complexes. The most noticeable variability in our spectra is that the blue continuum becomes weaker and almost disappears in the final X-shooter spectrum at +1868 rest-frame days after the outburst. Perhaps even more striking are the Fe II optical lines, in the blue (around H$\beta$) and red (around H$\alpha$) line, that become weaker as time progresses (see the zoomed-in view around the H$\beta$ line in Fig.~\ref{Hbeta}). This is a first time such strong Fe II emission is seen in a TDE, even when compared to other nuclear transients for which Fe II emission has been claimed, such as J123359.12+084211.5 in \citet{MacLeod_2019} and PS1-10adi \citep{2017ApJ...850...63J,He_2021}, AT~2018fyk \citep{Wevers_2019} and AT~2019dsg \citep{Cannizzaro2021}. We show the spectroscopic comparison of the iron-rich TDE candidates in Fig.~\ref{fig:iron_strong}. Some TDE and flaring AGN spectra show O III and N III lines which have been explained as due to the mechanism of Bowen fluorescence \citep{2019ApJ...873...92B,Leloudas_2019,Trakhtenbrot_2019,2019MNRAS.489.1463O}. We also searched for these lines, but in PS16dtm these are difficult to resolve due to strong blending with Fe II emission. 
An extremely weak feature is resolved on the place of a N III $\lambda$4640 line in the X-shooter spectrum (see later discussion).
We also detected strong features in the red part of the spectrum, in the range 7000-8000 \AA, which is fading with time, and completely disappearing in the X-shooter spectrum. These are most likely blended Fe II emission, as already noted by B17, with some probable contribution from the other nearby lines, such as O I $\lambda$8446, is clearly identified in the +1514 spectrum (see Fig.~\ref{fig_spectra}).

\subsubsection{Iron emission-line model for the spectral fitting}\label{sec:iron_model}

Next, we proceed to identify the emission lines and understand their temporal evolution, despite that, given the richness of features and their strong variability, their behavior is difficult to follow and disentangle since all emission lines are strongly affected by blending (see Fig.~\ref{Hbeta}). The complex Fe II ion can produce thousands of line transitions, thus these lines are typically blended and difficult to identify in the spectra. With this caveat in mind, we fit all spectra using a python-based tool called Fully Automated pythoN tool for AGN Spectra analYsis (FANTASY)\footnote{\url{https://fantasy-agn.readthedocs.io/en/latest/
}}, that was initially developed for fitting AGN optical spectra (\citealt{Ilic_2020,Rakic_2022}, Ili\'{c} et al. 2022, in prep.). 
The code fits the underlying broken power-law continuum and sets of emission lines, with the Fe II model based on the atomic parameters of Fe II.  In contrast to widely used fully empirical iron templates (for recent discussion on different Fe II templates, see \citealt{Park_2022} and references therein), such as that by
\citet{Boroson_1992}, our approach is slightly different and it was first developed by  \citet{2010ApJS..189...15K}. The semi-empirical model of \citet{2010ApJS..189...15K}  relies not only on the observed properties of AGN spectra, but also on atomic properties of the transitions, so the Fe II line sets are grouped according to the same lower energy level in the transition with line ratios connected through the line transition oscillatory strengths. \citet{2010ApJS..189...15K} produced a multicomponent template covering 4000-5500 \AA. Building up on that work, we extended the Fe II model to include the wavelength range up to 7000 $\AA$ to cover the area around H$\alpha$ line where Fe II emission is also very strong, using the atomic data from Kurucz database\footnote{https://lweb.cfa.harvard.edu/amp/ampdata/kurucz23/sekur.html} (Ili\'c et al., in prep).
 This approach can be useful to understand newly discovered transients, which show strong and complex Fe II emission.  
We assume the following constraints for the emission line ratios, widths and shifts:
\begin{enumerate}[(i)]
    \item the broad Balmer emission lines (H$\alpha$, H$\beta$, H$\gamma$) are fixed to have the same width and shift; from other strong broad lines in this range, we included He I $\lambda$5876, He II $\lambda$4686, the Na I doublet $\lambda\lambda$5890, 5896, and O I $\lambda$6046;
    \item the model of broad Fe II, with all lines having the same widths and shifts, and line ratios connected as described above. The broad Fe II lines are assumed to have the same profiles as broad Balmer lines \citep[see e.g.,][and references therein]{Dong_2011}, so this component is set to have the same width and shift as other broad lines;
    \item the AGN narrow emission lines, such as H$\alpha$, H$\beta$, H$\delta$, [O III], [N II], [S II], Ti II, Cr II (see \citealt{Veron_Cetty2006,Park_2022}, for the list of significant narrow lines) are constrained to have the same width and shift. In addition to fixing the ratio of [O III] and [N II] doublets to the theoretical values of $\approx 3$ \citep{Dimitrijevic_2007}; 
 
    \item the model of narrow Fe II, plus the forbidden Fe II lines, have the same width as narrow lines \citep{Veron_Cetty2006, Park_2022}.

\end{enumerate}

\subsubsection{Results of the modeling of the spectra in the $4100 - 7000$ $\AA$ range}\label{sec:results_iron_fit}

\begin{figure*}
\includegraphics[width=17.5cm]{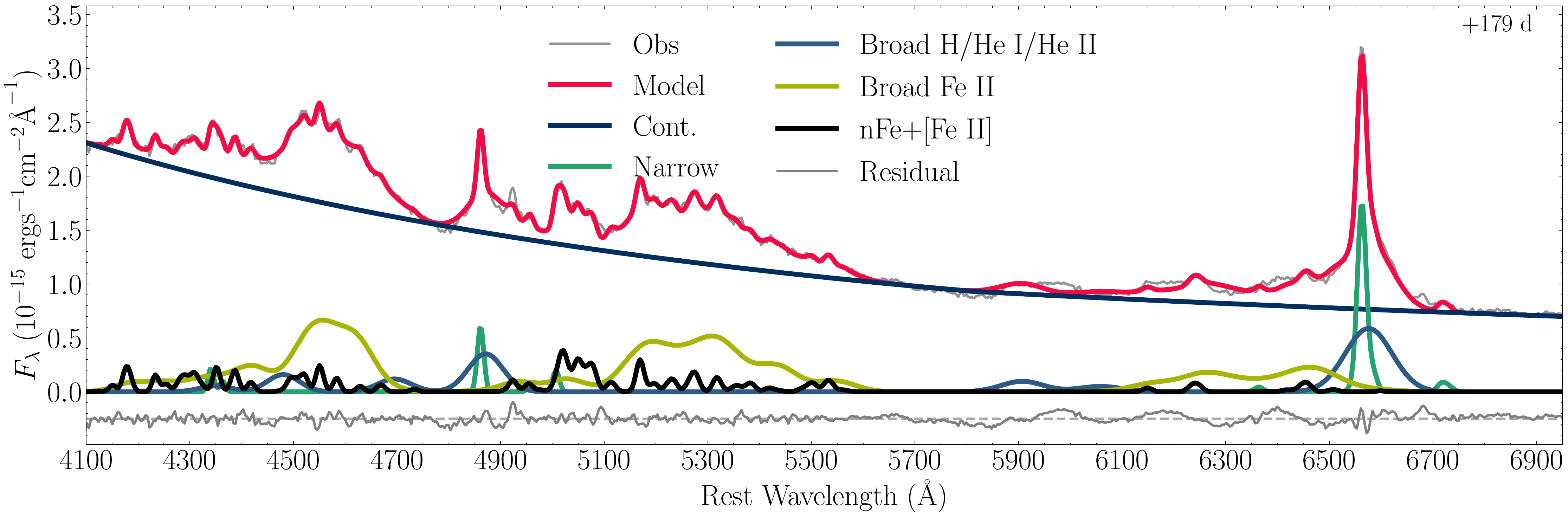}
\caption{Examples of multicomponent spectral fitting in the $4100 - 7000$ $\AA$ rest-frame range for the +179 days spectrum (solid gray line). 
The underlying continuum emission and all emission-line components are plotted with different colors, as indicated in the legend. For details on the assumed emission-line components see text. The final model (solid red line) which is the sum of all components is also shown, whereas the residual spectrum, is plotted below.}
\label{Fantasy_fit}
\end{figure*}

In Fig.~\ref{Fantasy_fit} we show an example of the results of the multicomponent spectral fitting in the $4100 - 7000  $ $\AA$ rest-frame wavelength range. Notably, our model was able to reconstruct the observed spectra, but only when assuming a broad component of Fe II multiplets. The presence of the broad Fe II component has been previously detected in AGN \citep[e.g.,][]{Dong_2011,Park_2022}; however, since this component is typically very weak, it is often merged with the continuum emission, especially in poor spectral resolution data. We also detected narrow features on top of the broad Fe II components, which our code usually identified as narrow Fe II emission, despite the limited instrumental resolution.  We were also able to identify the broad features blueward from H$\alpha$ region with the Fe II model (see Fig.~\ref{Fantasy_fit}).
We performed a series of tests with other emission-line models (such as no broad Fe II, no forbidden Fe II lines, etc.), but none were able to produce reasonable fit based on the fitting residuals. The goodness of the fit was evaluated using the $\chi^2$ parameter. 

\begin{figure}
\includegraphics[width=8.5cm]{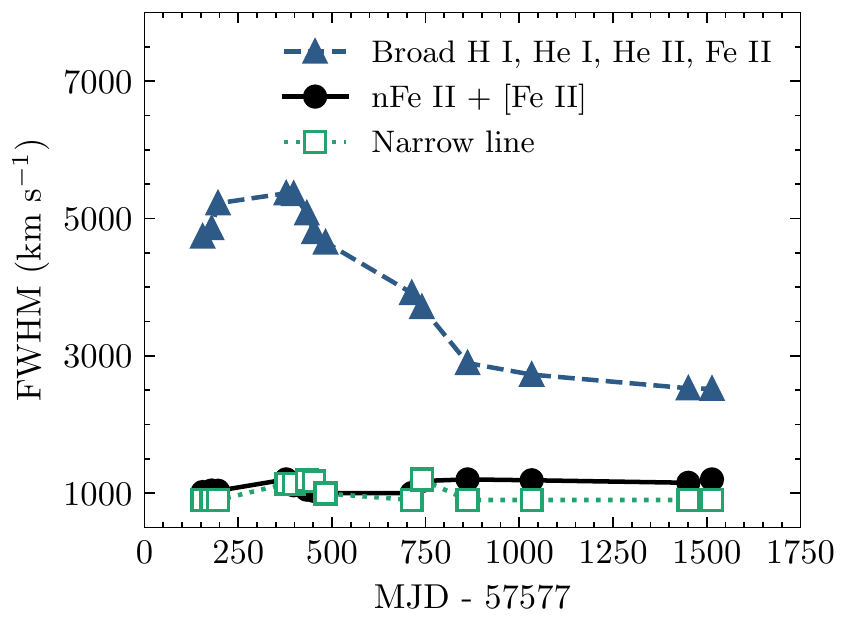}
\caption{Evolution of the FWHM of different line components (denoted in the upper right corner) during the spectroscopic campaign.}
\label{fwhm}
\end{figure}

\begin{figure}
\includegraphics[width=8.5cm]{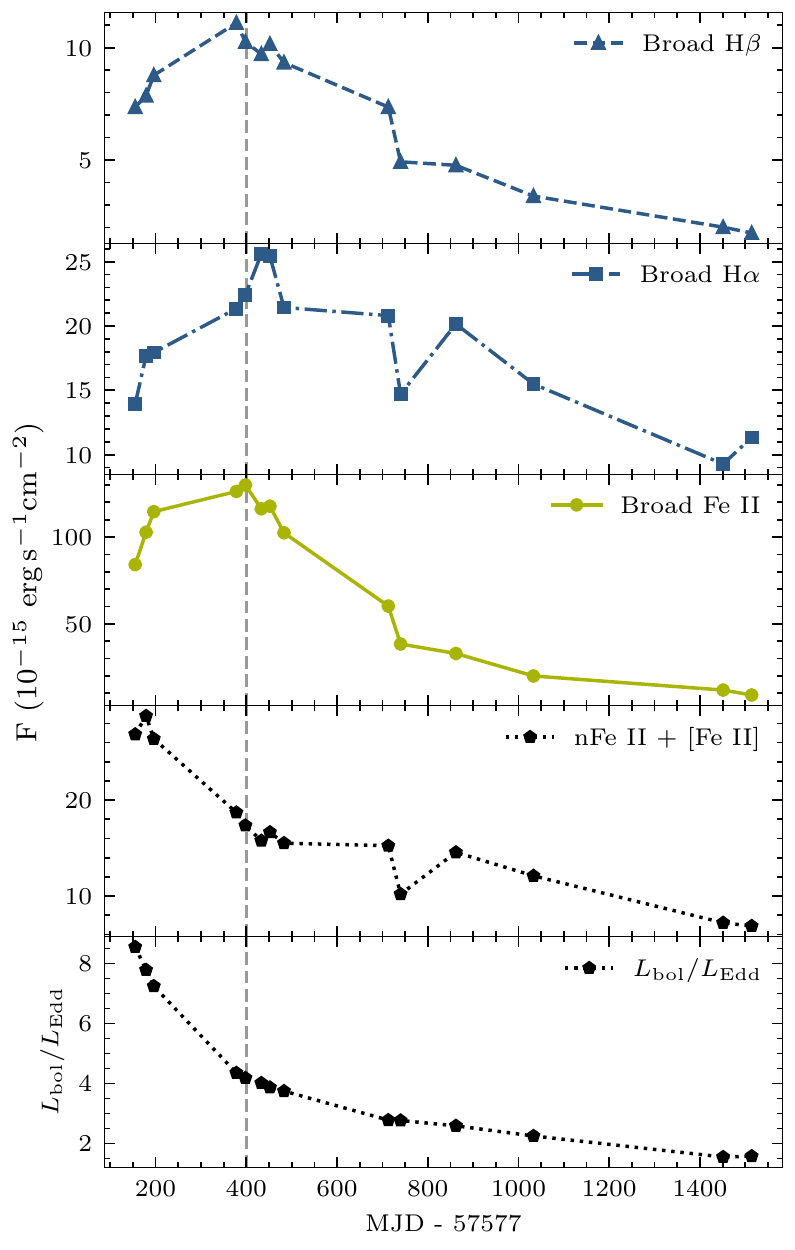}
\caption{Temporal evolution of the extracted emission lines and line-blends during the campaign. The bottom panel shows the evolution of $L_{\rm bol}/L_{\rm Edd}$ ratio, in which the bolometric luminosity is estimated from the spectral continuum at 5100 \AA. The vertical dashed line indicates +400 days after the estimated outburst time to guide the eye.}
\label{fitting_lc}
\end{figure}

In Fig.~\ref{fwhm} we show the time evolution of the FWHM of the most prominent emission lines, broad H$\alpha$ and H$\beta$, broad Fe II, and all narrow lines. It is evident that the strength of the lines slowly decreases with time.  The width of the narrow lines remains the same, but this is  constrained by the instrumental resolution. 

In Fig.~\ref{fitting_lc} we show the time evolution of the integrated emission-line features, that is, for the broad H$\alpha$ and H$\beta$ line, total broad Fe II, total narrow and forbidden Fe II lines. The strength of broad line-blends (H$\alpha$, H$\beta$, total Fe II broad) are showing similar trends, slight increase followed by long decrease with time. A similar evolution is seen in the narrow Fe II blend. The exact time of peak is difficult to identify due the fluctuations in line fluxes. Furthermore, the  extraction of single lines H$\alpha$ and H$\beta$ from the low-resolution spectra is difficult; it depends on the possibility for the code to identify the narrow [O III] $\lambda$5007 line. Nevertheless, the general similar trend is evident in all light curves, especially of line-blends. 

\begin{figure}
\includegraphics[width=4.2cm]{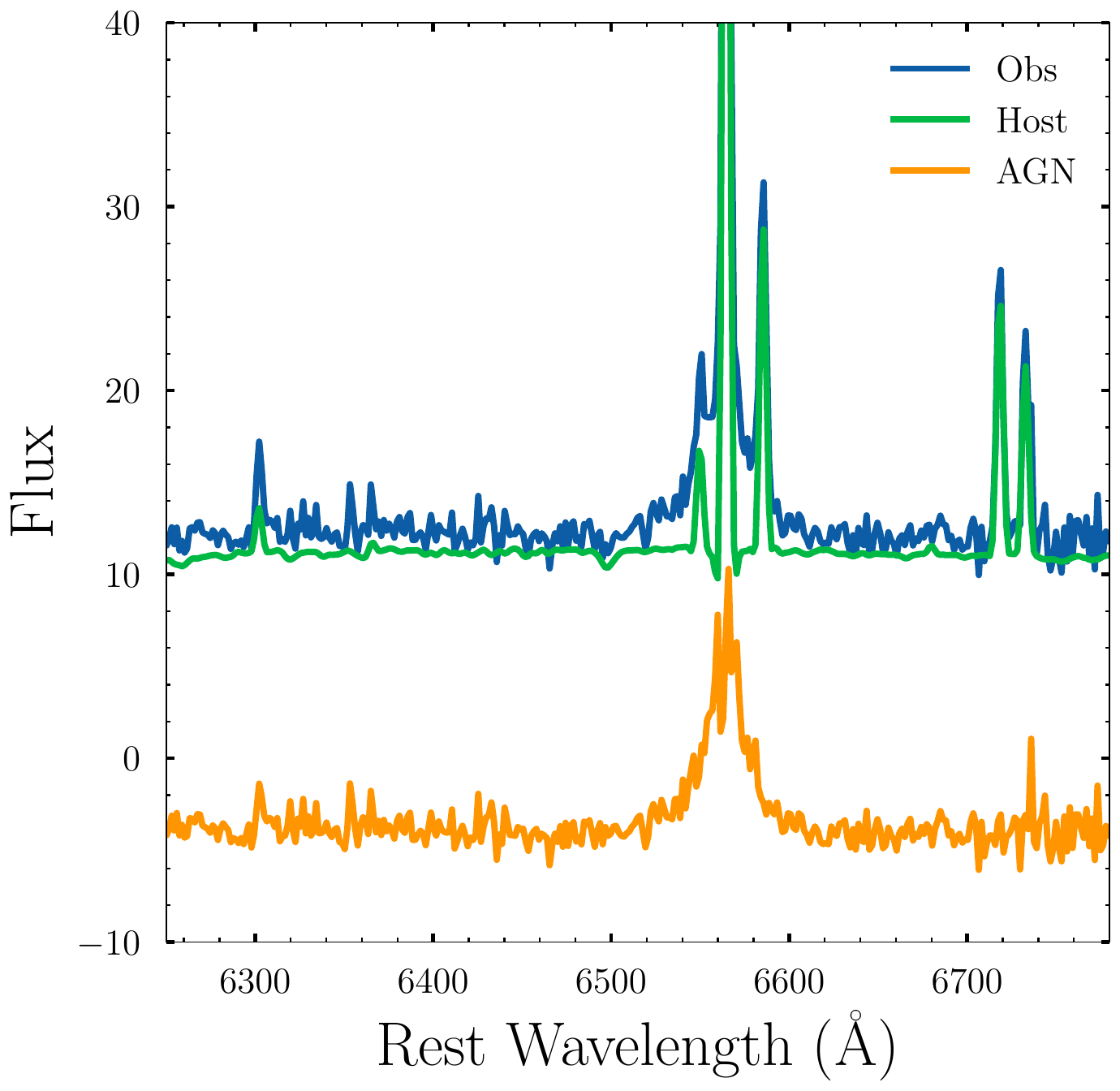}
\includegraphics[width=4.2cm]{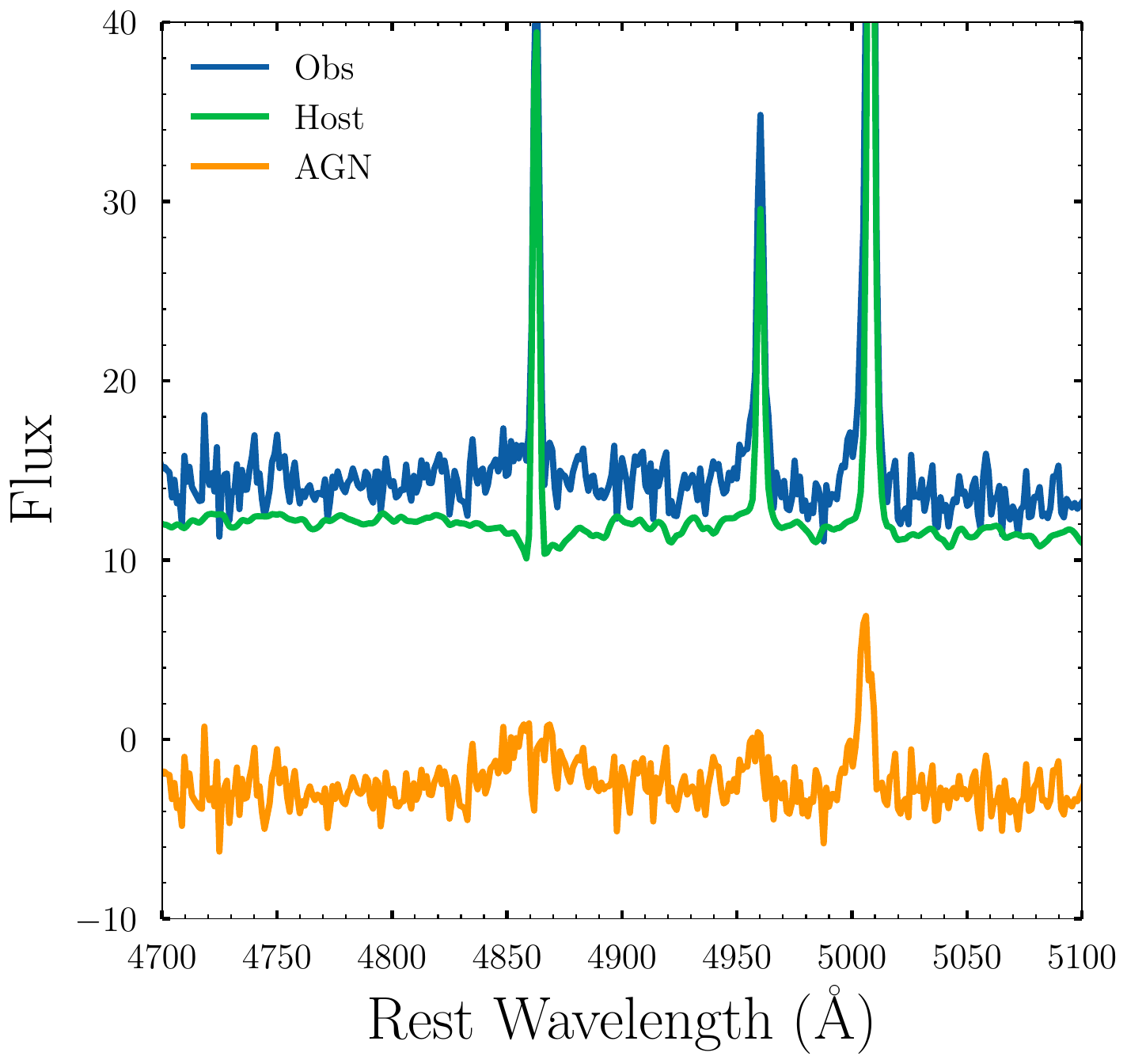}
\caption{The pure AGN components for H$\alpha$ (left) and H$\beta$ (right) region are shown, obtained when the host-galaxy contribution is subtracted from the observed total spectrum in the archival, preoutburst, SDSS spectrum.}
\label{host_SDSS}
\end{figure}

In Fig.~\ref{fitting_lc} we also show the ratio of the bolometric luminosity and the Eddington luminosity, $L_{\rm bol}/L_{\rm Edd}$.
The Eddington luminosity was estimated as $L_{\rm Edd} = 1.26 \times 10^{38} (M_{\rm BH}/M_\odot)$ $\rm{erg\, s}^{-1}$ with the SMBH mass of $M_{\rm BH} \sim 10^6 M_\odot$, which we determine in Sect.~\ref{sec:smbh_mass}. For the bolometric luminosity $L_{\rm bol}$ we used a different approach than the one used in Sect.~\ref{sec:blackbody}.
Here, to obtain the bolometric luminosity we used a standard procedure for quasars, where $L_{\rm bol} = k_{\rm bol} \lambda L_\lambda $ by applying the mean quasar bolometric correction $k_{\rm bol}\approx 10$ \citep[e.g.,][]{2006ApJS..166..470R, 2012MNRAS.422..478R} to the continuum luminosity at 5100 \AA\, extracted from the multicomponent fitting.  This assumes that the underlying continuum originates from the accretion disk of an AGN, and it remains to be investigated if this is a valid approximation for an AGN hosting a TDE. We subtracted the host-galaxy contribution to the $L_{5100}$ luminosity before calculating the Eddington ratios (see Fig. \ref{host_SDSS}). The host-galaxy spectrum was extracted from the SDSS spectrum, using the principal component analysis \citep{VandenBerk_2006}, as explained in \citet{Ilic_2020} (see Sect.~\ref{sec:CLAGN} for further discussion). A word of caution is given for the flux at 5100 \AA\, measured from the fits, since it is sensitive to the spectral multicomponent fitting, but even more to the absolute spectral calibration. Here all spectra are calibrated using photometry, however, for more accurate absolute calibration in the AGN monitoring, one usually applies more precise inter-calibration based on the constant narrow-line flux, such as [O III] or [S II] lines \citep{1992PASP..104..700V, 2017PASP..129b4007F}.

\begin{figure*}
\includegraphics[width=6cm]{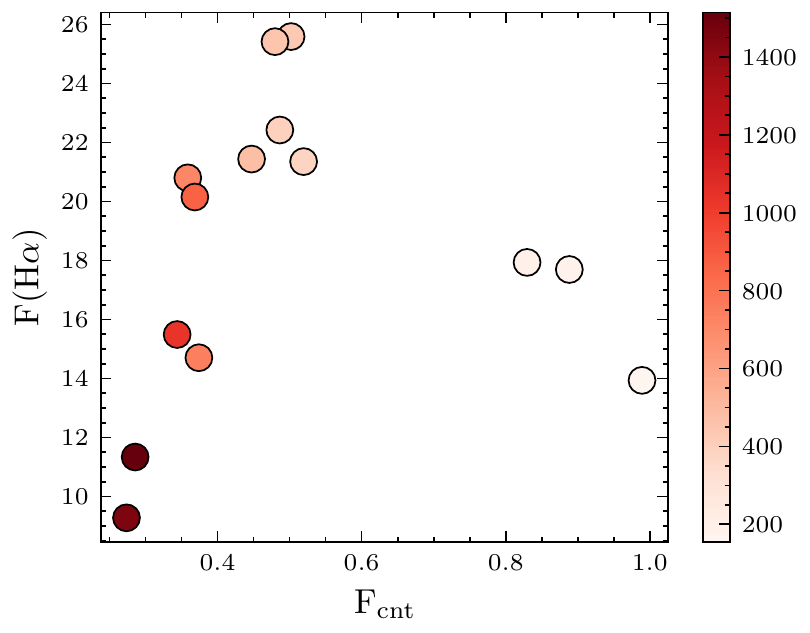}
\includegraphics[width=6cm]{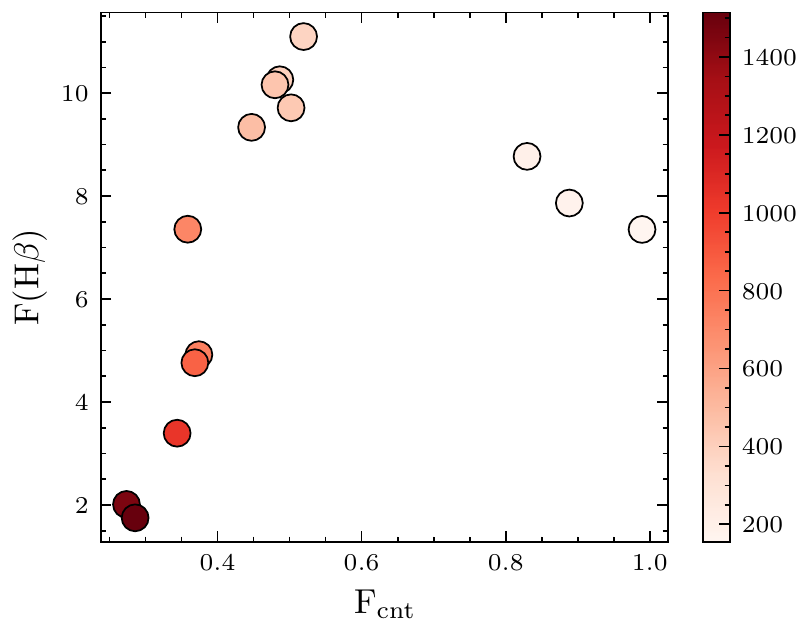}
\includegraphics[width=6cm]{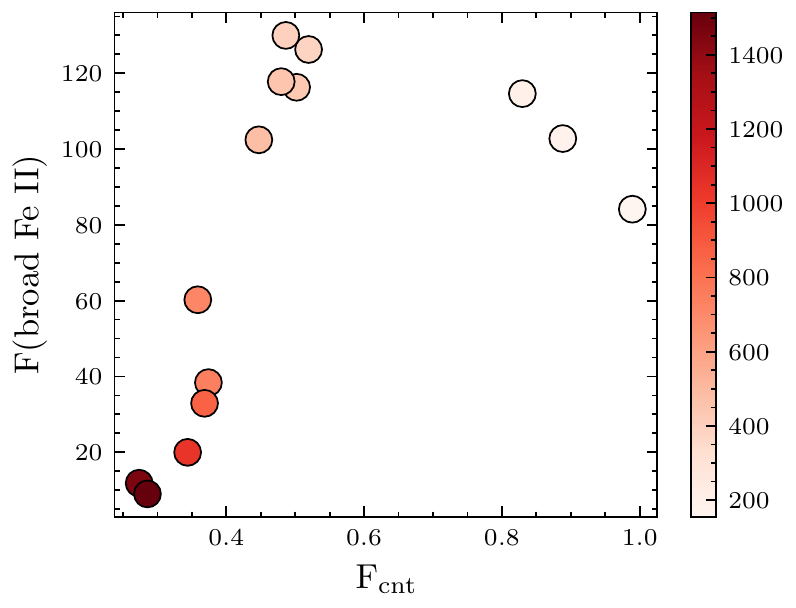}
\caption{Correlations between the continuum flux at 5800 $\AA$,  and the flux of H$\alpha$ (left), H$\beta$ (middle), and total broad Fe II blend (right) in units of $10^{-15}$ $\rm{erg\, s}^{-1}\rm{cm}^{-2}$. The color-bar indicates the day after MJD 57577. The clear correlation of the continuum and line fluxes (apart from the first three epochs) supports the photoionization as the main heating source (see text for details). }
\label{correl}
\end{figure*}

In order to assess the source of ionization of the broad lines and blends, in Fig.~\ref{correl} we plot the correlation between the continuum flux at 5800 \AA\ ($F_{\rm cnt}$), and the flux of H$\alpha$ (left), H$\beta$ (middle), and total Fe II blend (right).  $F_{\rm cnt}$ is measured from the observed spectra as the median of 5790-5810 \AA\, range, since this part is identified as free of emission lines. Apart from the first three epochs (+155, +179, +196) the line fluxes correlate with the continuum flux, supporting that photoionization by the central continuum source is the main heating source of the line emitting gas \citep[see e.g.,][]{2006LNP...693....1N}. 
The outliers are seen in each plot, indicating that in the first period the emission may be induced by other mechanisms, such as shock ionization. This is also supported by the higher gas velocities measured from the line widths. It is important to note that for consistency, all epochs were uniformly fitted with the same initial constraints, listed above.

\subsubsection{Spectral features in the $7000-9000$ \AA \, range}

In the spectral region redward of the H$\alpha$ line, there are broad features visible in the spectra shown in Fig.~\ref{fig_spectra}. These are most notably He I lines, namely He I 6680 (which is attached to the H$\alpha$ red wing), He I $\lambda$7065, He I $\lambda$7281. There are also Ca II triplet ($\lambda$8498, $\lambda$8542, $\lambda$8662), well known oxygen lines O I $\lambda$7774 and O I $\lambda$8446, as well as the Mg II $\lambda$7892 line present. All of these broad features are of similar width as the broad H$\alpha$ component, being at the maximum at a similar time,  and decreasing in intensity and width across time in the similar fashion. A broad feature around 7300 \AA \, was identified as [O II] $\lambda$7320, $\lambda$7330, [Ca II] $\lambda\lambda$7291, 7324 by B17, but the authors suggested that these feature could be associated to Fe II emission lines. However, we believe that this a mixture of Fe II blend and He I $\lambda$7281, as well as the feature centered around 8200 \AA. As it will be shown in the next section, all broad features disappeared in the last VLT/X-shooter spectrum, with possibly only weak Mg II and O I being present.

\subsubsection{Analysis of the medium-resolution VLT/X-shooter spectrum}
\begin{figure*}
\centering
\includegraphics[width=17.5cm]{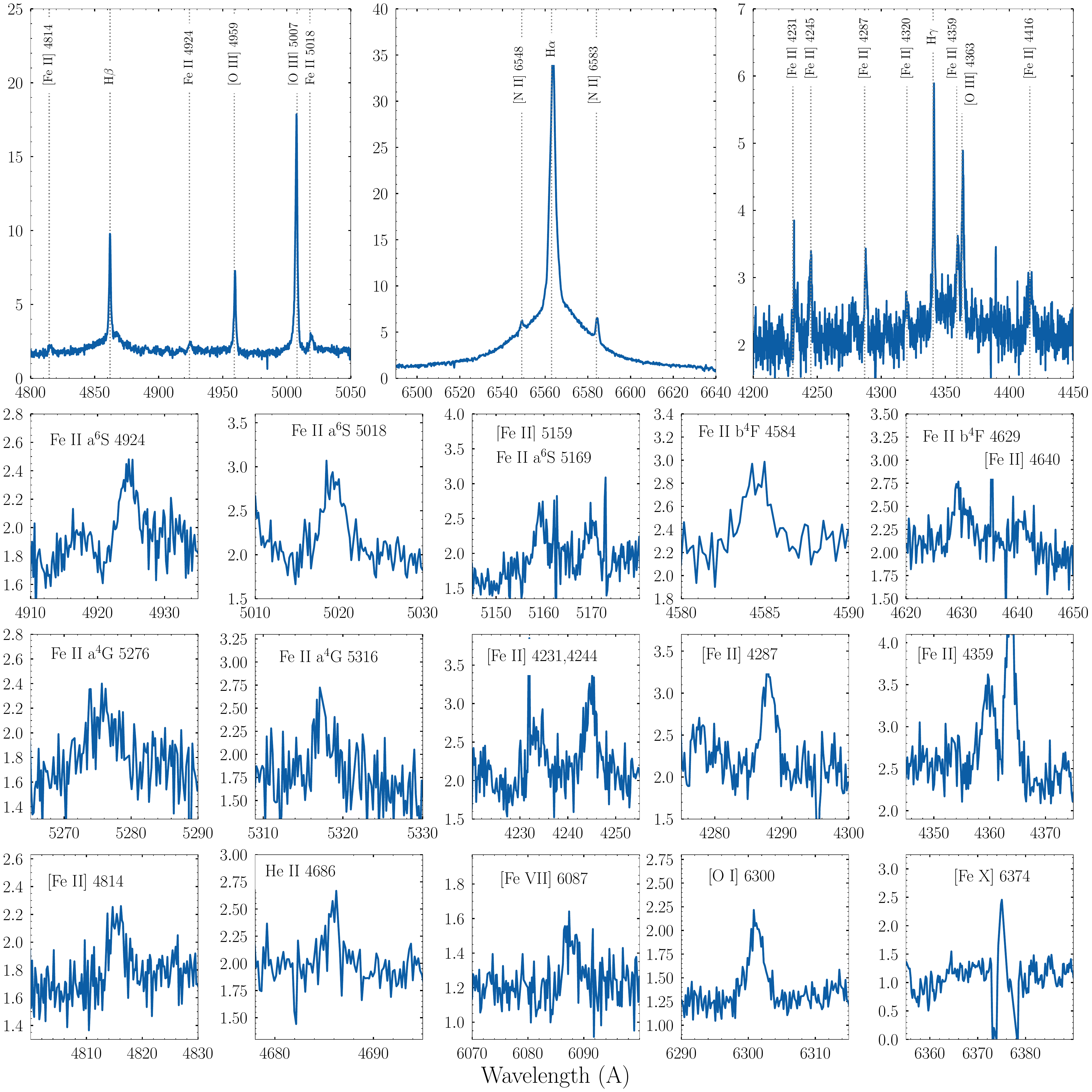}
\caption{The upper panels show the Balmer H$\beta$, H$\alpha$, and $H\gamma$ lines from the medium-spectral resolution VLT/X-shooter spectrum. The lower panels zoom in at the strongest identified Fe II and [Fe II] lines, as well as He II, [O I] and [S II]. The weak coronal line [Fe VII] $\lambda$6087 is also detected.}
\label{xsh_lines}
\end{figure*}

Given that the resolution of the VLT/X-shooter spectrum is much higher compared to those taken with NTT/EFOSC2, we were able to clearly resolve the broad component in the hydrogen Balmer H$\alpha$ and H$\beta$ lines (Fig.~\ref{xsh_lines}), whereas the broad component of the Balmer H$\gamma$ line is too weak to be detected). In the X-shooter spectrum, taken at +1868 days, it is noticeable that the broad Fe II emission has fully faded, and that the galaxy is most likely returning to its preoutburst state. Nevertheless, we were able to identify the strongest narrow Fe II lines coming from the a$^6$S, a$^4$G, and b$^4$F multiplets, as well as forbidden [Fe II] lines, displayed in the zoom-in plots in Fig.~\ref{xsh_lines}. The narrow emission lines can be now unambiguously identified and used for the diagnostics of physical conditions in the ionized gas. We have extracted the following narrow lines: H$\alpha$, H$\beta$, [O I] 6300, [S II] 6716, 6731, and [O II] 3727, 3729.

\begin{figure}
\includegraphics[width=8cm]{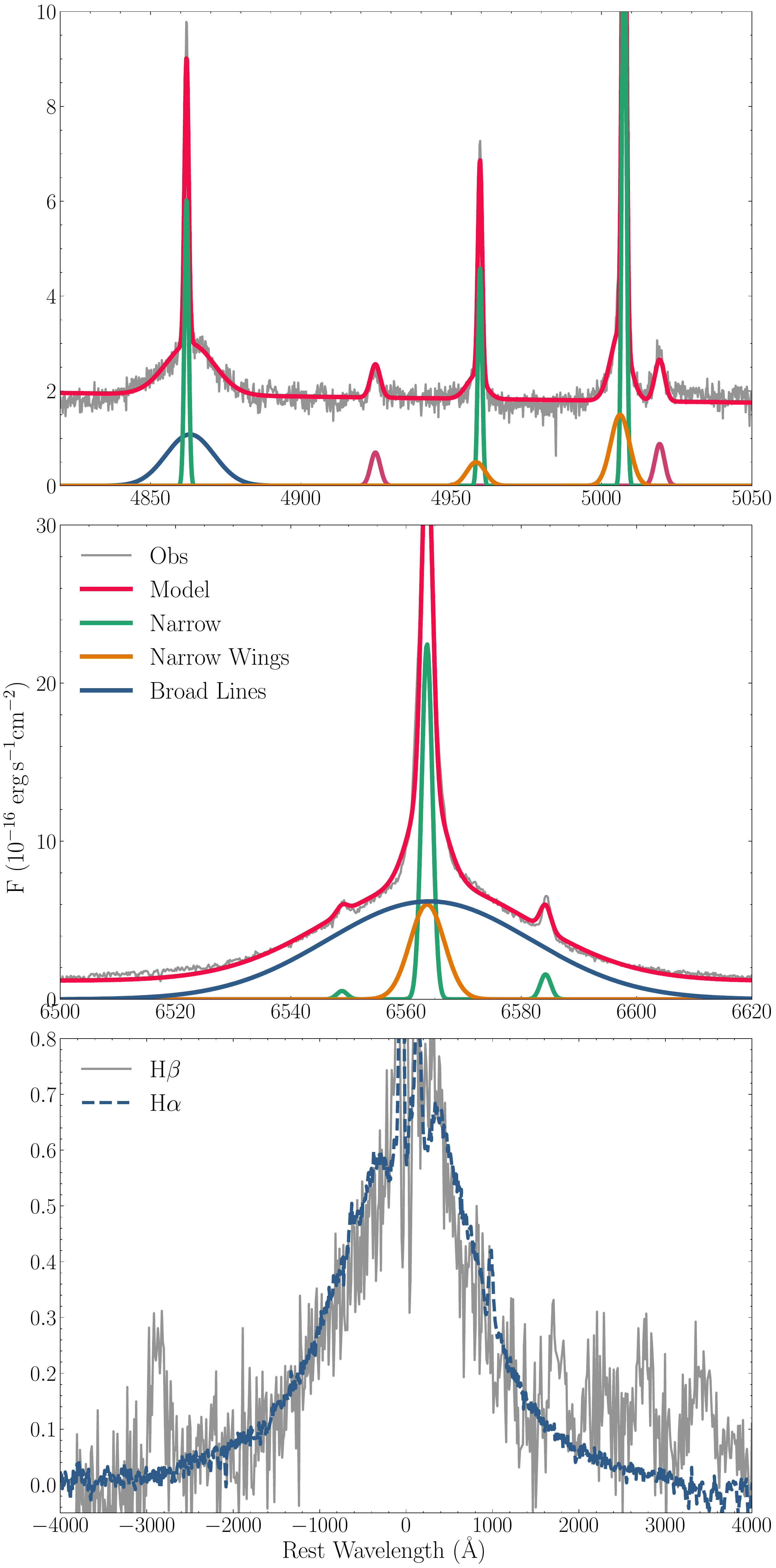}
\caption{Fits of the $\rm H_\beta$ (top) and $\rm H_\alpha$ (middle) line regions in the X-shooter spectrum. The bottom panel compares the broad components (obtained after subtracting the narrow lines) of the $\rm H_\beta$ and $\rm H_\alpha$ in the velocity space.}
\label{xsh_fit}
\end{figure}

We fit the H$\alpha$ and H$\beta$ line region with the multicomponent model in the same manner as described in Sect.~\ref{sec:iron_model} to measure the narrow emission line fluxes, as well as to extract the broad component of the Balmer lines. The strongest narrow emission lines, two [O III] lines and H$\alpha$ are fitted with two Gaussians, one for the core and the other for the line wings (Fig.~\ref{xsh_fit}). These are all constrained to have the same width and shift, and two a$^6$S Fe II lines ratios are fixed according to our model (see Sect.~\ref{sec:iron_model}). 

The width of Fe II lines is $\sim$220 $\rm km\,s^{-1}$, larger than the narrow lines, which have widths that are closer to $\sim$100 $\rm km\,s^{-1}$. This is evident for all iron lines, including the coronal [Fe VII] $\lambda$6087 line. The extracted  broad H$\alpha$ and H$\beta$ components (obtained after subtracting narrow lines) have similar profiles (Fig.~\ref{xsh_fit}, bottom panel), with H$\alpha$ being slightly broader with the width of $\sim$1900 $\rm km\,s^{-1}$. 
Both broad lines show slightly asymmetric profiles, with possibility that the line peak is a slightly shifted redward.

\subsubsection{The presence of coronal lines in the X-shooter spectrum}

The spectral lines coming from forbidden transitions of highly ionized ions (ionization potential larger than 54.4 eV) are typically referred to as coronal lines, and are known to exist in the optical spectra of Seyfert galaxies \citep[see for example][]{Korista_1989,Gelbord_2009}. \citet{Komossa_2008} and \citet{Wang_2012} discovered several coronal line emitters which can be interpreted as light echos from past TDEs. 
The high ionization potential of coronal lines indicates that soft X-rays are required for their production. These lines are typically of somewhat larger width than the narrow lines, with higher critical densities ($\sim10^{7.5} {\rm cm}^{-3}$ \citealt{2006agna.book.....O}).  It is assumed that they come from either the inner narrow-line region or from the inner edge of the torus \citep[e.g.,][]{2015MNRAS.448.2900R}. 
Their strengthening has been associated with some sort of transition event, such as the awaking of a dormant AGN \citep{Ilic_2020} or a TDE in a gas rich environment \citep{Wang_2012}. Recently, \citet{Onori_2022} reported the detection of coronal lines in the TDE AT~2017gge 1700 days after the initial outburst.

In the X-shooter spectrum, we identified a weak [Fe VII] $\lambda$6087 line (see bottom mid-panel in Fig.~\ref{xsh_lines}). Moreover, two other [Fe VII] lines at 5159 \AA\, and 5276 \AA\, could be also misidentified as Fe II lines. Another well-known coronal line [Fe X] 6374 seemed to be present as well, but some telluric disturbance makes its detection quite hard. We examined all previous spectra and the [Fe VII] $\lambda$6087 is either not present or too weak to be resolved in the low-resolution spectra. We also retrieved the spectrum with higher resolution from B17 at +192 taken with the MagE spectrograph to check if the aforementioned line was present, but it is also either too weak to be seen or not present. We note that there could be many other coronal lines \citep[for a review of coronal lines, see e.g.,][]{2015MNRAS.448.2900R} hidden in the strong iron blends throughout the evolution of transient events such as PS16dtm. Future monitoring campaigns of near-by transient events with high resolution instruments and high S/N spectra could shed light on the variability of these lines and on their connection with the occurrence of such transient events. It remains unclear what could be the origin of coronal lines in PS16dtm, since the X-ray emission remains weak in this object (see Sect.~\ref{sec:xray}) contrary to the case for AT~2017gge and AT~2019qiz which experienced considerable late-time X-ray brightening. 

\subsubsection{SMBH mass and AGN bolometric luminosity estimation}\label{sec:smbh_mass}

\begin{figure}[h]
\centering
\includegraphics[width=4.2cm]{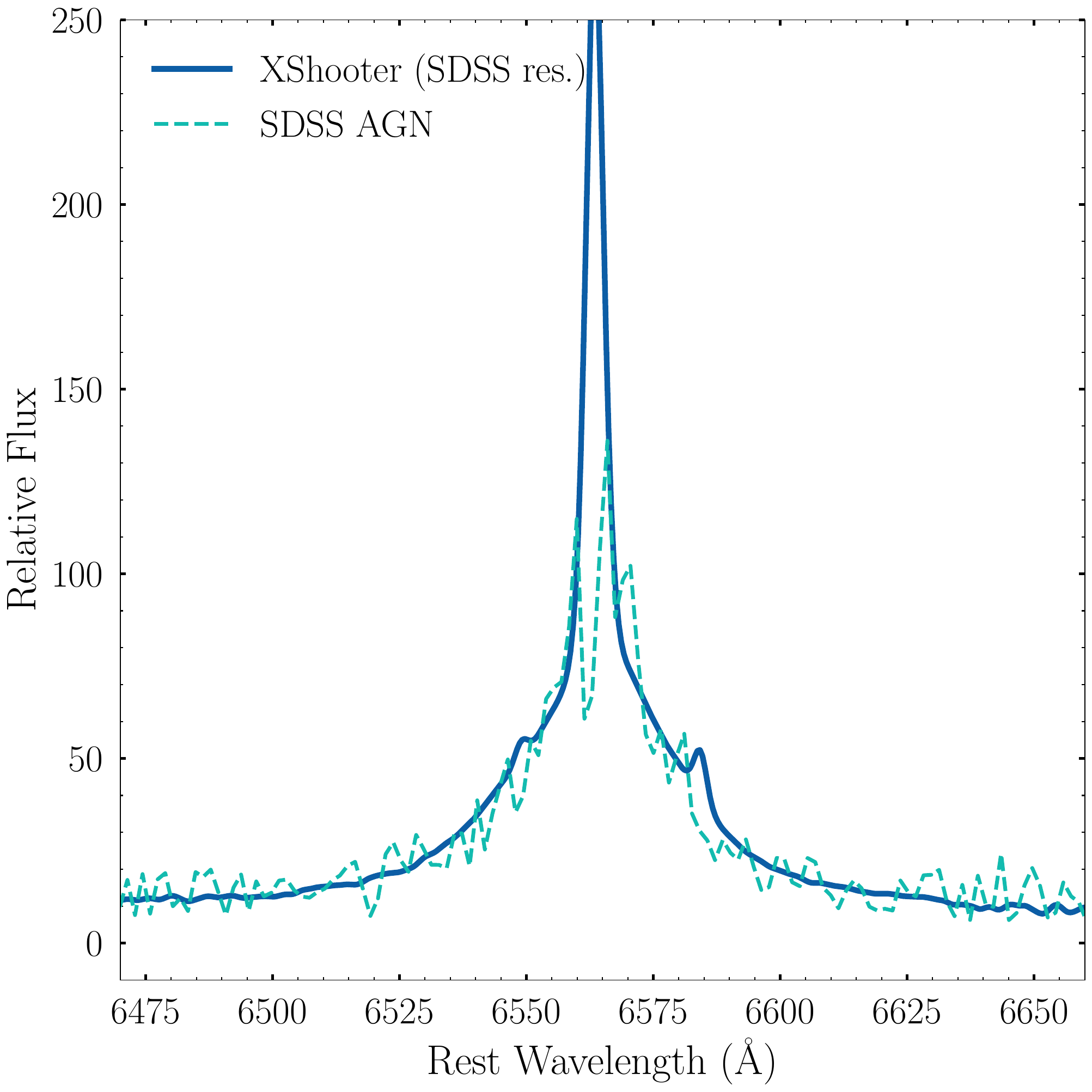}
\includegraphics[width=4.2cm]{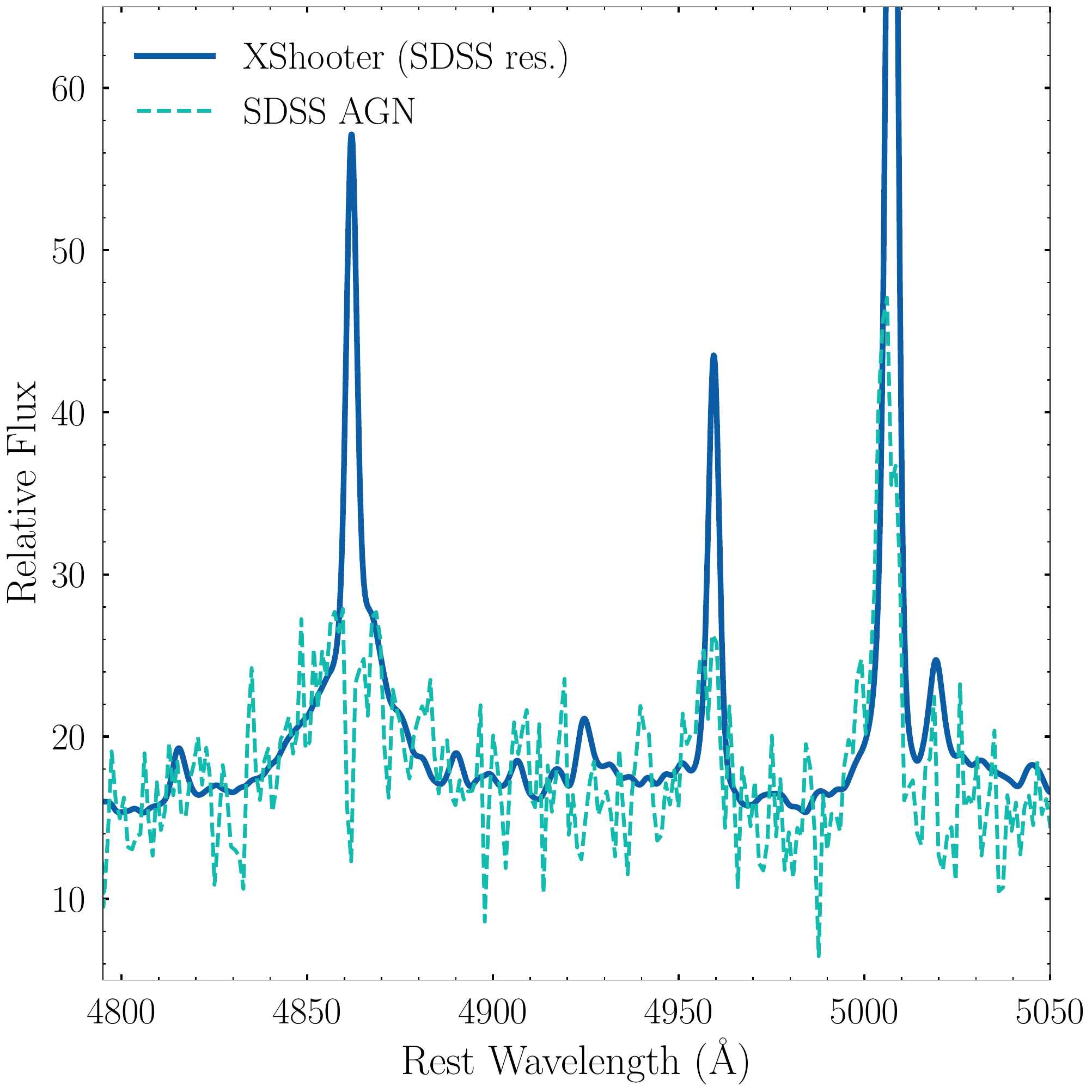}
\caption{X-shooter spectrum, scaled to the SDSS spectrum of poorer spectral resolution, compared to the SDSS AGN component spectra (from Fig.~\ref{host_SDSS}), which was scaled-up so that the broad line wings fit to the broad line of H$\alpha$ (left) and H$\beta$ (right).}
\label{xsh_sdss}
\end{figure}

To be able to measure the SMBH mass from the X-shooter spectrum, we first needed to check if the transition to the preoutburst state, has occurred.  Therefore, we compared the X-shooter with the archival, preoutburst SDSS spectrum. We attempted to scale the X-shooter spectrum to have the same spectral resolution as the SDSS spectrum, for which we used the [O I] 6300 \AA\, that is clearly seen in both spectra and is not contaminated with satellite lines. However, since the SDSS and X-shooter spectra are taken with different apertures (2\arcsec and $0.9 -1$\arcsec, respectively), the absolute re-scaling of the spectra using [O III] or [S II] emission line fluxes was not possible. This was due to the fact that the narrow emission lines are predominantly originating from the host galaxy (see Fig.~\ref{host_SDSS}), and the host-galaxy contribution was significantly different for the two used apertures.
Therefore, we compared the X-shooter spectrum smoothed to the SDSS spectral resolution, to the pure AGN component extracted from the SDSS spectrum, which was scaled-up so that the H$\alpha$ or H$\beta$ have the same broad line intensity, as shown in Fig.~\ref{xsh_sdss}. We then concluded that the broad line profiles of H$\alpha$ or H$\beta$ in the archival SDSS spectrum and in the X-shooter spectrum at +1868 days, are similar.

Using the approach from Sect.~\ref{sec:results_iron_fit} where the bolometric luminosity is expected to scale with the luminosity at 5100$\,\AA\,$, we estimated the AGN bolometric luminosity from the X-shooter spectrum to be $L_{\rm bol} = 3.7 \times 10^{43}$ $\rm{erg\, s}^{-1}$, which is almost reaching the preoutburst value, which we extracted from the SDSS spectrum to be $L_{\rm bol} = 1.6 \times 10^{43}$ $\rm{erg\, s}^{-1}$. Both values are host-corrected using the same constant host contribution estimated from the SDSS spectrum.

Given that the broad H$\beta$ line can be clearly extracted from the X-shooter spectrum, we calculated the mass of the SMBH using the single-epoch method \citep[see e.g.,][and references therein]{2009NewAR..53..140G} and the FWHM of the H$\beta$ line of $1160\pm190$ $\rm km\,s^{-1}$. The radius of the BLR is estimated using the radius-luminosity relation from \cite{2013ApJ...767..149B} applied to the AGN continuum luminosity at $5100\,\AA$. The M$_{\rm BH}$ is then calculated using the virial theorem assumption, with the virial factor of $f=0.75$ \citep[the same as in][who estimated the mass for this object]{2011ApJ...739...28X}. The obtained mass is $M_{\rm BH}= 10^{6.07 \pm 0.18} M_\odot$, similar to the previous estimates by B17 and \citet{2011ApJ...739...28X}. We note that the obtained value is dependent on the assumption for the virial factor $f$  \citep[see discussion in e.g.,][and references therein]{2020ApJ...903..112D}.

\subsection{Spectral classification of the host galaxy }
Previous works concluded that the host of PS16dtm is a NLSy1 galaxy (see Sect.~2 in B17), based on the detection of a weak broad-line component in the H$\alpha$ line \citep{Greene_2007,2011ApJ...739...28X} and high X-ray luminosity of $L_{\rm 2 - 10 keV}\sim 10^{42}$ $\rm{erg\, s}^{-1}$ \citep{Pons_2014}. The later could not be easily accounted for by star formation only, as it would require star formation rates of at least 200 $\rm M_\odot$ yr$^{-1}$ \citep{Ranalli_2003}. Some of the most common features of NLSy1 galaxies are strong and rich Fe II multiplets, however we note that in the quiet phase, the archival optical SDSS spectrum of the host shows no (or very weak) Fe II emission, which classifies this object as a rare type of NLSy1 \citep{Zhou_2006}. It is interesting to note that the spectrum of the host galaxy of the nuclear transient CSS100217 (SDSS J102912.58+404219.7) shows more typical NLSy1 features, such as stronger broad hydrogen Balmer lines, He I, and very prominent Fe II emission around H$\beta$ \citep{Drake_2011}. However, both host galaxies are not typical NLSy1 objects. Their location on the Baldwin-Phillips-Terlevich (BPT) diagrams \citep{Baldwin_1981, 2006MNRAS.372..961K} is either left to or at the border of the AGN–starburst composite objects location, suggesting the presence of significant star formation.

\begin{figure*}
\centering
\includegraphics[width=18cm]{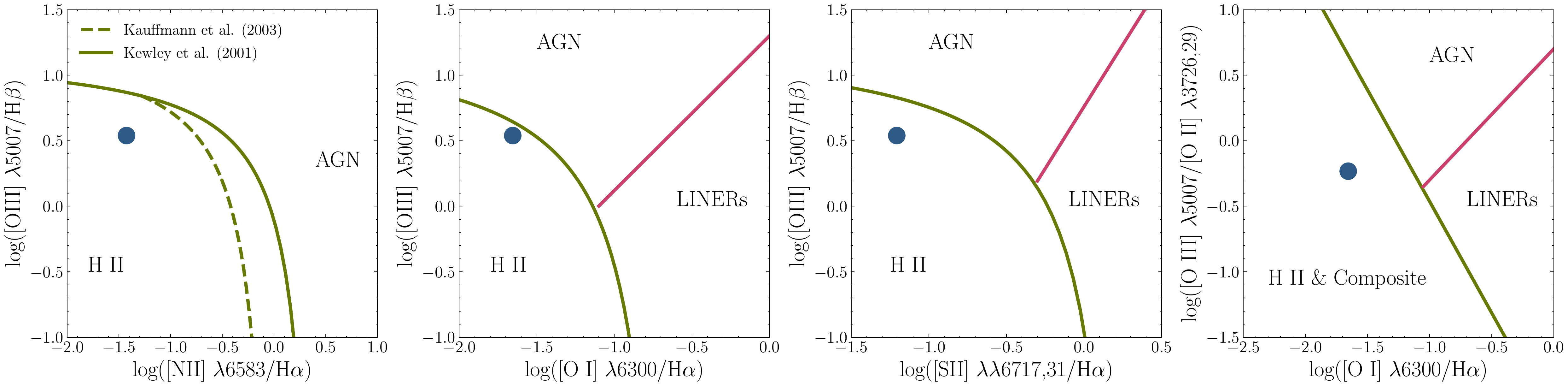}
\caption{Diagnostic diagrams based on the line ratios of [O III] $\lambda$5007/H$\beta$, [N II] $\lambda$6583/H$\alpha$, [O I] $\lambda$6300/H$\alpha$, [S II] $\lambda\lambda$6717,31/H$\alpha$, and [O III] $\lambda$5007/[O II] $\lambda\lambda$3726,3729 \citep{Baldwin_1981, 2006MNRAS.372..961K}. The blue filled circle denotes position of the host galaxy of PS16dtm measured from the X-shooter spectrum. }
\label{BPT}
\end{figure*}

 B17 placed the host AGN more on the border of the starburst-AGN division based on the line-ratios measured from the archival SDSS spectrum. Given that the X-shooter spectrum is showing that is returning to its preoutburst state, we can make use of its higher resolution to perform the diagnostics using the emission line ratios of [O III] $\lambda$5007/H$\beta$, [N II] $\lambda$6583/H$\alpha$, [O I] $\lambda$6300/H$\alpha$, [S II] $\lambda\lambda$6717,31/H$\alpha$, and [O III] $\lambda$5007/[O II] $\lambda\lambda$3726,3729 \citep{Baldwin_1981, 2006MNRAS.372..961K}. In Fig.~\ref{BPT} we plot the diagnostics diagrams using the narrow emission lines measured from the X-shooter spectrum. The location on the BPT diagrams, place the host galaxy in the area of AGN–starburst composite objects. This means that  the narrow-line emission has significant contribution to the ionization from stellar sources, much more than previously shown in B17.

\subsection{PS16dtm in terms of normal AGN variability}

B17 discarded the possibility that PS16dtm can be explained in terms of normal AGN variability. After six years of monitoring this nuclear transient, we can re-open the same question. The variability of an AGN can be assessed through the mean fractional variability parameter $F_{\rm var}$ \citep{Peterson_2001}, that can be approximated with the ratio of the variance and mean value. NLSy1 are known to be low-level variability AGN \citep{Ai_2010} also on longer time-scales \citep{Shapovalova_2012}, except for the radio-loud NLSy1 which exhibit blazar like variability \citep{Berton_2015}. For PS16dtm, the $F_{\rm var}$ is 0.43, 0.26, 0.58 for H$\beta$, H$\alpha$ and total broad Fe II, respectively. Again, we can confirm, with data that spans over six years that such high variability up to 50\% indicates that the outburst is not due to regular AGN activity.  

\begin{figure}
\includegraphics[width=8.5cm]{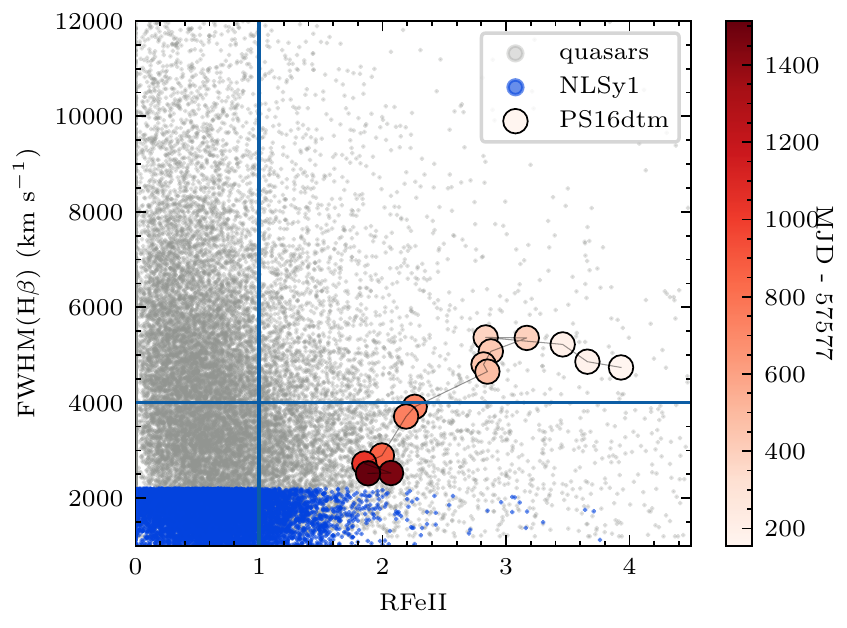}
\caption{Location of PS16dtm (full circles) on the AGN main-sequence plane, that is the FWHM H$\beta$ line vs. $ R_{\rm FeII}$, where $R_{\rm FeII}$ is the ratio of the equivalent width of the FeII, measured in the 4434–4684 \AA\, wavelength range to H$\beta$. The SDSS quasars from \citet{Shen_2011} catalog (gray dots) as well as the NLSy1 from \citet{Rakshit_2017} catalog (blue dots) are also shown. Solid lines at 4000 $\rm km\,s^{-1}$ and $R_{\rm FeII}=1$ indicate separations between different populations of AGN \citep{Marziani_2018}.  The color-bar indicates the day after MJD 57577. }
\label{MS_quasars}
\end{figure}

In Fig.~\ref{MS_quasars}, we investigated the location of the PS16dtm outburst on the AGN main-sequence plane, defined by two measured quantities: the FWHM of H$\beta$ and $R_{\rm FeII}$ \citep{Sulentic_2009,Marziani_2014,Shen_Ho_2014}, where $R_{\rm FeII}$ is the ratio of the equivalent width of the Fe II (we used here only the broad component since it is dominant, measured in the 4434–4684 \AA\, wavelength range) to H$\beta$. Typical AGN have  $R_{\rm FeII}\sim0.4$ with $\sim90\%$ of objects in the range 0.1 to 1. The PS16dtm outburst is located in the extreme right of the main sequence, in the location of highly accreting population A objects with $R_{\rm FeII} > 1$ \citep[see e.g.,][] {Marziani_2014}.

During 1600 days of observations, $R_{\rm FeII}$ and FWHM(H$\beta$) are slowly decreasing toward the preoutburst state, however the location on the main-sequence remains within the extreme accreators, far from the majority of AGN (indicated by SDSS quasars from \citet{Shen_2011} catalog, gray dots) and extreme end of NLSy1 (data from \citet{Rakshit_2017} catalog, blue dots). Such strong $R_{\rm FeII}$ indicates high-accretion rates, most likely super-Eddington accretion \citep{Marziani_2018,Panda_2018}.  This is further supported with our estimated $L_{\rm bol}/L_{\rm Edd}$ ratio, shown in the bottom panel of Fig.~\ref{fitting_lc}.

The strength of Fe II emission in PS16dtm is remarkably high, even when we consider that there are other TDE candidates with Fe II emission lines such as AT~2018fyk \citep{Wevers_2019} and PS1-10adi \citep{2017NatAs...1..865K,He_2021}, with their spectroscopic comparison is shown in Fig.~\ref{fig:iron_strong}. What is striking is that the $R_{\rm FeII}$ is $\sim$5 times larger than most AGN from the literature. We note that fitted spectra are photometrically calibrated, thus, the absolute fluxes should be taken with caution, however, the $R_{\rm FeII}$ is representative of the line ratio, therefore it should be more reliable.

\section{Discussion}\label{sec:disc}

\subsection{Dust echo from MIR emission and the time lags with respect to the optical peak}
 There are several optically discovered TDE candidates with MIR flares \citep[see e.g.,][]{Jiang_2021}. They are detected with a delay with respect to the optical ones, so they are interpreted as dust echoes of the TDE optical emission \citep{Velzen2016}. The “reverberation time lag,” due to light-travel times, between the optical/UV/X-ray variations and the IR response is a commonly used technique in AGN science. Such a time delay can be directly inferred by the light curves and it is presumed to provide an estimate of the distance to the outer radius of the dust torus. 
 However, many parameters influence the IR reverberation response, including the convolution of the UV/optical light curve with a transfer function that contains information about the geometry and structure of the torus \citep[see][for a detailed discussion]{2017ApJ...843....3A}.
 
 For galaxies with AGN, it is possible to estimate the inner radius of the dusty torus from the sublimation radius as  \citep[see e.g.,][]{2016MNRAS.460..980N,Jiang_2019}: 
\begin{equation}\label{eq.}
    R_{sub} = 0.121 \; L_{45}^{0.5}  \; T_{1800}^{-2.804}  \; a_{0.1}^{-0.51} \; (pc)
\end{equation}
where $L_{45}$ is the bolometric luminosity of the AGN in units of $10^{45}$~erg\,s$^{-1}$, $a_{0.1}$ is the grain size in units of 0.1~$\mu$m, and $T_{1800}$ is the temperature of the dust, normalized to the expected sublimation temperature of 1800~K. 
For PS16dtm, it is possible to make this calculation using both the pre- and post-outburst bolometric luminosity, which will yield different values, corresponding to PS16dtm clearing out a larger radius by sublimating the dust closer to the nucleus (J17). Using our estimates of the bolometric luminosity ($1.6 \times 10^{43}$ $\rm{erg\, s}^{-1}$ and $\sim 10^{45}$ $\rm{erg\, s}^{-1}$ pre- and post-outburst), we calculated the inner torus radius to change from $\sim$18 and $\sim$144 light days, respectively. 
It is worth noting that several works have found that the innermost torus radii based on dust reverberation were systematically smaller than the theoretical prediction of Equation \ref{eq.} by a factor of few  \citep[see the recent review][]{2021SSRv..217...63V}.

The MIR light curve was steeply rising already at +9 days with respect to the estimated optical outburst (Fig.~\ref{wise_plus_optical}), although the $W1-W2$ color at this early phase is bluer and the MIR could be consistent with the Rayleigh-Jeans tail of the UV/optical blackbody (Fig.~\ref{two_bb}). 
The MIR light curve is still rising at the end of our monitoring campaign, albeit slowly (see Fig.~\ref{wise_plus_optical}) suggesting that the time delay with respect to the optical must be at least $\gtrsim1500$ rest-frame days, implying for the outer radius of the dust medium to be $R\gtrsim4\times10^{18}$ cm, or $\gtrsim1.3$ pc.

Some authors have used the duration of the MIR echo as a discriminator between the CLAGN and TDE scenario \citep[see e.g.,][]{Kool_2020}, suggesting that CLAGN MIR echos evolve on much longer timescales of several years \citep[see the work by][]{Sheng_2017,Sheng_2020}, compared to TDE echos that evolve on timescales of months. In more recent works that gathered even more TDE and CLAGN candidates, this distinction is not clear anymore.
This is because there are CLAGN candidates that have shorter dust echos and vice versa, there are TDE candidates with longer MIR echos. \citet{Jiang_2021} looked at 23 TDE candidates from the literature and found 11 with NEOWISE emission; the time of the MIR peak ranges from 0 to 800 days after the optical peak, and their duration are from tens to more than 1000 days. In another TDE in a luminous infrared galaxy, a MIR echo lasting for at least 13 years and still going today, is observed \citep{Mattila_2018}. \citet{2022ApJ...927..227L} looked at the WISE data of 13 CLAGN and found the time lags of the variation in the mid-IR behind that in the optical band for 13 CLAGN with strong mid-IR variability, from tens to hundreds of days. We have also inspected the NEOWISE light curve of AT~2018dyk, which was classified as CLAGN by \citet{Frederick_2019}, despite that initially was suspected as TDE candidate \citep{2018ATel11953....1A}. The AT~2018dyk MIR light curve has rosen more that a magnitude above the host level and it is still declining at more than 4 years after the peak. The nuclear transient Gaia16aax which could be explained by both the TDE scenario or some variation in the acretion rate of the active SMBH, showed 140 days of time lag of the NEOWISE light curve compared to the optical \citep{Cannizzaro_2020}. In conclusion, at the moment the duration of the MIR echo cannot be used as a discriminator between the physical cause of the sudden optical flare.

Another proposed discriminator between TDEs and CLAGN is the covering factor of the dust, which can be determined from the ratio of the total energy radiated by the MIR by the energy absorbed by the dust \citep{Gezari_2021}. This has been justified by the work of \citet{Jiang_2021}, who found values on the order of $1\%$ for TDEs in quiescent galaxies\footnote{with the exception of the superluminous TDE ASSASN-15lh.}, while the characteristic  values of CLAGN covering factors   are two orders of magnitude higher.
The dust covering factor of  PS1-10adi is $\sim 40$\% \citep{Jiang_2019}, while for PS16dtm this is estimated to be $>10\% $ \citep[J17,][]{Jiang_2021} as the MIR peak has not been reached yet.  
We note that, since it is possible that TDEs in AGN have similar covering factors with CLAGN, the 
dust covering factor is not a discriminator for TDE vs. CLAGN scenario, but perhaps can indicate the presence of preexisting dusty torus in the cases where the AGN has not been already known.

\begin{figure}
\includegraphics[width=9cm]{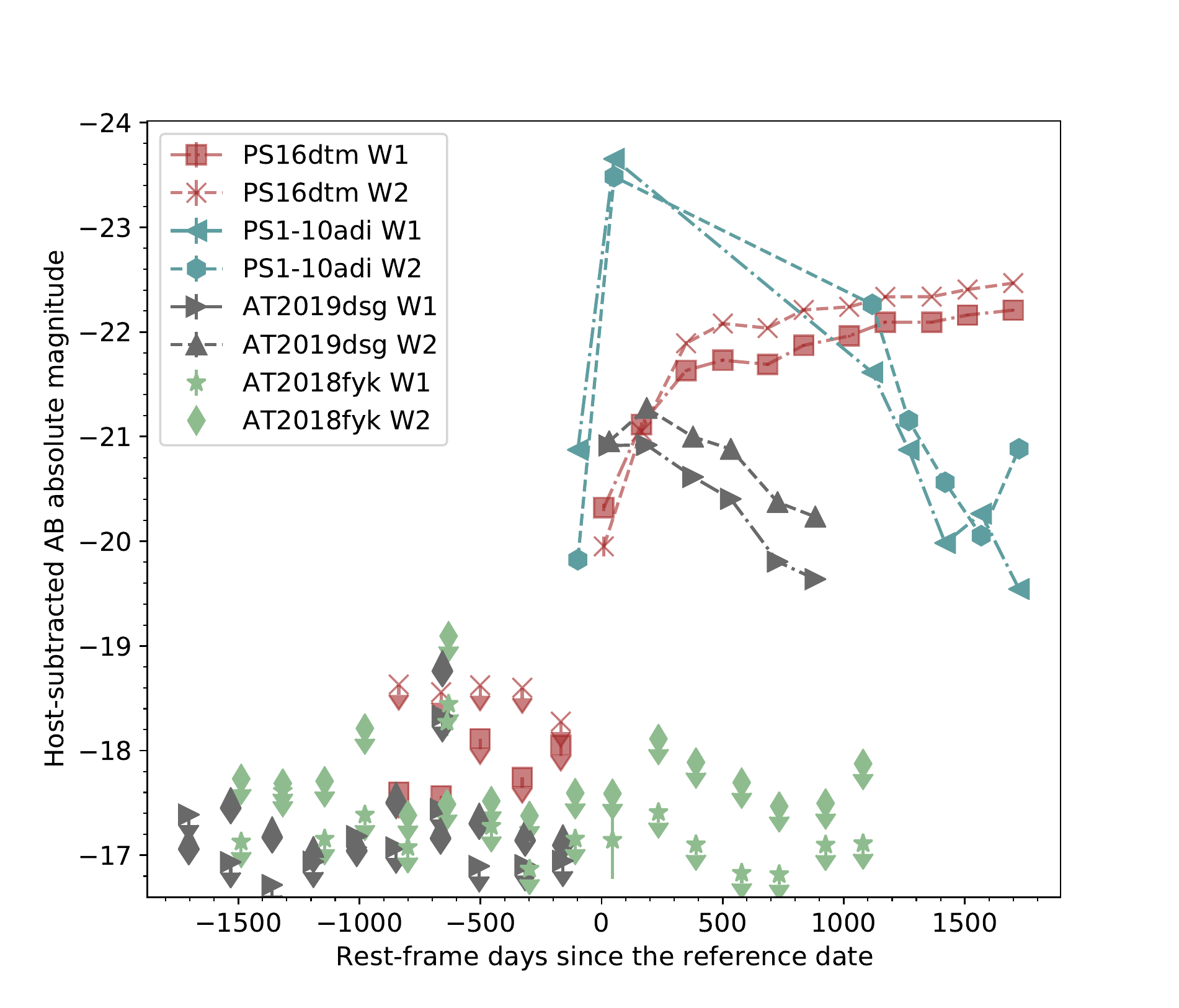}
\caption{Host-subtracted MIR light curves of TDEs from the literature where the Fe II have been identified, PS1-10adi \citep{2017NatAs...1..865K,Jiang_2019,He_2021}, AT~2018fyk \citet{Wevers_2019}, and AT~2019dsg \citep{Cannizzaro2021}. }
\label{fig:iron_strong_MIR}
\end{figure}

In the case of PS16dtm we measure a time lag  between the peak of the broad Fe II component and the continuum light curves (Fig.~\ref{fitting_lc}). The nominal value of this lag is about $\sim$300 days using the nominal peak of the Fe II ($\sim$400 days) and the continuum light curves (70 days for the first peak), although this is uncertain and could could be smaller as the timescales between 200 -- 400 days were not well probed by our spectroscopy. 
This could be consistent with the inner radius of the dust torus after PS16dtm sublimated the dust closest to the black hole, and the idea proposed by \citet{He_2021} for PS1-10adi that Fe II emission is related to the torus inner radius and to Fe that was initially locked in dust, which was evaporated by the TDE.
Given the large uncertainties in the time scales, it is not possible to directly confirm their hypothesis. 
We therefore investigated among TDEs with reported Fe II emission (see Fig.~\ref{fig:iron_strong}) and we found that nearly all of them have strong associated MIR flares (Fig.~\ref{fig:iron_strong_MIR}), with the exception of AT~2018fyk, which is also the one with the weakest Fe II.
This is seems to strengthen the potential relation between Fe II and dust, which needs to be investigated further in the future. 
We note, however, that the Fe II and the MIR emissions in PS16dtm evolve on very different time scales (dropping after 400 days and increasing steadily for 1500 days), so it is not obvious how to establish a  straightforward relation between them.

We finally note that a time lag is also seen between the continuum and H$\alpha$ (see Fig. \ref{fitting_lc}), similar to what has been observed for other TDEs, but a factor of 10 longer \citep{Charalampopoulos_2022}. This time lag is more evident than for Fe II and might or might not be of similar value, with both light curves peaking $\sim400$ days after the outburst. 

\subsection{AGN extreme variability and the TDE scenario}\label{sec:CLAGN}

The nuclear outburst PS16dtm may shed light on an important question on what triggers or maintains the AGN, especially in the cases where extreme variability has been observed. In fact, TDEs is one of the suggested processes that can fuel gas to the SMBH \citep{1975Natur.254..295H,1976MNRAS.176..633F}, and it could work for the low-luminosity end of AGNs where the average accretion rate
from tidal disruptions is enough to account for the luminosity \citep[see also][for a review on this subject]{2001sac..conf..223C}. After the publication of B17, $\sim 200$ CLAGN have been identified, with the detection of the event after it has already occurred, so detailed studies have not been possible \citep{Yang_2018,Frederick_2019,MacLeod_2019,Graham_2020,2021AJ....162..206S, 2022MNRAS.513L..57L,2022ApJ...933..180G,2022MNRAS.511...54H}. The classification into a CLAGN is based on the extreme changes in magnitude  or emission line intensities, with a disappearance or appearance of the broad component in emission lines \citep[see e.g.,][]{1984SvAL...10..335L,1985A&A...146L..11K,2019MNRAS.483..558O,Ilic_2020}. For example, \citet{MacLeod_2019} searched for  $| \Delta g| > 1$ mag, $|\Delta r| > 0.5$ mag variability over any of the available time baselines probed by the SDSS and Pan-STARRS1 surveys.
 The timescales of the variability of these CLAGN cannot be pinpointed precisely since they do not have good coverage, but typical timescales ranges from a roughly a year to 13 years in the rest-frame. 

There is still a vivid debate about what could cause the extreme variability in AGN, with many authors leaning toward the change in the accretion rate as the most plausible scenario \citep{2018MNRAS.480.3898N,2020A&A...641A.167S}, despite that it is difficult to explain the cases with rapid (less than a year) rise in the light curve, as the time scales that govern the dynamics of an accretion disk are much longer in the classical theoretical framework of accretion disk physics \citep{Cannizzaro_2020}. Other possible explanations are the obscuration of the broad-line region \citep[e.g.,][]{2012ApJ...747L..33E}, though questioned by many works \citep[e.g.,][]{Stern_2018,2022ApJ...925...84M},  close binary super-massive black holes \citep{2020A&A...643L...9W}, and supernova explosions \citep[][]{2015A&A...581A..17C}.  Growing number of studies are suggesting that some CLAGN are due to change in the accretion state in the SMBH triggered by a TDE \citep{2017IAUS..324..168K,2019ApJ...883...94T,Zhang_2021}. In fact, the observational signatures of CLAGN can be similar to TDEs that happen in galaxies with AGN, in sense that there is a transition phase that can be accompanied by a drastic change in the AGN continuum flux, so the optical and MIR become brighter when the CLAGN turn on.   

PS16dtm could have been classified as an extreme CLAGN with the change of brightness of several magnitudes (see Figs.~\ref{photometry} and \ref{wise_plus_optical}) according to the criteria in \citet{MacLeod_2019}, but it would have been clearly distinguished from an ordinary AGN variability \citep[e.g.,][]{2004ApJ...601..692V}. Given that PS16dtm was discovered with an all-sky survey which is biased toward the live selection of the most variable object, it is natural that PS16dtm may not be representative of a large population.
However, the rise of PS16dtm luminosity on such short time-scales ($<50$ days), makes it difficult to explain in terms of the intrinsic change to the AGN \citep[e.g., the changes in the accretion disk structure, see e.g.,][]{Stern_2018, Cannizzaro_2020}. From the spectroscopic point of view, the boost in the intensities of broad emission lines, as well as the width of the broad lines, may have also classified as CLAGN. The most striking change is the boost of Fe II emission, especially the appearance of the broad Fe II emission, that completely disappears in time (or gets so weak that it blends with the underlying continuum emission). Interestingly, the nuclear transient SDSS J123359.12+084211.5 in NLSy1 reported by \citet[][]{MacLeod_2019} and classified as a CLAGN, also shows a remarkable change in the Fe II emission. However, that event is a noticeable outlier in the CLAGN sample, especially based on the large Eddington ratio compared to the other CLAGN which are typically seen in lower Eddington ratio AGN \citep{MacLeod_2019}. Despite that the data of SDSS J123359.12+084211.5 are very limited, it is quite possible that PS16dtm and SDSS J123359.12+084211.5 are powered by the same physical mechanism.  We conclude that PS16dtm, without the extensive follow-up campaign shown here, might easily have been easily classified as a CLAGN, especially if the rising part of the light curve was missed. This implies that some CLAGN, such as SDSS J123359.12+084211.5, may actually be powered by a TDE. 

\section{Conclusions} \label{sec:conc}
Here, we studied PS16dtm, which is a nuclear outburst in a NLSy1 galaxy, with photometric and spectroscopic data that spans up to six years after its discovery. These are the main conclusions from these observations:
\begin{itemize}
    \item The NUV/optical light curve (see Fig.~\ref{photometry}), after an initial rise of $\sim$ 50 days, displays another peak after $\sim50$ days. After a plateau phase, it started to decay slowly. In the first $\sim270$ days, the NUV light curve scales with a $t^{-5/7}$ decay, then afterward with a shallower $t^{-1/6}$ decline. The NUV/optical photometry shows very little color evolution. A single blackbody model is not a good fit to the NUV/optical data. We also attempted to correct for extinction with a SMC-like extinction law, for which some authors have suggested that can be applied to the line of sight to AGN, but this did not yield satisfactory results.
    \item The MIR light curve which precedes  the first ATLAS optical detection by $~\sim 4$ days, has steeply risen in the first $\sim 500$ days  and it is still slowly rising  $\sim 1700$ days after the first outburst, but the maximum value has not been reached already. The MIR color in the first epochs compared to the later epochs are quite different. A blackbody fit to the MIR data indicates that  the temperature during the rising part of the light curve was $\sim 2300$ K, it then dropped to $\sim 1200$ K, and then settled to $\sim 900$ K.
    \item Our spectra taken between +155 and +1868 days past the outburst, show the following main properties. The blue continuum has dropped over time, the spectral evolution over the years proceeded with a very slow rate and almost returned to preoutburst level in our last spectrum. 
    The most striking spectroscopic features, except the broad Balmer lines, are the Fe II emission lines.
    We highlight that this broad Fe II emission is transient, in the sense that it was not present in the archival preoutburst spectrum and almost completely disappeared in our last spectrum at +1868 days after the outburst.   We made a spectroscopic comparison with other TDEs with reported broad Fe II emission in the optical, and conclude that PS16dtm is the strongest iron emitter so far. We have also investigated the NEOWISE light curves of these TDE candidates and found that three out of four of them have also shown exceptional MIR flaring.
    \item Thanks to our semi-empirical model which we developed here, we were able to identify and study the evolution of the emission lines. 
  The flux of the broad H$\alpha$, H$\beta$, broad Fe II and narrow Fe II lines, after a rise in the first $\sim200$ days after the outburst, their strength dropped over time after $\sim400$ days after the outburst. We also found that the majority of the flux of the broad Balmer and Fe II lines is consistent with being produced by photoionization at least starting 200 days past the outburst.

      \item We have also studied the MIR light curve and the light curve of the broad Fe II emission lines in order to explore the surrounding region. Given that the quasi-simultaneous rise of the MIR, optical and Fe II emission, it is tempting to assume that the regions from which they originate are nearby and that even the Fe II emitting region might be in the inner radius of the AGN torus.
 On one hand, there are the broad H$\alpha$, Fe II emission line that reach their peaks at $\sim350$ days after the optical data, even though this is uncertain given the gaps in the data. On the other hand, we estimated that the sublimation radius of the host AGN is $\sim 18$ light days. In addition, after the rise, the MIR, optical and Fe II light curve evolve on different timescales. Therefore, it was not possible to establish a direct link between the dust and the iron emitting regions.
     \item We found a weak coronal [Fe VII] 6087 line in our X-shooter spectrum at +1868 days. There are possibly other coronal lines, but their detection is not certain.
    
    \item  B17 predicted that the X-ray emission, which dimmed after the PS16dtm outburst, will reappear when the TDE accretion rate declines, that should be approximately a decade after the outburst. We found only weak X-ray emission in the 0.5-8 keV band at the location of PS16dtm, at +848, +1130 and +1429 days after the outburst. This is consistent with the levels observed by B17, so this reappearance has still not occurred yet, despite that the optical photometry and spectroscopy indicate that the emission is returning to the preoutburst level.
 
\item We have reexamined the host galaxy properties and found that the host galaxy, beyond the AGN emission, has also an important contribution from the ongoing star formation.

\end{itemize}

We have further discussed whether it is possible to use the duration of the dust echo or the dust covering factor as a indicator of the CLAGN or TDE scenario, and concluded that at this point this is not feasible given the overlap of the values of the two. We have reopened the question of whether PS16dtm can fit within the category of what is considered a "normal" AGN variability, and found that the answer is no. However, the extensive PS16dtm data shown here, suggests that perhaps some AGN extreme variability can be explained as a change in the accretion state in the SMBH triggered by a TDE.

The Vera C. Rubin Observatory Legacy Survey of Space and Time \citep[LSST;][]{Ivezic_2019}, will provide a sizeable sample of nuclear outbursts, such as TDEs, supernovae, and variable AGN, during its 10 years of surveying the sky \citep[see e.g.,][]{bricman2019prospects, 2021ApJ...910...93R}. Therefore, a large-scale systematic study in real time from the newly discovered nuclear candidates with the Vera Rubin Observatory will be needed in order to make progress in understanding CLAGN and TDEs.

\begin{acknowledgements}
TP acknowledges the financial support from the Slovenian Research Agency (grants I0-0033, P1-0031, J1-8136 and Z1-1853). This work was supported with travel grants by the Royal Swedish Academy of Sciences and the COST Action CA16104 GWverse.

GL, PC and MP were supported by a research grant (19054) from VILLUM FONDEN.

DI acknowledges funding from the grant 451-03-68/2022-14/200104 of the Ministry of
Education, Science, and Technological Development of the Republic of Serbia,
and the support of the Alexander von Humboldt Foundation.  FO acknowledges support from MIUR, PRIN 2017 (grant 20179ZF5KS) "The new frontier of the Multi-Messenger Astrophysics: follow-up of electromagnetic transient counterparts of gravitational wave sources" and the support of AHEAD2020 grant agreement n.871158. TEMB acknowledges financial support from the Spanish Ministerio de Ciencia e Innovaci\'on (MCIN), the Agencia Estatal de Investigaci\'on (AEI) 10.13039/501100011033 under the PID2020-115253GA-I00 HOSTFLOWS project, from Centro Superior de Investigaciones Cient\'ificas (CSIC) under the PIE project 20215AT016 and the I-LINK 2021 LINKA20409, and the program Unidad de Excelencia Mar\'ia de Maeztu CEX2020-001058-M. MF is supported by a Royal Society - Science Foundation Ireland University Research Fellowship. MG is supported by the EU Horizon 2020 research and innovation programme under grant agreement No 101004719. TMR acknowledges the financial support of the Vilho, Yrj{\"o} and Kalle V{\"a}is{\"a}l{\"a} Foundation of the Finnish academy of Science and Letters. KM is funded by the EU H2020 ERC grant no. 758638. AG acknowledges the financial support from the Slovenian Research Agency (research core funding P1-0031, infrastructure program I0-0033, project grants J1-8136, J1-2460). NI was partially supported by Polish NCN DAINA grant No. 2017/27/L/ST9/03221.

GD was supported by the Ministry of Education, Science and Technological Development of the Republic of Serbia (contract No 451-03-68/2022-14/200002) and by the observing grant support from the Institute of Astronomy and Rozhen NAO BAS through the bilateral joint research project "Gaia Celestial Reference Frame (CRF) and fast variable astronomical objects." \L{}W was partially supported from the Polish NCN grants: Harmonia No. 2018/30/M/ST9/00311, Daina No. 2017/27/L/ST9/03221, MNiSW grant DIR/WK/2018/12 and European Commission's H2020 OPTICON RadioNet Pilot grant No. 101004719.

The Liverpool Telescope is operated on the island of La Palma by Liverpool John Moores University in the Spanish Observatorio del Roque de los Muchachos of the Instituto de Astrofisica de Canarias with financial support from the UK Science and Technology Facilities Council.

This research made use of Astropy\footnote{http://www.astropy.org}, a community-developed core Python package for Astronomy (Astropy Collaboration et al. 2013, 2018).
\end{acknowledgements}

\bibliographystyle{aa} 
\bibliography{references} 

\begin{thebibliography}{153}
\expandafter\ifx\csname natexlab\endcsname\relax\def\natexlab#1{#1}\fi

\bibitem[{{Ai} {et~al.}(2010){Ai}, {Yuan}, {Zhou}, {Wang}, {Dong}, {Wang}, \&
  {Lu}}]{Ai_2010}
{Ai}, Y.~L., {Yuan}, W., {Zhou}, H.~Y., {et~al.} 2010, \apjl, 716, L31

\bibitem[{{Almeyda} {et~al.}(2017){Almeyda}, {Robinson}, {Richmond}, {Vazquez},
  \& {Nikutta}}]{2017ApJ...843....3A}
{Almeyda}, T., {Robinson}, A., {Richmond}, M., {Vazquez}, B., \& {Nikutta}, R.
  2017, \apj, 843, 3

\bibitem[{{Arcavi} {et~al.}(2018){Arcavi}, {Burke}, {French}, {Zabludoff},
  {Hiramatsu}, {McCully}, {Howell}, {Hosseinzadeh}, \&
  {Valenti}}]{2018ATel11953....1A}
{Arcavi}, I., {Burke}, J., {French}, K.~D., {et~al.} 2018, The Astronomer's
  Telegram, 11953, 1

\bibitem[{{Arnaud}(1996)}]{Arnaud1996ASPC..101...17A}
{Arnaud}, K.~A. 1996, in Astronomical Society of the Pacific Conference Series,
  Vol. 101, Astronomical Data Analysis Software and Systems V, ed. G.~H.
  {Jacoby} \& J.~{Barnes}, 17

\bibitem[{{Baldwin} {et~al.}(1981){Baldwin}, {Phillips}, \&
  {Terlevich}}]{Baldwin_1981}
{Baldwin}, J.~A., {Phillips}, M.~M., \& {Terlevich}, R. 1981, \pasp, 93, 5

\bibitem[{{Bellm} {et~al.}(2019){Bellm}, {Kulkarni}, {Barlow}, {Feindt},
  {Graham}, {Goobar}, {Kupfer}, {Ngeow}, {Nugent}, {Ofek}, {Prince}, {Riddle},
  {Walters}, \& {Ye}}]{2019PASP..131f8003B}
{Bellm}, E.~C., {Kulkarni}, S.~R., {Barlow}, T., {et~al.} 2019, \pasp, 131,
  068003

\bibitem[{{Bentz} {et~al.}(2013){Bentz}, {Denney}, {Grier}, {Barth},
  {Peterson}, {Vestergaard}, {Bennert}, {Canalizo}, {De Rosa}, {Filippenko},
  {Gates}, {Greene}, {Li}, {Malkan}, {Pogge}, {Stern}, {Treu}, \&
  {Woo}}]{2013ApJ...767..149B}
{Bentz}, M.~C., {Denney}, K.~D., {Grier}, C.~J., {et~al.} 2013, \apj, 767, 149

\bibitem[{{Berton} {et~al.}(2015){Berton}, {Foschini}, {Ciroi}, {Cracco}, {La
  Mura}, {Lister}, {Mathur}, {Peterson}, {Richards}, \&
  {Rafanelli}}]{Berton_2015}
{Berton}, M., {Foschini}, L., {Ciroi}, S., {et~al.} 2015, \aap, 578, A28

\bibitem[{{Blagorodnova} {et~al.}(2019){Blagorodnova}, {Cenko}, {Kulkarni},
  {Arcavi}, {Bloom}, {Duggan}, {Filippenko}, {Fremling}, {Horesh},
  {Hosseinzadeh}, {Karamehmetoglu}, {Levan}, {Masci}, {Nugent}, {Pasham},
  {Veilleux}, {Walters}, {Yan}, \& {Zheng}}]{2019ApJ...873...92B}
{Blagorodnova}, N., {Cenko}, S.~B., {Kulkarni}, S.~R., {et~al.} 2019, \apj,
  873, 92

\bibitem[{{Blanchard} {et~al.}(2017){Blanchard}, {Nicholl}, {Berger},
  {Guillochon}, {Margutti}, {Chornock}, {Alexand er}, {Leja}, \&
  {Drout}}]{Blanchard_2017}
{Blanchard}, P.~K., {Nicholl}, M., {Berger}, E., {et~al.} 2017, \apj, 843, 106

\bibitem[{{Boroson} \& {Green}(1992)}]{Boroson_1992}
{Boroson}, T.~A. \& {Green}, R.~F. 1992, \apjs, 80, 109

\bibitem[{{Bricman} \& {Gomboc}(2020)}]{bricman2019prospects}
{Bricman}, K. \& {Gomboc}, A. 2020, \apj, 890, 73

\bibitem[{{Calzetti}(2001)}]{Calzetti2001}
{Calzetti}, D. 2001, \pasp, 113, 1449

\bibitem[{{Campana} {et~al.}(2015){Campana}, {Mainetti}, {Colpi}, {Lodato},
  {D'Avanzo}, {Evans}, \& {Moretti}}]{2015A&A...581A..17C}
{Campana}, S., {Mainetti}, D., {Colpi}, M., {et~al.} 2015, \aap, 581, A17

\bibitem[{{Cannizzaro} {et~al.}(2020){Cannizzaro}, {Fraser}, {Jonker},
  {Pringle}, {Mattila}, {Hewett}, {Wevers}, {Kankare}, {Kostrzewa-Rutkowska},
  {Wyrzykowski}, {Onori}, {Harmanen}, {Ford}, {McKernan}, \&
  {Nixon}}]{Cannizzaro_2020}
{Cannizzaro}, G., {Fraser}, M., {Jonker}, P.~G., {et~al.} 2020, \mnras, 493,
  477

\bibitem[{{Cannizzaro} {et~al.}(2022){Cannizzaro}, {Levan}, {van Velzen}, \&
  {Brown}}]{Cannizzaro_2022}
{Cannizzaro}, G., {Levan}, A.~J., {van Velzen}, S., \& {Brown}, G. 2022,
  \mnras, 516, 529

\bibitem[{{Cannizzaro} {et~al.}(2021){Cannizzaro}, {Wevers}, {Jonker},
  {P{\'e}rez-Torres}, {Moldon}, {Mata-S{\'a}nchez}, {Leloudas}, {Pasham},
  {Mattila}, {Arcavi}, {Decker French}, {Onori}, {Inserra}, {Nicholl},
  {Gromadzki}, {Chen}, {M{\"u}ller-Bravo}, {Short}, {Anderson}, {Young},
  {Gendreau}, {Arzoumanian}, {L{\"o}wenstein}, {Remillard}, {Roy}, \&
  {Hiramatsu}}]{Cannizzaro2021}
{Cannizzaro}, G., {Wevers}, T., {Jonker}, P.~G., {et~al.} 2021, \mnras, 504,
  792

\bibitem[{{Chambers} {et~al.}(2016{\natexlab{a}}){Chambers}, {Huber},
  {Flewelling}, {Magnier}, {Primak}, {Schultz}, {Smartt}, {Smith}, {Tonry},
  {Waters}, {Wright}, \& {Young}}]{2016TNSTR.562....1C}
{Chambers}, K.~C., {Huber}, M.~E., {Flewelling}, H., {et~al.}
  2016{\natexlab{a}}, Transient Name Server Discovery Report, 2016-562, 1

\bibitem[{{Chambers} {et~al.}(2016{\natexlab{b}}){Chambers}, {Magnier},
  {Metcalfe}, {Flewelling}, {Huber}, {Waters}, {Denneau}, {Draper}, {Farrow},
  {Finkbeiner}, {Holmberg}, {Koppenhoefer}, {Price}, {Rest}, {Saglia},
  {Schlafly}, {Smartt}, {Sweeney}, {Wainscoat}, {Burgett}, {Chastel}, {Grav},
  {Heasley}, {Hodapp}, {Jedicke}, {Kaiser}, {Kudritzki}, {Luppino}, {Lupton},
  {Monet}, {Morgan}, {Onaka}, {Shiao}, {Stubbs}, {Tonry}, {White},
  {Ba{\~n}ados}, {Bell}, {Bender}, {Bernard}, {Boegner}, {Boffi}, {Botticella},
  {Calamida}, {Casertano}, {Chen}, {Chen}, {Cole}, {Deacon}, {Frenk},
  {Fitzsimmons}, {Gezari}, {Gibbs}, {Goessl}, {Goggia}, {Gourgue}, {Goldman},
  {Grant}, {Grebel}, {Hambly}, {Hasinger}, {Heavens}, {Heckman}, {Henderson},
  {Henning}, {Holman}, {Hopp}, {Ip}, {Isani}, {Jackson}, {Keyes}, {Koekemoer},
  {Kotak}, {Le}, {Liska}, {Long}, {Lucey}, {Liu}, {Martin}, {Masci}, {McLean},
  {Mindel}, {Misra}, {Morganson}, {Murphy}, {Obaika}, {Narayan},
  {Nieto-Santisteban}, {Norberg}, {Peacock}, {Pier}, {Postman}, {Primak},
  {Rae}, {Rai}, {Riess}, {Riffeser}, {Rix}, {R{\"o}ser}, {Russel}, {Rutz},
  {Schilbach}, {Schultz}, {Scolnic}, {Strolger}, {Szalay}, {Seitz}, {Small},
  {Smith}, {Soderblom}, {Taylor}, {Thomson}, {Taylor}, {Thakar}, {Thiel},
  {Thilker}, {Unger}, {Urata}, {Valenti}, {Wagner}, {Walder}, {Walter},
  {Watters}, {Werner}, {Wood-Vasey}, \& {Wyse}}]{2016arXiv161205560C}
{Chambers}, K.~C., {Magnier}, E.~A., {Metcalfe}, N., {et~al.}
  2016{\natexlab{b}}, arXiv e-prints, arXiv:1612.05560

\bibitem[{{Chan} {et~al.}(2021){Chan}, {Piran}, \& {Krolik}}]{Chan2021}
{Chan}, C.-H., {Piran}, T., \& {Krolik}, J.~H. 2021, \apj, 914, 107

\bibitem[{{Charalampopoulos} {et~al.}(2022){Charalampopoulos}, {Leloudas},
  {Malesani}, {Wevers}, {Arcavi}, {Nicholl}, {Pursiainen}, {Lawrence},
  {Anderson}, {Benetti}, {Cannizzaro}, {Chen}, {Galbany}, {Gromadzki},
  {Guti{\'e}rrez}, {Inserra}, {Jonker}, {M{\"u}ller-Bravo}, {Onori}, {Short},
  {Sollerman}, \& {Young}}]{Charalampopoulos_2022}
{Charalampopoulos}, P., {Leloudas}, G., {Malesani}, D.~B., {et~al.} 2022, \aap,
  659, A34

\bibitem[{{Combes}(2001)}]{2001sac..conf..223C}
{Combes}, F. 2001, in Advanced Lectures on the Starburst-AGN, ed.
  I.~{Aretxaga}, D.~{Kunth}, \& R.~{M{\'u}jica}, 223

\bibitem[{{Dalla Bont{\`a}} {et~al.}(2020){Dalla Bont{\`a}}, {Peterson},
  {Bentz}, {Brandt}, {Ciroi}, {De Rosa}, {Fonseca Alvarez}, {Grier}, {Hall},
  {Hern{\'a}ndez Santisteban}, {Ho}, {Homayouni}, {Horne}, {Kochanek}, {Li},
  {Morelli}, {Pizzella}, {Pogge}, {Schneider}, {Shen}, {Trump}, \&
  {Vestergaard}}]{2020ApJ...903..112D}
{Dalla Bont{\`a}}, E., {Peterson}, B.~M., {Bentz}, M.~C., {et~al.} 2020, \apj,
  903, 112

\bibitem[{{Dimitrijevi{\'c}} {et~al.}(2007){Dimitrijevi{\'c}}, {Popovi{\'c}},
  {Kova{\v{c}}evi{\'c}}, {Da{\v{c}}i{\'c}}, \& {Ili{\'c}}}]{Dimitrijevic_2007}
{Dimitrijevi{\'c}}, M.~S., {Popovi{\'c}}, L.~{\v{C}}., {Kova{\v{c}}evi{\'c}},
  J., {Da{\v{c}}i{\'c}}, M., \& {Ili{\'c}}, D. 2007, \mnras, 374, 1181

\bibitem[{{Dong} {et~al.}(2016){Dong}, {Chen}, {Bose}, {Stanek}, {Kochanek},
  {Holoien}, {Shappee}, {Prieto}, {Brown}, \& {Milne}}]{2016ATel.9843....1D}
{Dong}, S., {Chen}, P., {Bose}, S., {et~al.} 2016, The Astronomer's Telegram,
  9843, 1

\bibitem[{{Dong} {et~al.}(2011){Dong}, {Wang}, {Ho}, {Wang}, {Fan}, {Wang},
  {Zhou}, \& {Yuan}}]{Dong_2011}
{Dong}, X.-B., {Wang}, J.-G., {Ho}, L.~C., {et~al.} 2011, \apj, 736, 86

\bibitem[{{Drake} {et~al.}(2011){Drake}, {Djorgovski}, {Mahabal}, {Anderson},
  {Roy}, {Mohan}, {Ravindranath}, {Frail}, {Gezari}, {Neill}, {Ho}, {Prieto},
  {Thompson}, {Thorstensen}, {Wagner}, {Kowalski}, {Chiang}, {Grove},
  {Schinzel}, {Wood}, {Carrasco}, {Recillas}, {Kewley}, {Archana}, {Basu},
  {Wadadekar}, {Kumar}, {Myers}, {Phinney}, {Williams}, {Graham}, {Catelan},
  {Beshore}, {Larson}, \& {Christensen}}]{Drake_2011}
{Drake}, A.~J., {Djorgovski}, S.~G., {Mahabal}, A., {et~al.} 2011, \apj, 735,
  106

\bibitem[{{Elitzur}(2012)}]{2012ApJ...747L..33E}
{Elitzur}, M. 2012, \apjl, 747, L33

\bibitem[{{Fausnaugh}(2017)}]{2017PASP..129b4007F}
{Fausnaugh}, M.~M. 2017, \pasp, 129, 024007

\bibitem[{{Fitch} {et~al.}(1967){Fitch}, {Pacholczyk}, \&
  {Weymann}}]{1967ApJ...150L..67F}
{Fitch}, W.~S., {Pacholczyk}, A.~G., \& {Weymann}, R.~J. 1967, \apjl, 150, L67

\bibitem[{{Foreman-Mackey} {et~al.}(2013){Foreman-Mackey}, {Hogg}, {Lang}, \&
  {Goodman}}]{ForemanMackey2013}
{Foreman-Mackey}, D., {Hogg}, D.~W., {Lang}, D., \& {Goodman}, J. 2013, \pasp,
  125, 306

\bibitem[{{Frank} \& {Rees}(1976)}]{1976MNRAS.176..633F}
{Frank}, J. \& {Rees}, M.~J. 1976, \mnras, 176, 633

\bibitem[{Frederick {et~al.}(2019)Frederick, Gezari, Graham, Cenko, van Velzen,
  Stern, Blagorodnova, Kulkarni, Yan, De, \& et~al.}]{Frederick_2019}
Frederick, S., Gezari, S., Graham, M.~J., {et~al.} 2019, The Astrophysical
  Journal, 883, 31

\bibitem[{{Frederick} {et~al.}(2021){Frederick}, {Gezari}, {Graham},
  {Sollerman}, {van Velzen}, {Perley}, {Stern}, {Ward}, {Hammerstein}, {Hung},
  {Yan}, {Andreoni}, {Bellm}, {Duev}, {Kowalski}, {Mahabal}, {Masci},
  {Medford}, {Rusholme}, {Smith}, \& {Walters}}]{2021ApJ...920...56F}
{Frederick}, S., {Gezari}, S., {Graham}, M.~J., {et~al.} 2021, \apj, 920, 56

\bibitem[{{Freudling} {et~al.}(2013){Freudling}, {Romaniello}, {Bramich},
  {Ballester}, {Forchi}, {Garc{\'{\i}}a-Dabl{\'o}}, {Moehler}, \&
  {Neeser}}]{2013A&A...559A..96F}
{Freudling}, W., {Romaniello}, M., {Bramich}, D.~M., {et~al.} 2013, \aap, 559,
  A96

\bibitem[{{Fruscione} {et~al.}(2006){Fruscione}, {McDowell}, {Allen},
  {Brickhouse}, {Burke}, {Davis}, {Durham}, {Elvis}, {Galle}, {Harris},
  {Huenemoerder}, {Houck}, {Ishibashi}, {Karovska}, {Nicastro}, {Noble},
  {Nowak}, {Primini}, {Siemiginowska}, {Smith}, \&
  {Wise}}]{2006SPIE.6270E..1VF}
{Fruscione}, A., {McDowell}, J.~C., {Allen}, G.~E., {et~al.} 2006, in Society
  of Photo-Optical Instrumentation Engineers (SPIE) Conference Series, Vol.
  6270, Society of Photo-Optical Instrumentation Engineers (SPIE) Conference
  Series, ed. D.~R. {Silva} \& R.~E. {Doxsey}, 62701V

\bibitem[{{Gafton} \& {Rosswog}(2019)}]{2019MNRAS.487.4790G}
{Gafton}, E. \& {Rosswog}, S. 2019, \mnras, 487, 4790

\bibitem[{{Gaskell}(2009)}]{2009NewAR..53..140G}
{Gaskell}, C.~M. 2009, \nar, 53, 140

\bibitem[{{Gehrels} {et~al.}(2004){Gehrels}, {Chincarini}, {Giommi}, {Mason},
  {Nousek}, {Wells}, {White}, {Barthelmy}, {Burrows}, {Cominsky}, {Hurley},
  {Marshall}, {M{\'e}sz{\'a}ros}, {Roming}, {Angelini}, {Barbier}, {Belloni},
  {Campana}, {Caraveo}, {Chester}, {Citterio}, {Cline}, {Cropper}, {Cummings},
  {Dean}, {Feigelson}, {Fenimore}, {Frail}, {Fruchter}, {Garmire}, {Gendreau},
  {Ghisellini}, {Greiner}, {Hill}, {Hunsberger}, {Krimm}, {Kulkarni}, {Kumar},
  {Lebrun}, {Lloyd-Ronning}, {Markwardt}, {Mattson}, {Mushotzky}, {Norris},
  {Osborne}, {Paczynski}, {Palmer}, {Park}, {Parsons}, {Paul}, {Rees},
  {Reynolds}, {Rhoads}, {Sasseen}, {Schaefer}, {Short}, {Smale}, {Smith},
  {Stella}, {Tagliaferri}, {Takahashi}, {Tashiro}, {Townsley}, {Tueller},
  {Turner}, {Vietri}, {Voges}, {Ward}, {Willingale}, {Zerbi}, \&
  {Zhang}}]{2004ApJ...611.1005G}
{Gehrels}, N., {Chincarini}, G., {Giommi}, P., {et~al.} 2004, \apj, 611, 1005

\bibitem[{Gelbord {et~al.}(2009)Gelbord, Mullaney, \& Ward}]{Gelbord_2009}
Gelbord, J.~M., Mullaney, J.~R., \& Ward, M.~J. 2009, Monthly Notices of the
  Royal Astronomical Society, 397, 172

\bibitem[{{Gezari}(2021)}]{Gezari_2021}
{Gezari}, S. 2021, \araa, 59 [\eprint[arXiv]{2104.14580}]

\bibitem[{{Gezari} {et~al.}(2009){Gezari}, {Heckman}, {Cenko}, {Eracleous},
  {Forster}, {Gon{\c{c}}alves}, {Martin}, {Morrissey}, {Neff}, {Seibert},
  {Schiminovich}, \& {Wyder}}]{2009ApJ...698.1367G}
{Gezari}, S., {Heckman}, T., {Cenko}, S.~B., {et~al.} 2009, \apj, 698, 1367

\bibitem[{{Gordon} {et~al.}(2003){Gordon}, {Clayton}, {Misselt}, {Landolt}, \&
  {Wolff}}]{2003ApJ...594..279G}
{Gordon}, K.~D., {Clayton}, G.~C., {Misselt}, K.~A., {Landolt}, A.~U., \&
  {Wolff}, M.~J. 2003, \apj, 594, 279

\bibitem[{{Graham} {et~al.}(2020){Graham}, {Ross}, {Stern}, {Drake},
  {McKernan}, {Ford}, {Djorgovski}, {Mahabal}, {Glikman}, {Larson}, \&
  {Christensen}}]{Graham_2020}
{Graham}, M.~J., {Ross}, N.~P., {Stern}, D., {et~al.} 2020, \mnras, 491, 4925

\bibitem[{{Green} {et~al.}(2022){Green}, {Pulgarin-Duque}, {Anderson},
  {MacLeod}, {Eracleous}, {Ruan}, {Runnoe}, {Graham}, {Roulston}, {Schneider},
  {Ahlf}, {Bizyaev}, {Brownstein}, {del Casal}, {Dodd}, {Hoover}, {Matt},
  {Merloni}, {Pan}, {Ramirez}, \& {Ridder}}]{2022ApJ...933..180G}
{Green}, P.~J., {Pulgarin-Duque}, L., {Anderson}, S.~F., {et~al.} 2022, \apj,
  933, 180

\bibitem[{{Greene} \& {Ho}(2007)}]{Greene_2007}
{Greene}, J.~E. \& {Ho}, L.~C. 2007, \apj, 670, 92

\bibitem[{{Guillochon} \& {Ramirez-Ruiz}(2013)}]{2013ApJ...767...25G}
{Guillochon}, J. \& {Ramirez-Ruiz}, E. 2013, \apj, 767, 25

\bibitem[{{He} {et~al.}(2021){He}, {Jiang}, {Wang}, {Liu}, {Sun}, {Guo},
  {Shen}, {Cai}, {Shu}, {Sheng}, {Liang}, \& {Xu}}]{He_2021}
{He}, Z., {Jiang}, N., {Wang}, T., {et~al.} 2021, \apjl, 907, L29

\bibitem[{{Hills}(1975)}]{1975Natur.254..295H}
{Hills}, J.~G. 1975, \nat, 254, 295

\bibitem[{{Hinkle} {et~al.}(2021){Hinkle}, {Holoien}, {Shappee}, \&
  {Auchettl}}]{2021ApJ...910...83H}
{Hinkle}, J.~T., {Holoien}, T. W.~S., {Shappee}, B.~J., \& {Auchettl}, K. 2021,
  \apj, 910, 83

\bibitem[{{Hinkle} {et~al.}(2022){Hinkle}, {Holoien}, {Shappee}, {Neustadt},
  {Auchettl}, {Vallely}, {Shahbandeh}, {Kluge}, {Kochanek}, {Stanek}, {Huber},
  {Post}, {Bersier}, {Ashall}, {Tucker}, {Williams}, {de Jaeger}, {Do},
  {Fausnaugh}, {Gruen}, {Hopp}, {Myles}, {Obermeier}, {Payne}, \&
  {Thompson}}]{2022ApJ...930...12H}
{Hinkle}, J.~T., {Holoien}, T. W.~S., {Shappee}, B.~J., {et~al.} 2022, \apj,
  930, 12

\bibitem[{{Hodgkin} {et~al.}(2021){Hodgkin}, {Harrison}, {Breedt}, {Wevers},
  {Rixon}, {Delgado}, {Yoldas}, {Kostrzewa-Rutkowska}, {Wyrzykowski}, {van
  Leeuwen}, {Blagorodnova}, {Campbell}, {Eappachen}, {Fraser}, {Ihanec},
  {Koposov}, {Kruszy{\'n}ska}, {Marton}, {Rybicki}, {Brown}, {Burgess},
  {Busso}, {Cowell}, {De Angeli}, {Diener}, {Evans}, {Gilmore}, {Holland},
  {Jonker}, {van Leeuwen}, {Mignard}, {Osborne}, {Portell}, {Prusti},
  {Richards}, {Riello}, {Seabroke}, {Walton}, {{\'A}brah{\'a}m}, {Altavilla},
  {Baker}, {Bastian}, {O'Brien}, {de Bruijne}, {Butterley}, {Carrasco},
  {Casta{\~n}eda}, {Clark}, {Clementini}, {Copperwheat}, {Cropper},
  {Damljanovic}, {Davidson}, {Davis}, {Dennefeld}, {Dhillon}, {Dolding},
  {Dominik}, {Esquej}, {Eyer}, {Fabricius}, {Fridman}, {Froebrich}, {Garralda},
  {Gomboc}, {Gonz{\'a}lez-Vidal}, {Guerra}, {Hambly}, {Hardy}, {Holl},
  {Hourihane}, {Japelj}, {Kann}, {Kiss}, {Knigge}, {Kolb}, {Komossa},
  {K{\'o}sp{\'a}l}, {Kov{\'a}cs}, {Kun}, {Leto}, {Lewis}, {Littlefair},
  {Mahabal}, {Mundell}, {Nagy}, {Padeletti}, {Palaversa}, {Pigulski},
  {Pretorius}, {van Reeven}, {Ribeiro}, {Roelens}, {Rowell}, {Schartel},
  {Scholz}, {Schwope}, {Sip{\H{o}}cz}, {Smartt}, {Smith}, {Serraller},
  {Steeghs}, {Sullivan}, {Szabados}, {Szegedi-Elek}, {Tisserand}, {Tomasella},
  {van Velzen}, {Whitelock}, {Wilson}, \& {Young}}]{2021A&A...652A..76H}
{Hodgkin}, S.~T., {Harrison}, D.~L., {Breedt}, E., {et~al.} 2021, \aap, 652,
  A76

\bibitem[{{Hon} {et~al.}(2022){Hon}, {Wolf}, {Onken}, {Webster}, \&
  {Auchettl}}]{2022MNRAS.511...54H}
{Hon}, W.~J., {Wolf}, C., {Onken}, C.~A., {Webster}, R., \& {Auchettl}, K.
  2022, \mnras, 511, 54

\bibitem[{{Huber} {et~al.}(2015){Huber}, {Chambers}, {Flewelling}, {Willman},
  {Primak}, {Schultz}, {Gibson}, {Magnier}, {Waters}, {Tonry}, {Wainscoat},
  {Smith}, {Wright}, {Smartt}, {Foley}, {Jha}, {Rest}, \&
  {Scolnic}}]{2015ATel.7153....1H}
{Huber}, M., {Chambers}, K.~C., {Flewelling}, H., {et~al.} 2015, The
  Astronomer's Telegram, 7153, 1

\bibitem[{{Ili{\'c}} {et~al.}(2020){Ili{\'c}}, {Oknyansky}, {Popovi{\'c}},
  {Tsygankov}, {Belinski}, {Tatarnikov}, {Dodin}, {Shatsky}, {Ikonnikova},
  {Raki{\'c}}, {Kova{\v{c}}evi{\'c}}, {Mar{\v{c}}eta-Mandi{\'c}}, {Burlak},
  {Mishin}, {Metlova}, {Potanin}, \& {Zheltoukhov}}]{Ilic_2020}
{Ili{\'c}}, D., {Oknyansky}, V., {Popovi{\'c}}, L.~{\v{C}}., {et~al.} 2020,
  \aap, 638, A13

\bibitem[{{Ivezi{\'c}} {et~al.}(2019){Ivezi{\'c}}, {Kahn}, {Tyson}, {Abel},
  {Acosta}, {Allsman}, {Alonso}, {AlSayyad}, {Anderson}, {Andrew}, {Angel},
  {Angeli}, {Ansari}, {Antilogus}, {Araujo}, {Armstrong}, {Arndt}, {Astier},
  {Aubourg}, {Auza}, {Axelrod}, {Bard}, {Barr}, {Barrau}, {Bartlett}, {Bauer},
  {Bauman}, {Baumont}, {Bechtol}, {Bechtol}, {Becker}, {Becla}, {Beldica},
  {Bellavia}, {Bianco}, {Biswas}, {Blanc}, {Blazek}, {Blandford}, {Bloom},
  {Bogart}, {Bond}, {Booth}, {Borgland}, {Borne}, {Bosch}, {Boutigny},
  {Brackett}, {Bradshaw}, {Brandt}, {Brown}, {Bullock}, {Burchat}, {Burke},
  {Cagnoli}, {Calabrese}, {Callahan}, {Callen}, {Carlin}, {Carlson},
  {Chandrasekharan}, {Charles-Emerson}, {Chesley}, {Cheu}, {Chiang}, {Chiang},
  {Chirino}, {Chow}, {Ciardi}, {Claver}, {Cohen-Tanugi}, {Cockrum}, {Coles},
  {Connolly}, {Cook}, {Cooray}, {Covey}, {Cribbs}, {Cui}, {Cutri}, {Daly},
  {Daniel}, {Daruich}, {Daubard}, {Daues}, {Dawson}, {Delgado}, {Dellapenna},
  {de Peyster}, {de Val-Borro}, {Digel}, {Doherty}, {Dubois},
  {Dubois-Felsmann}, {Durech}, {Economou}, {Eifler}, {Eracleous}, {Emmons},
  {Fausti Neto}, {Ferguson}, {Figueroa}, {Fisher-Levine}, {Focke}, {Foss},
  {Frank}, {Freemon}, {Gangler}, {Gawiser}, {Geary}, {Gee}, {Geha}, {Gessner},
  {Gibson}, {Gilmore}, {Glanzman}, {Glick}, {Goldina}, {Goldstein}, {Goodenow},
  {Graham}, {Gressler}, {Gris}, {Guy}, {Guyonnet}, {Haller}, {Harris},
  {Hascall}, {Haupt}, {Hernandez}, {Herrmann}, {Hileman}, {Hoblitt}, {Hodgson},
  {Hogan}, {Howard}, {Huang}, {Huffer}, {Ingraham}, {Innes}, {Jacoby}, {Jain},
  {Jammes}, {Jee}, {Jenness}, {Jernigan}, {Jevremovi{\'c}}, {Johns}, {Johnson},
  {Johnson}, {Jones}, {Juramy-Gilles}, {Juri{\'c}}, {Kalirai}, {Kallivayalil},
  {Kalmbach}, {Kantor}, {Karst}, {Kasliwal}, {Kelly}, {Kessler}, {Kinnison},
  {Kirkby}, {Knox}, {Kotov}, {Krabbendam}, {Krughoff}, {Kub{\'a}nek},
  {Kuczewski}, {Kulkarni}, {Ku}, {Kurita}, {Lage}, {Lambert}, {Lange},
  {Langton}, {Le Guillou}, {Levine}, {Liang}, {Lim}, {Lintott}, {Long},
  {Lopez}, {Lotz}, {Lupton}, {Lust}, {MacArthur}, {Mahabal}, {Mandelbaum},
  {Markiewicz}, {Marsh}, {Marshall}, {Marshall}, {May}, {McKercher}, {McQueen},
  {Meyers}, {Migliore}, {Miller}, {Mills}, {Miraval}, {Moeyens}, {Moolekamp},
  {Monet}, {Moniez}, {Monkewitz}, {Montgomery}, {Morrison}, {Mueller},
  {Muller}, {Mu{\~n}oz Arancibia}, {Neill}, {Newbry}, {Nief}, {Nomerotski},
  {Nordby}, {O'Connor}, {Oliver}, {Olivier}, {Olsen}, {O'Mullane}, {Ortiz},
  {Osier}, {Owen}, {Pain}, {Palecek}, {Parejko}, {Parsons}, {Pease},
  {Peterson}, {Peterson}, {Petravick}, {Libby Petrick}, {Petry},
  {Pierfederici}, {Pietrowicz}, {Pike}, {Pinto}, {Plante}, {Plate}, {Plutchak},
  {Price}, {Prouza}, {Radeka}, {Rajagopal}, {Rasmussen}, {Regnault}, {Reil},
  {Reiss}, {Reuter}, {Ridgway}, {Riot}, {Ritz}, {Robinson}, {Roby}, {Roodman},
  {Rosing}, {Roucelle}, {Rumore}, {Russo}, {Saha}, {Sassolas}, {Schalk},
  {Schellart}, {Schindler}, {Schmidt}, {Schneider}, {Schneider}, {Schoening},
  {Schumacher}, {Schwamb}, {Sebag}, {Selvy}, {Sembroski}, {Seppala}, {Serio},
  {Serrano}, {Shaw}, {Shipsey}, {Sick}, {Silvestri}, {Slater}, {Smith},
  {Smith}, {Sobhani}, {Soldahl}, {Storrie-Lombardi}, {Stover}, {Strauss},
  {Street}, {Stubbs}, {Sullivan}, {Sweeney}, {Swinbank}, {Szalay}, {Takacs},
  {Tether}, {Thaler}, {Thayer}, {Thomas}, {Thornton}, {Thukral}, {Tice},
  {Trilling}, {Turri}, {Van Berg}, {Vanden Berk}, {Vetter}, {Virieux},
  {Vucina}, {Wahl}, {Walkowicz}, {Walsh}, {Walter}, {Wang}, {Wang}, {Warner},
  {Wiecha}, {Willman}, {Winters}, {Wittman}, {Wolff}, {Wood-Vasey}, {Wu},
  {Xin}, {Yoachim}, \& {Zhan}}]{Ivezic_2019}
{Ivezi{\'c}}, {\v{Z}}., {Kahn}, S.~M., {Tyson}, J.~A., {et~al.} 2019, \apj,
  873, 111

\bibitem[{{Jiang} {et~al.}(2021){Jiang}, {Wang}, {Hu}, {Sun}, {Dou}, \&
  {Xiao}}]{Jiang_2021}
{Jiang}, N., {Wang}, T., {Hu}, X., {et~al.} 2021, \apj, 911, 31

\bibitem[{{Jiang} {et~al.}(2019){Jiang}, {Wang}, {Mou}, {Liu}, {Dou}, {Sheng},
  \& {Wang}}]{Jiang_2019}
{Jiang}, N., {Wang}, T., {Mou}, G., {et~al.} 2019, \apj, 871, 15

\bibitem[{{Jiang} {et~al.}(2017){Jiang}, {Wang}, {Yan}, {Xiao}, {Yang}, {Dou},
  {Wang}, {Cutri}, \& {Mainzer}}]{2017ApJ...850...63J}
{Jiang}, N., {Wang}, T., {Yan}, L., {et~al.} 2017, \apj, 850, 63

\bibitem[{{Kankare} {et~al.}(2017){Kankare}, {Kotak}, {Mattila}, {Lundqvist},
  {Ward}, {Fraser}, {Lawrence}, {Smartt}, {Meikle}, {Bruce}, {Harmanen},
  {Hutton}, {Inserra}, {Kangas}, {Pastorello}, {Reynolds},
  {Romero-Ca{\~n}izales}, {Smith}, {Valenti}, {Chambers}, {Hodapp}, {Huber},
  {Kaiser}, {Kudritzki}, {Magnier}, {Tonry}, {Wainscoat}, \&
  {Waters}}]{2017NatAs...1..865K}
{Kankare}, E., {Kotak}, R., {Mattila}, S., {et~al.} 2017, Nature Astronomy, 1,
  865

\bibitem[{{Kewley} {et~al.}(2006){Kewley}, {Groves}, {Kauffmann}, \&
  {Heckman}}]{2006MNRAS.372..961K}
{Kewley}, L.~J., {Groves}, B., {Kauffmann}, G., \& {Heckman}, T. 2006, \mnras,
  372, 961

\bibitem[{{Kollatschny} \& {Fricke}(1985)}]{1985A&A...146L..11K}
{Kollatschny}, W. \& {Fricke}, K.~J. 1985, \aap, 146, L11

\bibitem[{{Komossa} {et~al.}(2017){Komossa}, {Grupe}, {Schartel}, {Gallo},
  {Gomez}, {Kollatschny}, {Kriss}, {Leighly}, {Longinotti}, {Parker},
  {Santos-Lleo}, {Wilkins}, \& {Zetzl}}]{2017IAUS..324..168K}
{Komossa}, S., {Grupe}, D., {Schartel}, N., {et~al.} 2017, in New Frontiers in
  Black Hole Astrophysics, ed. A.~{Gomboc}, Vol. 324, 168--171

\bibitem[{{Komossa} {et~al.}(2008){Komossa}, {Zhou}, {Wang}, {Ajello}, {Ge},
  {Greiner}, {Lu}, {Salvato}, {Saxton}, {Shan}, {Xu}, \& {Yuan}}]{Komossa_2008}
{Komossa}, S., {Zhou}, H., {Wang}, T., {et~al.} 2008, \apjl, 678, L13

\bibitem[{{Kool} {et~al.}(2020){Kool}, {Reynolds}, {Mattila}, {Kankare},
  {P{\'e}rez-Torres}, {Efstathiou}, {Ryder}, {Romero-Ca{\~n}izales}, {Lu},
  {Heikkil{\"a}}, {Anderson}, {Berton}, {Bright}, {Cannizzaro}, {Eappachen},
  {Fraser}, {Gromadzki}, {Jonker}, {Kuncarayakti}, {Lundqvist}, {Maeda},
  {McDermid}, {Medling}, {Moran}, {Reguitti}, {Shahbandeh}, {Tsygankov}, {U},
  \& {Wevers}}]{Kool_2020}
{Kool}, E.~C., {Reynolds}, T.~M., {Mattila}, S., {et~al.} 2020, \mnras, 498,
  2167

\bibitem[{{Korista} \& {Ferland}(1989)}]{Korista_1989}
{Korista}, K.~T. \& {Ferland}, G.~J. 1989, \apj, 343, 678

\bibitem[{{Kova{\v{c}}evi{\'c}} {et~al.}(2010){Kova{\v{c}}evi{\'c}},
  {Popovi{\'c}}, \& {Dimitrijevi{\'c}}}]{2010ApJS..189...15K}
{Kova{\v{c}}evi{\'c}}, J., {Popovi{\'c}}, L.~{\v{C}}., \& {Dimitrijevi{\'c}},
  M.~S. 2010, \apjs, 189, 15

\bibitem[{{Kulkarni}(2013)}]{2013ATel.4807....1K}
{Kulkarni}, S.~R. 2013, The Astronomer's Telegram, 4807, 1

\bibitem[{{Leloudas} {et~al.}(2019){Leloudas}, {Dai}, {Arcavi}, {Vreeswijk},
  {Mockler}, {Roy}, {Malesani}, {Schulze}, {Wevers}, {Fraser}, {Ramirez-Ruiz},
  {Auchettl}, {Burke}, {Cannizzaro}, {Charalampopoulos}, {Chen}, {Cikota},
  {Della Valle}, {Galbany}, {Gromadzki}, {Heintz}, {Hiramatsu}, {Jonker},
  {Kostrzewa-Rutkowska}, {Maguire}, {Mandel}, {Nicholl}, {Onori}, {Roth},
  {Smartt}, {Wyrzykowski}, \& {Young}}]{Leloudas_2019}
{Leloudas}, G., {Dai}, L., {Arcavi}, I., {et~al.} 2019, \apj, 887, 218

\bibitem[{{Leloudas} {et~al.}(2016){Leloudas}, {Fraser}, {Stone}, {van Velzen},
  {Jonker}, {Arcavi}, {Fremling}, {Maund}, {Smartt}, {Kr{\`\i}hler},
  {Miller-Jones}, {Vreeswijk}, {Gal-Yam}, {Mazzali}, {De Cia}, {Howell},
  {Inserra}, {Patat}, {de Ugarte Postigo}, {Yaron}, {Ashall}, {Bar},
  {Campbell}, {Chen}, {Childress}, {Elias-Rosa}, {Harmanen}, {Hosseinzadeh},
  {Johansson}, {Kangas}, {Kankare}, {Kim}, {Kuncarayakti}, {Lyman}, {Magee},
  {Maguire}, {Malesani}, {Mattila}, {McCully}, {Nicholl}, {Prentice},
  {Romero-Ca{\~n}izales}, {Schulze}, {Smith}, {Sollerman}, {Sullivan},
  {Tucker}, {Valenti}, {Wheeler}, \& {Young}}]{Leloudas_2016}
{Leloudas}, G., {Fraser}, M., {Stone}, N.~C., {et~al.} 2016, Nature Astronomy,
  1, 0002

\bibitem[{{Lodato} \& {Rossi}(2011)}]{2011MNRAS.410..359L}
{Lodato}, G. \& {Rossi}, E.~M. 2011, \mnras, 410, 359

\bibitem[{{L{\'o}pez-Navas} {et~al.}(2022){L{\'o}pez-Navas},
  {Mart{\'\i}nez-Aldama}, {Bernal}, {S{\'a}nchez-S{\'a}ez}, {Ar{\'e}valo},
  {Graham}, {Hern{\'a}ndez-Garc{\'\i}a}, {Lira}, \& {Rojas
  Lobos}}]{2022MNRAS.513L..57L}
{L{\'o}pez-Navas}, E., {Mart{\'\i}nez-Aldama}, M.~L., {Bernal}, S., {et~al.}
  2022, \mnras, 513, L57

\bibitem[{{Lyu} {et~al.}(2022){Lyu}, {Wu}, {Yan}, {Yu}, \&
  {Liu}}]{2022ApJ...927..227L}
{Lyu}, B., {Wu}, Q., {Yan}, Z., {Yu}, W., \& {Liu}, H. 2022, \apj, 927, 227

\bibitem[{{Lyutyj} {et~al.}(1984){Lyutyj}, {Oknyanskij}, \&
  {Chuvaev}}]{1984SvAL...10..335L}
{Lyutyj}, V.~M., {Oknyanskij}, V.~L., \& {Chuvaev}, K.~K. 1984, Soviet
  Astronomy Letters, 10, 335

\bibitem[{{MacLeod} {et~al.}(2019){MacLeod}, {Green}, {Anderson}, {Bruce},
  {Eracleous}, {Graham}, {Homan}, {Lawrence}, {LeBleu}, {Ross}, {Ruan},
  {Runnoe}, {Stern}, {Burgett}, {Chambers}, {Kaiser}, {Magnier}, \&
  {Metcalfe}}]{MacLeod_2019}
{MacLeod}, C.~L., {Green}, P.~J., {Anderson}, S.~F., {et~al.} 2019, \apj, 874,
  8

\bibitem[{{Magnier} {et~al.}(2013){Magnier}, {Schlafly}, {Finkbeiner}, {Juric},
  {Tonry}, {Burgett}, {Chambers}, {Flewelling}, {Kaiser}, {Kudritzki},
  {Morgan}, {Price}, {Sweeney}, \& {Stubbs}}]{Magnier2013}
{Magnier}, E.~A., {Schlafly}, E., {Finkbeiner}, D., {et~al.} 2013, \apjs, 205,
  20

\bibitem[{{Mainzer} {et~al.}(2014){Mainzer}, {Bauer}, {Cutri}, {Grav},
  {Masiero}, {Beck}, {Clarkson}, {Conrow}, {Dailey}, {Eisenhardt}, {Fabinsky},
  {Fajardo-Acosta}, {Fowler}, {Gelino}, {Grillmair}, {Heinrichsen}, {Kendall},
  {Kirkpatrick}, {Liu}, {Masci}, {McCallon}, {Nugent}, {Papin}, {Rice},
  {Royer}, {Ryan}, {Sevilla}, {Sonnett}, {Stevenson}, {Thompson}, {Wheelock},
  {Wiemer}, {Wittman}, {Wright}, \& {Yan}}]{2014ApJ...792...30M}
{Mainzer}, A., {Bauer}, J., {Cutri}, R.~M., {et~al.} 2014, \apj, 792, 30

\bibitem[{{Marziani} {et~al.}(2018){Marziani}, {Dultzin}, {Sulentic}, {Del
  Olmo}, {Negrete}, {Mart{\'\i}nez-Aldama}, {D'Onofrio}, {Bon}, {Bon}, \&
  {Stirpe}}]{Marziani_2018}
{Marziani}, P., {Dultzin}, D., {Sulentic}, J.~W., {et~al.} 2018, Frontiers in
  Astronomy and Space Sciences, 5, 6

\bibitem[{{Marziani} \& {Sulentic}(2014)}]{Marziani_2014}
{Marziani}, P. \& {Sulentic}, J.~W. 2014, \mnras, 442, 1211

\bibitem[{{Mattila} {et~al.}(2018){Mattila}, {P{\'e}rez-Torres}, {Efstathiou},
  {Mimica}, {Fraser}, {Kankare}, {Alberdi}, {Aloy}, {Heikkil{\"a}}, {Jonker},
  {Lundqvist}, {Mart{\'\i}-Vidal}, {Meikle}, {Romero-Ca{\~n}izales}, {Smartt},
  {Tsygankov}, {Varenius}, {Alonso-Herrero}, {Bondi}, {Fransson},
  {Herrero-Illana}, {Kangas}, {Kotak}, {Ram{\'\i}rez-Olivencia},
  {V{\"a}is{\"a}nen}, {Beswick}, {Clements}, {Greimel}, {Harmanen},
  {Kotilainen}, {Nandra}, {Reynolds}, {Ryder}, {Walton}, {Wiik}, \&
  {{\"O}stlin}}]{Mattila_2018}
{Mattila}, S., {P{\'e}rez-Torres}, M., {Efstathiou}, A., {et~al.} 2018,
  Science, 361, 482

\bibitem[{{Mehdipour} {et~al.}(2022){Mehdipour}, {Kriss}, {Brenneman},
  {Costantini}, {Kaastra}, {Branduardi-Raymont}, {Di Gesu}, {Ebrero}, \&
  {Mao}}]{2022ApJ...925...84M}
{Mehdipour}, M., {Kriss}, G.~A., {Brenneman}, L.~W., {et~al.} 2022, \apj, 925,
  84

\bibitem[{{Mockler} {et~al.}(2019){Mockler}, {Guillochon}, \&
  {Ramirez-Ruiz}}]{2019ApJ...872..151M}
{Mockler}, B., {Guillochon}, J., \& {Ramirez-Ruiz}, E. 2019, \apj, 872, 151

\bibitem[{{Moriya} {et~al.}(2017){Moriya}, {Tanaka}, {Morokuma}, \&
  {Ohsuga}}]{Moriya_2017}
{Moriya}, T.~J., {Tanaka}, M., {Morokuma}, T., \& {Ohsuga}, K. 2017, \apjl,
  843, L19

\bibitem[{{Namekata} \& {Umemura}(2016)}]{2016MNRAS.460..980N}
{Namekata}, D. \& {Umemura}, M. 2016, \mnras, 460, 980

\bibitem[{{Netzer}(2006)}]{2006LNP...693....1N}
{Netzer}, H. 2006, in Physics of Active Galactic Nuclei at all Scales, ed.
  D.~{Alloin}, Vol. 693, 1

\bibitem[{{Noda} \& {Done}(2018)}]{2018MNRAS.480.3898N}
{Noda}, H. \& {Done}, C. 2018, \mnras, 480, 3898

\bibitem[{{Oknyansky} {et~al.}(2019){Oknyansky}, {Winkler}, {Tsygankov},
  {Lipunov}, {Gorbovskoy}, {van Wyk}, {Buckley}, \&
  {Tyurina}}]{2019MNRAS.483..558O}
{Oknyansky}, V.~L., {Winkler}, H., {Tsygankov}, S.~S., {et~al.} 2019, \mnras,
  483, 558

\bibitem[{{Onori} {et~al.}(2019){Onori}, {Cannizzaro}, {Jonker}, {Fraser},
  {Kostrzewa-Rutkowska}, {Martin-Carrillo}, {Benetti}, {Elias-Rosa},
  {Gromadzki}, {Harmanen}, {Mattila}, {Strizinger}, {Terreran}, \&
  {Wevers}}]{2019MNRAS.489.1463O}
{Onori}, F., {Cannizzaro}, G., {Jonker}, P.~G., {et~al.} 2019, \mnras, 489,
  1463

\bibitem[{{Onori} {et~al.}(2022){Onori}, {Cannizzaro}, {Jonker}, {Kim},
  {Nicholl}, {Mattila}, {Reynolds}, {Fraser}, {Wevers}, {Brocato}, {Anderson},
  {Carini}, {Charalampopoulos}, {Clark}, {Gromadzki}, {Guti{\'e}rrez},
  {Ihanec}, {Inserra}, {Lawrence}, {Leloudas}, {Lundqvist}, {M{\"u}ller-Bravo},
  {Piranomonte}, {Pursiainen}, {Rybicki}, {Somero}, {Young}, {Chambers}, {Gao},
  {de Boer}, \& {Magnier}}]{Onori_2022}
{Onori}, F., {Cannizzaro}, G., {Jonker}, P.~G., {et~al.} 2022, \mnras, 517, 76

\bibitem[{{Osterbrock} \& {Ferland}(2006)}]{2006agna.book.....O}
{Osterbrock}, D.~E. \& {Ferland}, G.~J. 2006, {Astrophysics of gaseous nebulae
  and active galactic nuclei}

\bibitem[{{Panda} {et~al.}(2018){Panda}, {Czerny}, {Adhikari}, {Hryniewicz},
  {Wildy}, {Kuraszkiewicz}, \& {{\'S}niegowska}}]{Panda_2018}
{Panda}, S., {Czerny}, B., {Adhikari}, T.~P., {et~al.} 2018, \apj, 866, 115

\bibitem[{{Park} {et~al.}(2022){Park}, {Barth}, {Ho}, \& {Laor}}]{Park_2022}
{Park}, D., {Barth}, A.~J., {Ho}, L.~C., \& {Laor}, A. 2022, \apjs, 258, 38

\bibitem[{{Peterson}(2001)}]{Peterson_2001}
{Peterson}, B.~M. 2001, in Advanced Lectures on the Starburst-AGN, ed.
  I.~{Aretxaga}, D.~{Kunth}, \& R.~{M{\'u}jica}, 3

\bibitem[{{Phinney}(1989)}]{Phinney_1989}
{Phinney}, E.~S. 1989, in The Center of the Galaxy, ed. M.~{Morris}, Vol. 136,
  543

\bibitem[{{Pons} \& {Watson}(2014)}]{Pons_2014}
{Pons}, E. \& {Watson}, M.~G. 2014, \aap, 568, A108

\bibitem[{{Raki{\'c}}(2022)}]{Rakic_2022}
{Raki{\'c}}, N. 2022, \mnras, 516, 1624

\bibitem[{{Rakshit} {et~al.}(2017){Rakshit}, {Stalin}, {Chand}, \&
  {Zhang}}]{Rakshit_2017}
{Rakshit}, S., {Stalin}, C.~S., {Chand}, H., \& {Zhang}, X.-G. 2017, \apjs,
  229, 39

\bibitem[{{Ranalli} {et~al.}(2003){Ranalli}, {Comastri}, \&
  {Setti}}]{Ranalli_2003}
{Ranalli}, P., {Comastri}, A., \& {Setti}, G. 2003, \aap, 399, 39

\bibitem[{{Rees}(1988)}]{1988Natur.333..523R}
{Rees}, M.~J. 1988, \nat, 333, 523

\bibitem[{{Ren} {et~al.}(2022){Ren}, {Wang}, {Cai}, \&
  {Guo}}]{2022ApJ...925...50R}
{Ren}, W., {Wang}, J., {Cai}, Z., \& {Guo}, H. 2022, \apj, 925, 50

\bibitem[{{Richards} {et~al.}(2006){Richards}, {Lacy}, {Storrie-Lombardi},
  {Hall}, {Gallagher}, {Hines}, {Fan}, {Papovich}, {Vanden Berk}, {Trammell},
  {Schneider}, {Vestergaard}, {York}, {Jester}, {Anderson}, {Budav{\'a}ri}, \&
  {Szalay}}]{2006ApJS..166..470R}
{Richards}, G.~T., {Lacy}, M., {Storrie-Lombardi}, L.~J., {et~al.} 2006, \apjs,
  166, 470

\bibitem[{{Roming} {et~al.}(2005){Roming}, {Kennedy}, {Mason}, {Nousek}, {Ahr},
  {Bingham}, {Broos}, {Carter}, {Hancock}, {Huckle}, {Hunsberger}, {Kawakami},
  {Killough}, {Koch}, {McLelland}, {Smith}, {Smith}, {Soto}, {Boyd},
  {Breeveld}, {Holland}, {Ivanushkina}, {Pryzby}, {Still}, \&
  {Stock}}]{2005SSRv..120...95R}
{Roming}, P. W.~A., {Kennedy}, T.~E., {Mason}, K.~O., {et~al.} 2005, \ssr, 120,
  95

\bibitem[{{Rose} {et~al.}(2015){Rose}, {Elvis}, \&
  {Tadhunter}}]{2015MNRAS.448.2900R}
{Rose}, M., {Elvis}, M., \& {Tadhunter}, C.~N. 2015, \mnras, 448, 2900

\bibitem[{{Roth} {et~al.}(2016){Roth}, {Kasen}, {Guillochon}, \&
  {Ramirez-Ruiz}}]{Roth_2016}
{Roth}, N., {Kasen}, D., {Guillochon}, J., \& {Ramirez-Ruiz}, E. 2016, \apj,
  827, 3

\bibitem[{{Roth} {et~al.}(2021){Roth}, {van Velzen}, {Cenko}, \&
  {Mushotzky}}]{2021ApJ...910...93R}
{Roth}, N., {van Velzen}, S., {Cenko}, S.~B., \& {Mushotzky}, R.~F. 2021, \apj,
  910, 93

\bibitem[{{Runnoe} {et~al.}(2012){Runnoe}, {Brotherton}, \&
  {Shang}}]{2012MNRAS.422..478R}
{Runnoe}, J.~C., {Brotherton}, M.~S., \& {Shang}, Z. 2012, \mnras, 422, 478

\bibitem[{{Salim} \& {Narayanan}(2020)}]{2020ARA&A..58..529S}
{Salim}, S. \& {Narayanan}, D. 2020, \araa, 58, 529

\bibitem[{{S{\'a}nchez-S{\'a}ez} {et~al.}(2021){S{\'a}nchez-S{\'a}ez}, {Lira},
  {Mart{\'\i}}, {S{\'a}nchez-Pi}, {Arredondo}, {Bauer}, {Bayo},
  {Cabrera-Vives}, {Donoso-Oliva}, {Est{\'e}vez}, {Eyheramendy}, {F{\"o}rster},
  {Hern{\'a}ndez-Garc{\'\i}a}, {Arancibia}, {P{\'e}rez-Carrasco},
  {Sep{\'u}lveda}, \& {Vergara}}]{2021AJ....162..206S}
{S{\'a}nchez-S{\'a}ez}, P., {Lira}, H., {Mart{\'\i}}, L., {et~al.} 2021, \aj,
  162, 206

\bibitem[{{S{\'a}nchez-S{\'a}ez} {et~al.}(2018){S{\'a}nchez-S{\'a}ez}, {Lira},
  {Mej{\'\i}a-Restrepo}, {Ho}, {Ar{\'e}valo}, {Kim}, {Cartier}, \&
  {Coppi}}]{Sanchez_2018}
{S{\'a}nchez-S{\'a}ez}, P., {Lira}, P., {Mej{\'\i}a-Restrepo}, J., {et~al.}
  2018, \apj, 864, 87

\bibitem[{{Schlafly} \& {Finkbeiner}(2011)}]{2011ApJ...737..103S}
{Schlafly}, E.~F. \& {Finkbeiner}, D.~P. 2011, \apj, 737, 103

\bibitem[{{Selsing} {et~al.}(2019){Selsing}, {Malesani}, {Goldoni}, {Fynbo},
  {Kr{\"u}hler}, {Antonelli}, {Arabsalmani}, {Bolmer}, {Cano}, {Christensen},
  {Covino}, {D'Avanzo}, {D'Elia}, {De Cia}, {de Ugarte Postigo}, {Flores},
  {Friis}, {Gomboc}, {Greiner}, {Groot}, {Hammer}, {Hartoog}, {Heintz},
  {Hjorth}, {Jakobsson}, {Japelj}, {Kann}, {Kaper}, {Ledoux}, {Leloudas},
  {Levan}, {Maiorano}, {Melandri}, {Milvang-Jensen}, {Palazzi}, {Palmerio},
  {Perley}, {Pian}, {Piranomonte}, {Pugliese}, {S{\'a}nchez-Ram{\'\i}rez},
  {Savaglio}, {Schady}, {Schulze}, {Sollerman}, {Sparre}, {Tagliaferri},
  {Tanvir}, {Th{\"o}ne}, {Vergani}, {Vreeswijk}, {Watson}, {Wiersema},
  {Wijers}, {Xu}, \& {Zafar}}]{2019A&A...623A..92S}
{Selsing}, J., {Malesani}, D., {Goldoni}, P., {et~al.} 2019, \aap, 623, A92

\bibitem[{{Shapovalova} {et~al.}(2012){Shapovalova}, {Popovi{\'c}}, {Burenkov},
  {Chavushyan}, {Ili{\'c}}, {Kova{\v{c}}evi{\'c}}, {Kollatschny},
  {Kova{\v{c}}evi{\'c}}, {Bochkarev}, {Valdes}, {Torrealba},
  {Le{\'o}n-Tavares}, {Mercado}, {Ben{\'\i}tez}, {Carrasco}, {Dultzin}, \& {de
  la Fuente}}]{Shapovalova_2012}
{Shapovalova}, A.~I., {Popovi{\'c}}, L.~{\v{C}}., {Burenkov}, A.~N., {et~al.}
  2012, \apjs, 202, 10

\bibitem[{{Shappee} {et~al.}(2014){Shappee}, {Prieto}, {Grupe}, {Kochanek},
  {Stanek}, {De Rosa}, {Mathur}, {Zu}, {Peterson}, {Pogge}, {Komossa}, {Im},
  {Jencson}, {Holoien}, {Basu}, {Beacom}, {Szczygie{\l}}, {Brimacombe},
  {Adams}, {Campillay}, {Choi}, {Contreras}, {Dietrich}, {Dubberley},
  {Elphick}, {Foale}, {Giustini}, {Gonzalez}, {Hawkins}, {Howell}, {Hsiao},
  {Koss}, {Leighly}, {Morrell}, {Mudd}, {Mullins}, {Nugent}, {Parrent},
  {Phillips}, {Pojmanski}, {Rosing}, {Ross}, {Sand}, {Terndrup}, {Valenti},
  {Walker}, \& {Yoon}}]{2014ApJ...788...48S}
{Shappee}, B.~J., {Prieto}, J.~L., {Grupe}, D., {et~al.} 2014, \apj, 788, 48

\bibitem[{{Shen} \& {Ho}(2014)}]{Shen_Ho_2014}
{Shen}, Y. \& {Ho}, L.~C. 2014, \nat, 513, 210

\bibitem[{{Shen} {et~al.}(2011){Shen}, {Richards}, {Strauss}, {Hall},
  {Schneider}, {Snedden}, {Bizyaev}, {Brewington}, {Malanushenko},
  {Malanushenko}, {Oravetz}, {Pan}, \& {Simmons}}]{Shen_2011}
{Shen}, Y., {Richards}, G.~T., {Strauss}, M.~A., {et~al.} 2011, \apjs, 194, 45

\bibitem[{{Sheng} {et~al.}(2020){Sheng}, {Wang}, {Jiang}, {Ding}, {Cai}, {Guo},
  {Sun}, {Dou}, \& {Yang}}]{Sheng_2020}
{Sheng}, Z., {Wang}, T., {Jiang}, N., {et~al.} 2020, \apj, 889, 46

\bibitem[{{Sheng} {et~al.}(2017){Sheng}, {Wang}, {Jiang}, {Yang}, {Yan}, {Dou},
  \& {Peng}}]{Sheng_2017}
{Sheng}, Z., {Wang}, T., {Jiang}, N., {et~al.} 2017, \apjl, 846, L7

\bibitem[{{Smartt} {et~al.}(2015){Smartt}, {Valenti}, {Fraser}, {Inserra},
  {Young}, {Sullivan}, {Pastorello}, {Benetti}, {Gal-Yam}, {Knapic},
  {Molinaro}, {Smareglia}, {Smith}, {Taubenberger}, {Yaron}, {Anderson},
  {Ashall}, {Balland}, {Baltay}, {Barbarino}, {Bauer}, {Baumont}, {Bersier},
  {Blagorodnova}, {Bongard}, {Botticella}, {Bufano}, {Bulla}, {Cappellaro},
  {Campbell}, {Cellier-Holzem}, {Chen}, {Childress}, {Clocchiatti},
  {Contreras}, {Dall'Ora}, {Danziger}, {de Jaeger}, {De Cia}, {Della Valle},
  {Dennefeld}, {Elias-Rosa}, {Elman}, {Feindt}, {Fleury}, {Gall},
  {Gonzalez-Gaitan}, {Galbany}, {Morales Garoffolo}, {Greggio}, {Guillou},
  {Hachinger}, {Hadjiyska}, {Hage}, {Hillebrandt}, {Hodgkin}, {Hsiao}, {James},
  {Jerkstrand}, {Kangas}, {Kankare}, {Kotak}, {Kromer}, {Kuncarayakti},
  {Leloudas}, {Lundqvist}, {Lyman}, {Hook}, {Maguire}, {Manulis}, {Margheim},
  {Mattila}, {Maund}, {Mazzali}, {McCrum}, {McKinnon}, {Moreno-Raya},
  {Nicholl}, {Nugent}, {Pain}, {Pignata}, {Phillips}, {Polshaw}, {Pumo},
  {Rabinowitz}, {Reilly}, {Romero-Ca{\~n}izales}, {Scalzo}, {Schmidt},
  {Schulze}, {Sim}, {Sollerman}, {Taddia}, {Tartaglia}, {Terreran},
  {Tomasella}, {Turatto}, {Walker}, {Walton}, {Wyrzykowski}, {Yuan}, \&
  {Zampieri}}]{2015A&A...579A..40S}
{Smartt}, S.~J., {Valenti}, S., {Fraser}, M., {et~al.} 2015, \aap, 579, A40

\bibitem[{{Smith} {et~al.}(2020){Smith}, {Smartt}, {Young}, {Tonry}, {Denneau},
  {Flewelling}, {Heinze}, {Weiland}, {Stalder}, {Rest}, {Stubbs}, {Anderson},
  {Chen}, {Clark}, {Do}, {F{\"o}rster}, {Fulton}, {Gillanders}, {McBrien},
  {O'Neill}, {Srivastav}, \& {Wright}}]{2020PASP..132h5002S}
{Smith}, K.~W., {Smartt}, S.~J., {Young}, D.~R., {et~al.} 2020, \pasp, 132,
  085002

\bibitem[{{Sniegowska} {et~al.}(2020){Sniegowska}, {Czerny}, {Bon}, \&
  {Bon}}]{2020A&A...641A.167S}
{Sniegowska}, M., {Czerny}, B., {Bon}, E., \& {Bon}, N. 2020, \aap, 641, A167

\bibitem[{{{\'S}niegowska} {et~al.}(2022){{\'S}niegowska}, {Grzedzielski},
  {Czerny}, \& {Janiuk}}]{2022AN....34310065S}
{{\'S}niegowska}, M., {Grzedzielski}, M., {Czerny}, B., \& {Janiuk}, A. 2022,
  Astronomische Nachrichten, 343, e210065

\bibitem[{{Steele} {et~al.}(2004){Steele}, {Smith}, {Rees}, {Baker}, {Bates},
  {Bode}, {Bowman}, {Carter}, {Etherton}, {Ford}, {Fraser}, {Gomboc}, {Lett},
  {Mansfield}, {Marchant}, {Medrano-Cerda}, {Mottram}, {Raback}, {Scott},
  {Tomlinson}, \& {Zamanov}}]{2004SPIE.5489..679S}
{Steele}, I.~A., {Smith}, R.~J., {Rees}, P.~C., {et~al.} 2004, in Society of
  Photo-Optical Instrumentation Engineers (SPIE) Conference Series, Vol. 5489,
  \procspie, ed. J.~{Oschmann}, Jacobus~M., 679--692

\bibitem[{{Stern} {et~al.}(2018){Stern}, {McKernan}, {Graham}, {Ford}, {Ross},
  {Meisner}, {Assef}, {Balokovi{\'c}}, {Brightman}, {Dey}, {Drake},
  {Djorgovski}, {Eisenhardt}, \& {Jun}}]{Stern_2018}
{Stern}, D., {McKernan}, B., {Graham}, M.~J., {et~al.} 2018, \apj, 864, 27

\bibitem[{{Sulentic} {et~al.}(2009){Sulentic}, {Marziani}, \&
  {Zamfir}}]{Sulentic_2009}
{Sulentic}, J.~W., {Marziani}, P., \& {Zamfir}, S. 2009, \nar, 53, 198

\bibitem[{{Terreran} {et~al.}(2016){Terreran}, {Berton}, {Benetti},
  {Cappellaro}, {Elias-Rosa}, {Ochner}, {Pastorello}, {Tomasella}, \&
  {Turatto}}]{2016ATel.9417....1T}
{Terreran}, G., {Berton}, M., {Benetti}, S., {et~al.} 2016, The Astronomer's
  Telegram, 9417, 1

\bibitem[{{Thomsen} {et~al.}(2022){Thomsen}, {Kwan}, {Dai}, {Wu}, {Roth}, \&
  {Ramirez-Ruiz}}]{2022ApJ...937L..28T}
{Thomsen}, L.~L., {Kwan}, T.~M., {Dai}, L., {et~al.} 2022, \apjl, 937, L28

\bibitem[{{Tonry}(2011)}]{2011PASP..123...58T}
{Tonry}, J.~L. 2011, \pasp, 123, 58

\bibitem[{{Tonry} {et~al.}(2018){Tonry}, {Denneau}, {Heinze}, {Stalder},
  {Smith}, {Smartt}, {Stubbs}, {Weiland}, \& {Rest}}]{2018PASP..130f4505T}
{Tonry}, J.~L., {Denneau}, L., {Heinze}, A.~N., {et~al.} 2018, \pasp, 130,
  064505

\bibitem[{{Tonry} {et~al.}(2012){Tonry}, {Stubbs}, {Lykke}, {Doherty},
  {Shivvers}, {Burgett}, {Chambers}, {Hodapp}, {Kaiser}, {Kudritzki},
  {Magnier}, {Morgan}, {Price}, \& {Wainscoat}}]{Tonry+2012}
{Tonry}, J.~L., {Stubbs}, C.~W., {Lykke}, K.~R., {et~al.} 2012, \apj, 750, 99

\bibitem[{{Trakhtenbrot} {et~al.}(2019){Trakhtenbrot}, {Arcavi}, {MacLeod},
  {Ricci}, {Kara}, {Graham}, {Stern}, {Harrison}, {Burke}, {Hiramatsu},
  {Hosseinzadeh}, {Howell}, {Smartt}, {Rest}, {Prieto}, {Shappee}, {Holoien},
  {Bersier}, {Filippenko}, {Brink}, {Zheng}, {Li}, {Remillard}, \&
  {Loewenstein}}]{2019ApJ...883...94T}
{Trakhtenbrot}, B., {Arcavi}, I., {MacLeod}, C.~L., {et~al.} 2019, \apj, 883,
  94

\bibitem[{Trakhtenbrot {et~al.}(2019)Trakhtenbrot, Arcavi, Ricci, Tacchella,
  Stern, Netzer, Jonker, Horesh, Mejía-Restrepo, Hosseinzadeh, \&
  et~al.}]{Trakhtenbrot_2019}
Trakhtenbrot, B., Arcavi, I., Ricci, C., {et~al.} 2019, Nature Astronomy, 3,
  242–250

\bibitem[{{van Groningen} \& {Wanders}(1992)}]{1992PASP..104..700V}
{van Groningen}, E. \& {Wanders}, I. 1992, \pasp, 104, 700

\bibitem[{van Velzen {et~al.}(2020)van Velzen, Holoien, Onori, Hung, \&
  Arcavi}]{van_Velzen_2020}
van Velzen, S., Holoien, T. W.-S., Onori, F., Hung, T., \& Arcavi, I. 2020,
  Space Science Reviews, 216

\bibitem[{van Velzen {et~al.}(2016)van Velzen, Mendez, Krolik, \&
  Gorjian}]{Velzen2016}
van Velzen, S., Mendez, A.~J., Krolik, J.~H., \& Gorjian, V. 2016, The
  Astrophysical Journal, 829, 19

\bibitem[{{van Velzen} {et~al.}(2021){van Velzen}, {Pasham}, {Komossa}, {Yan},
  \& {Kara}}]{2021SSRv..217...63V}
{van Velzen}, S., {Pasham}, D.~R., {Komossa}, S., {Yan}, L., \& {Kara}, E.~A.
  2021, \ssr, 217, 63

\bibitem[{{van Velzen} {et~al.}(2019){van Velzen}, {Stone}, {Metzger},
  {Gezari}, {Brown}, \& {Fruchter}}]{2019ApJ...878...82V}
{van Velzen}, S., {Stone}, N.~C., {Metzger}, B.~D., {et~al.} 2019, \apj, 878,
  82

\bibitem[{{Vanden Berk} {et~al.}(2006){Vanden Berk}, {Shen}, {Yip},
  {Schneider}, {Connolly}, {Burton}, {Jester}, {Hall}, {Szalay}, \&
  {Brinkmann}}]{VandenBerk_2006}
{Vanden Berk}, D.~E., {Shen}, J., {Yip}, C.-W., {et~al.} 2006, \aj, 131, 84

\bibitem[{{Vanden Berk} {et~al.}(2004){Vanden Berk}, {Wilhite}, {Kron},
  {Anderson}, {Brunner}, {Hall}, {Ivezi{\'c}}, {Richards}, {Schneider}, {York},
  {Brinkmann}, {Lamb}, {Nichol}, \& {Schlegel}}]{2004ApJ...601..692V}
{Vanden Berk}, D.~E., {Wilhite}, B.~C., {Kron}, R.~G., {et~al.} 2004, \apj,
  601, 692

\bibitem[{{Vernet} {et~al.}(2011){Vernet}, {Dekker}, {D'Odorico}, {Kaper},
  {Kjaergaard}, {Hammer}, {Randich}, {Zerbi}, {Groot}, {Hjorth}, {Guinouard},
  {Navarro}, {Adolfse}, {Albers}, {Amans}, {Andersen}, {Andersen}, {Binetruy},
  {Bristow}, {Castillo}, {Chemla}, {Christensen}, {Conconi}, {Conzelmann},
  {Dam}, {de Caprio}, {de Ugarte Postigo}, {Delabre}, {di Marcantonio},
  {Downing}, {Elswijk}, {Finger}, {Fischer}, {Flores}, {Fran{\c{c}}ois},
  {Goldoni}, {Guglielmi}, {Haigron}, {Hanenburg}, {Hendriks}, {Horrobin},
  {Horville}, {Jessen}, {Kerber}, {Kern}, {Kiekebusch}, {Kleszcz}, {Klougart},
  {Kragt}, {Larsen}, {Lizon}, {Lucuix}, {Mainieri}, {Manuputy}, {Martayan},
  {Mason}, {Mazzoleni}, {Michaelsen}, {Modigliani}, {Moehler}, {M{\o}ller},
  {Norup S{\o}rensen}, {N{\o}rregaard}, {P{\'e}roux}, {Patat}, {Pena}, {Pragt},
  {Reinero}, {Rigal}, {Riva}, {Roelfsema}, {Royer}, {Sacco}, {Santin},
  {Schoenmaker}, {Spano}, {Sweers}, {Ter Horst}, {Tintori}, {Tromp}, {van
  Dael}, {van der Vliet}, {Venema}, {Vidali}, {Vinther}, {Vola}, {Winters},
  {Wistisen}, {Wulterkens}, \& {Zacchei}}]{2011A&A...536A.105V}
{Vernet}, J., {Dekker}, H., {D'Odorico}, S., {et~al.} 2011, \aap, 536, A105

\bibitem[{{V{\'e}ron-Cetty} {et~al.}(2006){V{\'e}ron-Cetty}, {Joly},
  {V{\'e}ron}, {Boroson}, {Lipari}, \& {Ogle}}]{Veron_Cetty2006}
{V{\'e}ron-Cetty}, M.~P., {Joly}, M., {V{\'e}ron}, P., {et~al.} 2006, \aap,
  451, 851

\bibitem[{{Wang} \& {Bon}(2020)}]{2020A&A...643L...9W}
{Wang}, J.-M. \& {Bon}, E. 2020, \aap, 643, L9

\bibitem[{{Wang} {et~al.}(2012){Wang}, {Zhou}, {Komossa}, {Wang}, {Yuan}, \&
  {Yang}}]{Wang_2012}
{Wang}, T.-G., {Zhou}, H.-Y., {Komossa}, S., {et~al.} 2012, \apj, 749, 115

\bibitem[{{Wevers} {et~al.}(2019){Wevers}, {Pasham}, {van Velzen}, {Leloudas},
  {Schulze}, {Miller-Jones}, {Jonker}, {Gromadzki}, {Kankare}, {Hodgkin},
  {Wyrzykowski}, {Kostrzewa-Rutkowska}, {Moran}, {Berton}, {Maguire}, {Onori},
  {Mattila}, \& {Nicholl}}]{Wevers_2019}
{Wevers}, T., {Pasham}, D.~R., {van Velzen}, S., {et~al.} 2019, \mnras, 488,
  4816

\bibitem[{{Willott}(2005)}]{2005ApJ...627L.101W}
{Willott}, C.~J. 2005, \apjl, 627, L101

\bibitem[{{Wright} {et~al.}(2010){Wright}, {Eisenhardt}, {Mainzer}, {Ressler},
  {Cutri}, {Jarrett}, {Kirkpatrick}, {Padgett}, {McMillan}, {Skrutskie},
  {Stanford}, {Cohen}, {Walker}, {Mather}, {Leisawitz}, {Gautier}, {McLean},
  {Benford}, {Lonsdale}, {Blain}, {Mendez}, {Irace}, {Duval}, {Liu}, {Royer},
  {Heinrichsen}, {Howard}, {Shannon}, {Kendall}, {Walsh}, {Larsen}, {Cardon},
  {Schick}, {Schwalm}, {Abid}, {Fabinsky}, {Naes}, \&
  {Tsai}}]{2010AJ....140.1868W}
{Wright}, E.~L., {Eisenhardt}, P. R.~M., {Mainzer}, A.~K., {et~al.} 2010, \aj,
  140, 1868

\bibitem[{{Wyrzykowski} {et~al.}(2017){Wyrzykowski}, {Zieli{\'n}ski},
  {Kostrzewa-Rutkowska}, {Hamanowicz}, {Jonker}, {Arcavi}, {Guillochon},
  {Brown}, {Koz{\l}owski}, {Udalski}, {Szyma{\'n}ski}, {Soszy{\'n}ski},
  {Poleski}, {Pietrukowicz}, {Skowron}, {Mr{\'o}z}, {Ulaczyk}, {Pawlak},
  {Rybicki}, {Greiner}, {Kr{\"u}hler}, {Bolmer}, {Smartt}, {Maguire}, \&
  {Smith}}]{2017MNRAS.465L.114W}
{Wyrzykowski}, {\L}., {Zieli{\'n}ski}, M., {Kostrzewa-Rutkowska}, Z., {et~al.}
  2017, \mnras, 465, L114

\bibitem[{{Xiao} {et~al.}(2011){Xiao}, {Barth}, {Greene}, {Ho}, {Bentz},
  {Ludwig}, \& {Jiang}}]{2011ApJ...739...28X}
{Xiao}, T., {Barth}, A.~J., {Greene}, J.~E., {et~al.} 2011, \apj, 739, 28

\bibitem[{{Yang} {et~al.}(2018){Yang}, {Wu}, {Fan}, {Jiang}, {McGreer},
  {Shangguan}, {Yao}, {Wang}, {Joshi}, {Green}, {Wang}, {Feng}, {Fu}, {Yang},
  \& {Liu}}]{Yang_2018}
{Yang}, Q., {Wu}, X.-B., {Fan}, X., {et~al.} 2018, \apj, 862, 109

\bibitem[{{Yaron} \& {Gal-Yam}(2012)}]{Yaron_2012}
{Yaron}, O. \& {Gal-Yam}, A. 2012, \pasp, 124, 668

\bibitem[{{Zabludoff} {et~al.}(2021){Zabludoff}, {Arcavi}, {La Massa},
  {Perets}, {Trakhtenbrot}, {Zauderer}, {Auchettl}, {Dai}, {French}, {Hung},
  {Kara}, {Lodato}, {Maksym}, {Qin}, {Ramirez-Ruiz}, {Roth}, {Runnoe}, \&
  {Wevers}}]{2021SSRv..217...54Z}
{Zabludoff}, A., {Arcavi}, I., {La Massa}, S., {et~al.} 2021, \ssr, 217, 54

\bibitem[{{Zhang} {et~al.}(2022){Zhang}, {Shu}, {Sheng}, {Sun}, {Dou}, {Jiang},
  {Wang}, {Hu}, {Wang}, \& {Wang}}]{2022A&A...660A.119Z}
{Zhang}, W.~J., {Shu}, X.~W., {Sheng}, Z.~F., {et~al.} 2022, \aap, 660, A119

\bibitem[{{Zhang}(2021)}]{Zhang_2021}
{Zhang}, X. 2021, \apj, 919, 13

\bibitem[{{Zhou} {et~al.}(2006){Zhou}, {Wang}, {Yuan}, {Lu}, {Dong}, {Wang}, \&
  {Lu}}]{Zhou_2006}
{Zhou}, H., {Wang}, T., {Yuan}, W., {et~al.} 2006, \apjs, 166, 128

\end{thebibliography}

\end{document}